\documentclass{emulateapj}
\usepackage{graphics,graphicx,rotating}

\newcommand{\simgt}{\lower.5ex\hbox{$\; \buildrel > \over \sim \;$}}
\newcommand{\simlt}{\lower.5ex\hbox{$\; \buildrel < \over \sim \;$}}

\def\btheta{\mbox{\boldmath $\theta$}}

\shorttitle{Combining Lens Distortion and Depletion}
\shortauthors{Umetsu \& Broadhurst}

\begin{document}


\title{
Combining Lens Distortion and Depletion to Map the
Mass Distribution of A1689\altaffilmark{1}
}


\author{Keiichi Umetsu\altaffilmark{2,3}}
\email{keiichi@asiaa.sinica.edu.tw}
%\and
\author{Tom Broadhurst\altaffilmark{4}}
\email{tjb@wise.tau.ac.il}


\altaffiltext{1}{Based in part on data collected at the Subaru Telescope,
  which is operated by the National Astronomical Society of Japan}
\altaffiltext{2}
 {Institute of Astronomy and Astrophysics, Academia Sinica,  P.~O. Box
 23-141, Taipei 106,  Taiwan}
\altaffiltext{3}
 {Leung center for Cosmology and Particle Astrophysics,
  National Taiwan University, Taipei 106, Taiwan}
\altaffiltext{4}
 {
School of Physics and Astronomy, Tel Aviv University, Israel
}


\begin{abstract}
We derive a projected 2D mass map of the well studied galaxy cluster
A1689 based on an entropy-regularized maximum-likelihood combination
of the lens magnification and distortion of red background galaxies
registered in deep Subaru images. The method is not restricted to the
weak regime but applies to the whole area outside the tangential
critical curve, where non-linearity between the surface mass-density
and the observables extends to a radius of a few arcminutes. The known
strong lensing information is also readily incorporated in this
approach, represented as a central pixel with a mean surface density
close to the critical value. We also utilize the distortion
measurements to locally downweight the intrinsic clustering noise,
which otherwise perturbs the depletion signal. The resulting 2D map
shows that the surface density of A1689 is smoothly varying and
symmetric, similar to the distribution of cluster members, with no
apparent substructure at $r \simgt 130 {\rm kpc}/h$ ($\sim 1\arcmin$).
The projected mass profile continuously steepens with radius and is well
fitted by the Navarro-Frenk-White model, but with a surprising large
concentration $c_{\rm vir}=13.4^{+5.3}_{-3.3}$, lying far from the
predicted value of $c_{\rm vir}\sim 5$, corresponding to the measured
virial mass, $M_{\rm vir}=(2.1\pm 0.2)\times 10^{15}M_{\odot}$, 
posing a challenge to the standard assumptions defining the $\Lambda$CDM
model. We examine the consistency of our results with estimates
derived with the standard weak lensing estimators and by
comparison with the inner mass profile obtained from strong lensing.
All the reconstructions tested here imply a virial mass in the range,
$M_{\rm vir}=(1.5-2.1)\times 10^{15}M_{\odot}$, and the combined ACS
and Subaru-2D mass reconstruction yields a tight constraint on the
concentration parameter, $c_{\rm vir}=12.7\pm 1\pm2.8$
($c_{200}\sim 10$), improving upon the statistical accuracy of our
 earlier 1D analysis. 
Importantly, our best fitting profile properly reproduces
the observed Einstein radius of $45''$ ($z_s=1$), in contrast to
other weak lensing work, reporting lower concentration profiles,
which underestimate the observed Einstein radius.
\end{abstract}


%% subject header.  Objects should be in the appropriate "individual"
%% headers (e.g. quasars: individual, stars: individual, etc.) with the
%% additional provision that the total number of headers, including each
%% individual object, not exceed six.  The \objectname{} macro, and its
%% alias \object{}, is used to mark each object.  The macro takes the object
%% name as its primary argument.  This name will appear in the paper
%% and serve as the link's anchor in the electronic edition if the name
%% is recognized by the data centers.  The macro also takes an optional
%% argument in parentheses in cases where the data center identification
%% differs from what is to be printed in the paper.

%\keywords{globular clusters: general ---
%globular clusters: individual(\objectname{NGC 6397},
%\object{NGC 6624}, \objectname[M 15]{NGC 7078},
%\object[Cl 1938-341]{Terzan 8})}

%% Modified by KU (2007/07/21) 
\keywords{cosmology: observations --- dark matter --- galaxies: clusters:
individual (A1689) --- gravitational lensing}


\section{Introduction}

Weak gravitational lensing of background galaxies provides a unique,
direct way to study the mass distribution of galaxy clusters
(Bartelmann \& Schneider 2001) via the systematic shape distortion of
background galaxies (Tyson, Wenk, \& Valdes 1990; Kaiser \& Squires
1993; Schneider \& Seitz 1996; Umetsu, Tada, \& Futamase 1999) and
also to a lesser extent by the magnification of the background
(Broadhurst, Taylor, \& Peacock 1995; Taylor et al. 1998). We have
examined both these effects in our earlier work on A1689 (Broadhurst
et al. 2005a, hereafter B05a), where we found good consistency between
the magnitude of the radial depletion of background red galaxies
caused by lens magnification and the weak lensing distortion profile
of the same background galaxy population, which we then combined to
derive an improved mass profile.

A limitation of the magnification technique is the intrinsic
clustering of the background, which for red galaxies is mainly in the
form of localized groups of modest angular size in the background
field. In principle, with redshift information sharp overdensities can
be isolated in redshift and down weighted, or with sufficiently large
number of redshifts, the shift of the magnified luminosity function
can be utilized independently of density fluctuations 
(Broadhurst et al. 1995; Zhang \& Pen 2005). 
The combination of all lensing related effects is of course
desirable, leading to the derivation of the highest precision feasible
when constructing mass maps and density profiles. Furthermore, because
lensing effects depend on distance, the cosmological redshift
distance relation may be constrained via the geometric scaling of the lensing
signal with redshift (e.g., Taylor et al. 2007; Medezinski et
al. 2007).

Advances in the quality of imaging encourage a closer examination
of the empirical effects of lensing and the development of more
comprehensive techniques to extract reliable high resolution
information. From space, deep multi-color images of massive clusters
can be used to identify many sets of multiple images per cluster
(Broadhurst et al. 2005b, hereafter B05b; Gavazzi et al. 2003; Kneib et
al. 2004; Sand et al. 2004; Smith et al. 2005; 
Brada${\check {\rm c}}$ et al. 2006).  
From the ground the stable prime-focus wide-field cameras
of Subaru and CFHT are producing data of sufficient quality to examine
weak lensing distortions over a wide range of radius. 
%(e.g.,van Waerbeke et al. 2000; Miyazaki et al. 2002; Sato et al. 2003;
%Hamana et al. 2003; Gavazzi et al. 2004; Hoekstra et al. 2006; Mahdavi
%et al. 2007; Medezinski et al. 2007; Okabe \& Umetsu 2008).
%%% 
More recently, wide-field near IR cameras, such as MOIRCS on Subaru
WIRCAM on CFHT, OMEGA2000 on
Calar-Alto and WFCAM on UKIRT, may help improve the accuracy of
photometric redshifts for many faint galaxies.  However, it is still
the case that no set of deep high quality wide field images exists for
any massive cluster with full optical-IR coverage, despite all the
progress of field surveys.

A further motivation for pursuing accurate lensing maps is the
increased precision of model predictions for statistical properties
of cluster-sized mass halos in the standard Lambda cold dark matter
($\Lambda$CDM) model.  
Many of the free parameters of this model now rest on a firm empirical
foundation with relatively tight constraints on the index and
normalization of the power spectrum of density perturbation and the
background cosmological model (e.g., Spergel et al. 2003; Tegmark et
al. 2004; Spergel et al. 2007).
%%%
In this context $N$-body simulations have become impressively
comprehensive, in particular the recent Millennium simulation
(Springel et al. 2005) which simulates a huge volume of $500 {\rm
Mpc}/h$, and has been used to predict the mass function and evolution
of nearly 100,000 group and cluster sized CDM halos.
%%%%
This model is tightly defined and hence amenable to comparisons with
the real Universe, particularly for the case of clusters where baryons,
which are usually omitted from large scale simulation, are not expected 
to have a significant impact on the shape of gravitational potential of 
a cluster since the high temperature of the cluster gas prevents
efficient cooling, and hence the majority of baryons simply trace 
the gravitational potential of the dominant dark matter.
%This model is tightly proscribed and is amenable to comparisons with
%the real Universe, particularly for the case of clusters where baryons
%which are absent from this simulation are not expected to have a
%significant impact on the gravitational potential of a cluster due to
%the high temperature of cluster gas preventing efficient cooling, so
%that the majority of baryons simply trace the gravitational potential
%of the dominant dark matter.
%%%%%
Accurate $N$-body simulations based on the $\Lambda$CDM scenario predict
a relatively shallow, low-concentration mass profile for massive cluster
halos, where the logarithmic gradient flattens continuously toward the
center of mass (Navarro, Frenk, \& White 1997, hereafter NFW)
with a flatter central slope than a purely isothermal
body interior to the inner characteristic radius,
$r_s \simlt  100-200 {\rm kpc}/h$.
%%%
A useful index of the degree of concentration, $c_{\rm vir}$,
compares the virial radius, $r_{\rm vir}$, to 
$r_s$ of the NFW profile, $c_{\rm vir} \equiv r_{\rm vir}/r_s$.
%Massive clusters are of particular interest in the context of this
%model, because they are predicted to
%have a relatively shallow mass profile (or low concentration) 
%fitted described by the form proposed by Navarro, Frenk, \& White
%(1996). 
This prediction for the CDM halo $c_{\rm vir}$--$M_{\rm vir}$ relation 
has been established thoroughly with high resolution
simulations (e.g., Navarro et al. 1997; Bullock et al. 2001; Wechsler et  
al. 2002; Neto et al. 2007)
with some intrinsic variation related to the individual assembly
history of a cluster (e.g., Jing \& Suto 2000; Tasitsiomi et al. 2004;
Hennawi et al. 2007).
%%%
In particular, the detailed $N$-body millennium simulation (Springel et
al. 2005) predicts a simple relationship between the halo mass and 
concentration parameter for halo masses in the range of galaxy groups
to massive clusters, as quantified by Neto et al. (2007),
who found that the expected median value for cluster sized halos
of $M_{\rm vir}\sim 10^{15}M_\odot$ to be $c_{\rm vir}\sim 5$ at $z=0$
($c_{200}\sim 4$), 
with a spread of the order of $\Delta\log_{10}c_{\rm vir}=0.1$.
(see Johnston et al. 2007 for a good summary of the state of art in halo
concentrations based on Wechsler et al. 2006 and Neto et al. 2007).

In this paper we explore further methods designed to achieve the
maximum possible lensing precision by combining all lensing
information for A1689. This cluster is among the most massive clusters
with the largest known Einstein radius ($\sim 50\arcsec$), and is one of
the best studied clusters for lensing work (Tyson et al. 1990; 
 Tyson \& Fisher 1995; Taylor et al. 1998; King, Clowe, \& Schneider
 2002; Bardeaul et al. 2005; B05a; B05b; Oguri et al. 2005; Halkola,
 Seitz, \& Pannella 2006; Bardeau et al. 2007; Limousin et al. 2007;
 Medezinski et al. 2007;  Umetsu, Takada, \& Broadhurst 2007;
 Okura, Umetsu, \& Futamase 2008), located at a moderately low redshift 
of $z=0.183$. 
%%%
In B05a we developed a 
%{\bf maximum likelihood method for} 
``model-independent'' method\footnote{We remind the reader that model 
dependence is unavoidable to some extent in scientific analysis.
In this work we define the term ``model independent'' to refer to
those methods without prior assumptions about the functional form 
of the lensing profiles and distributions.} 
for reconstructing the cluster mass profile using azimuthally-averaged
weak-lensing shape distortion and magnification bias measurements, in
the wide-field, Subaru images. This together with many multiple images
identified in deep {\it Hubble Space Telescope} (HST) Advanced Camera
for Surveys (ACS) imaging defined a detailed lensing based
cluster mass profile out to the cluster virial radius ($r\simlt 2
h^{-1}$ Mpc). The combined strong and weak lensing mass profile is
well fitted by an NFW profile (Navarro et al. 1997) with
high concentration of $c_{\rm vir}\sim 13.7$, which is significantly
larger than theoretically expected ($c_{\rm vir}\sim 5$) for the
standard $\Lambda$CDM model (Bullock et al. 2001; Neto et al. 2007),
although the degree of concentration is still controversial (B05a;
Medezinski et al. 2007; Limousin et al. 2007).
%%%
Such a high concentration is also seen in other massive clusters
from careful lensing work, such as MS 2137-23 ($c_{200}\simeq 12$,
Gavazzi et al. 2003) and  CL0024+1654  ($c_{200}\simeq 22$, Kneib et
al. 2003).  These results could raise serious questions regarding the
basic assumptions behind the $\Lambda$CDM model. If clusters collapse
earlier than predicted then it is expected that denser and hence more
concentrated halos will develop in the context of CDM (Wechsler et
al. 2002). 
%%%
On the other hand, it has been argued that part of this discrepancy from
lensing observations could be reconciled by observational effects such
as triaxiality of CDM halos (Oguri et al. 2005;
Hennawi et al. 2007; Sereno 2007; Corless \& King 2007), and the
projection of structure along the line of sight (e.g., King \& Corless
2007), 
both of which boost the projected surface mass density and hence the
lensing  signal. Such observational biases in the lensing-based
concentration parameter have been explored in details by
Hennawi et al. (2007) on the basis of $N$-body simulations, 
indicating a positive bias of $\sim 30\%$ in the halo concentration
derived from 2D lensing measurements. Although A1689 is a very round
shaped cluster with evidence of only modest substructure (Teague,
Carter, \& Grey 1990; Girardi et al. 1997; Andersson \& Madejski 2004;
 Czoske 2004; B05b), such a chance alignment of structure could be a
 potential source of high concentrations.
%%%
Furthermore, for a reliable measurement of the cluster mass profile, 
systematic errors inherent in the lensing measurements, such
as the uncertainty in the background redshift distribution
and the dilution effect on the lensing signal due to contamination by
cluster  members (B05a; Medezinski et al. 2007), need to be taken into
account.

The paper is organized as follows.  
We briefly summarize in \S 2 the basis of cluster weak lensing.
%In \S 3 we describe the observational data, background sample selection,
%and joint weak lensing analysis of shape distortion and magnification
%bias data. 
In \S 3 we describe the observational data and 
the background sample selection for the weak
lensing analysis;
%%%
we then 
%%% outline
summarize
our joint weak lensing analysis of shape distortion and magnification
bias data.
%A combind weak lensing shape analysis of A1689 is presented in \S 3.
%In \S 4 we describe our magnification bias measurements in A1689.
%%%
In \S 4 we present a method for reconstructing the two-dimensional mass
distribution of A1689 from combined weak lensing shape distortion and
magnification bias measurements.
In \S 5 we derive mass profiles of A1689 from weak lensing data using
three different methods, and compare resulting mass profiles; we also
combine our weak lensing mass profiles with strong lensing constraints
from previous studies to test the CDM paradigm; 
then, we assess carefully various
sources of potential systematic error in the halo concentration
parameter derived from the lensing observations.
Finally, summary and
discussions are given in \S 6.

%%%

Throughout this paper, we use the AB magnitude system, and adopt a
concordance $\Lambda$CDM cosmology with ($\Omega_{\rm m0}=0.3$,
$\Omega_{\Lambda 0}=0.7$, $h=0.7$).
%%%
In this cosmology one arcminute corresponds to the physical scale
$129$kpc$/h$ for this cluster.  The reference center of our analysis
is fixed at the center of the cD galaxy: ${\rm RA} = 13:11:29.52, {\rm
Dec} = -01:20:27.59$ (J2000.0).

\section{Cluster Weak Lensing}
\label{sec:wl}

Weak gravitational lensing is responsible for the weak shape-distortion
and magnification of the images of background sources due to the
gravitational field of intervening foreground clusters of galaxies
and large scale structures in the universe.
The deformation of the image can be 
described by the $2\times 2$
Jacobian matrix $\cal{A}_{\alpha\beta}$
($\alpha,\beta=1,2$)
of the lens mapping.
The Jacobian ${\cal A}_{\alpha\beta}$ is real and symmetric, so that
it can be decomposed as
\begin{eqnarray}
\label{eq:jacob}
{\cal A}_{\alpha\beta} &=& (1-\kappa)\delta_{\alpha\beta}
 -\Gamma_{\alpha\beta},\\
\Gamma_{\alpha\beta}&=&
\left( 
\begin{array}{cc} 
+{\gamma}_1   & {\gamma}_2 \\
 {\gamma}_2  & -{\gamma}_1 
\end{array} 
\right),
\end{eqnarray}
where 
$\delta_{\alpha\beta}$ is Kronecker's delta,
$\Gamma_{\alpha\beta}$ is the trace-free, symmetric shear matrix
%defined as $\Gamma_{\alpha\beta}=\gamma_1\sigma_3+\gamma_2\sigma1$
with $\gamma_{\alpha}$ being the components of 
spin-2
complex gravitational
shear $\gamma:=\gamma_1+i\gamma_2$,
describing the anisotropic shape distortion,
%and $\sigma_{\alpha}$ being the $2\times 2$ Pauli matrices,
and $\kappa$ is the 
%$\kappa(\btheta)=\int\! d\Sigma\,\Sigma_{\rm crit}^{-1}$ is the
lensing convergence responsible for the 
trace-part of the Jacobian matrix, describing the isotropic area
distortion.
In the weak lensing limit where $\kappa,|\gamma|\ll 1$, 
$\Gamma_{\alpha\beta}$ induces a quadrupole anisotropy of the 
background image, which can be observed from ellipticities 
of background galaxy images.
%%%%
The flux  magnification due to gravitational lensing
is given by the inverse Jacobian
determinant,
\begin{equation}
\label{eq:mu}
\mu = 
\frac{1}{{\rm det}{\cal A}}
=
\frac{1}{(1-\kappa)^2-|\gamma|^2},
\end{equation}
where we assume subcritical lensing, i.e., 
${\rm det}{\cal A}(\btheta)>0$.

The lensing convergence is expressed as a line-of-sight projection
of the matter density contrast out to the source plane (${\rm S}$)
weighted by certain combination $g$ of co-moving angular diameter distances
(e.g., Jain et al. 2000),
\begin{equation} 
\label{eq:kappa}
\kappa =  
\frac{3H_0^2\Omega_m}{2c^2}
\int_0^{\chi_{\rm S}}\!d\chi\, g(\chi,\chi_{\rm
 S})\frac{\delta}{a} 
\equiv \int\!d\Sigma_m\,\Sigma_{\rm crit}^{-1}
\end{equation}
%$\kappa(\btheta)=\int\! d\Sigma\,\Sigma_{\rm crit}^{-1}$,
where $a$ is the cosmic scale factor, $\chi$ is the co-moving distance;
$\Sigma_m$ is the surface mass density of matter, $\Sigma_m
=\int_0^{\chi_{\rm S}}\!d\chi \, a(\rho_m-\bar{\rho})$,
with respect to the cosmic mean density $\bar{\rho}$, and
%where 
$\Sigma_{\rm crit}$ 
is the critical surface mass density for gravitational lensing,
\begin{equation}
\label{eq:sigmacrit}
\Sigma_{\rm crit} = \frac{c^2}{4\pi G}\frac{D_{s}}{D_d D_{ds}}
\end{equation}
with $D_s$, $D_d$, and $D_{ds}$ being the angular diameter distances
from the observer to the source, from the observer to the deflecting
lens, and from the lens to the source, respectively.
For a fixed background cosmology and a lens redshift $z_d$,
$\Sigma_{\rm crit}$ is a function of background source redshift
$z_s$.
%In the following we closely follow the standard notation of 
%Bartelmann \& Schneider (2001).
For a given mass distribution $\Sigma(\btheta)$, the lensing signal is
proportional to the angular diameter distance ratio,
$D_{ds}/D_s$.
%\begin{equation} 
%\label{eq:dratio}
%D(z_s) \equiv D_{ds}/D_s.
%\end{equation}

In the present weak lensing study we aim to reconstruct
the dimensionless surface mass density $\kappa$ 
from weak lensing distortion and magnification data.
To do this, we utilize the relation between the gradients of 
$\kappa$ and $\gamma$ (Kaiser 1995; Crittenden et al. 2002),
\begin{equation}
\label{eq:local}
\triangle \kappa (\btheta)
= \partial^{\alpha}\partial^{\beta}\Gamma_{\alpha\beta}(\btheta)
= 2\hat{\cal D}^*\gamma(\btheta)
\end{equation}
where 
$\hat{\cal D}$ is the complex differential operator
$\hat{\cal D}=(\partial_1^2-\partial_2^2)/2+i\partial_1\partial_2$.
The Green's function for the two-dimensional Poisson equation is
$\triangle^{-1}(\btheta,\btheta')=\ln|\btheta-\btheta'|/(2\pi)$,
so that equation (\ref{eq:local}) can be solved to yield the following
non-local relation between $\kappa$ and $\gamma$ (Kaiser \& Squires 1993):
\begin{equation}
\label{eq:gamma2kappa}
\kappa(\btheta) = 
\frac{1}{\pi}\int\!d^2\theta'\,D^*(\btheta-\btheta')\gamma(\btheta')
\end{equation}
where $D(\btheta)$ is the complex kernel defined as 
\begin{equation}
\label{eq:kerneld}
D(\btheta) = \frac{\theta_2^2-\theta_1^2-2i\theta_1\theta_2}{|\theta|^4}.
\end{equation}
Similarly, the spin-2 shear field can be expressed in terms of the
lensing convergence as
\begin{equation} 
\label{eq:kappa2gamma}
\gamma(\btheta) = 
\frac{1}{\pi}\int\!d^2\theta'\,D(\btheta-\btheta')\kappa(\btheta').
\end{equation} 
Note that adding a constant mass sheet to $\kappa$ in equation
(\ref{eq:kappa2gamma}) does not change the 
shear field
$\gamma(\btheta)$ which is observable in the weak lensing limit, 
leading to the so-called {\it mass-sheet degeneracy}
based solely on shape-distortion measurements (e.g., Bartelmann \&
Schneider 2001; Umetsu et al. 1999).
In general, the observable quantity is not the 
gravitational shear $\gamma$ but the {\it reduced} shear,
\begin{equation}
\label{eq:redshear}
g=\frac{\gamma}{1-\kappa}
\end{equation}
in the subcritical regime where ${\rm det}{\cal A}>0$
(or $1/g^*$ in the negative parity region with ${\rm det}{\cal A}<0$). 
We see that the reduced shear $g$ is invariant under the following
global transformation:
\begin{equation}
\label{eq:invtrans}
\kappa(\btheta) \to \lambda \kappa(\btheta) + 1-\lambda, \ \ \ 
\gamma(\btheta) \to \lambda \gamma(\btheta)
\end{equation}
with an arbitrary scalar constant $\lambda\ne 0$ 
(Schneider \& Seitz 1995). This transformation is equivalent to scaling 
the Jacobian matrix ${\cal A}(\btheta)$ with $\lambda$, 
$\cal {A}(\btheta) \to \lambda {\cal
A}(\btheta)$. This mass-sheet degeneracy can be unambiguously broken
by measuring the magnification effects, because the magnification $\mu$
transforms under the invariance transformation (\ref{eq:invtrans}) as
\begin{equation}
\mu(\btheta) \to \lambda^2 \mu(\btheta).
\end{equation}

\section{Data Analysis}
\label{sec:analysis}

In this section we present a full technical description of our weak
lensing distortion and magnification analyses on A1689 based on the
Subaru images, which were analysed in our earlier work of B05a and
Medezinski  et al. (2007).
This work is also based on the same weak lensing
shape and magnification measurements as used in
B05a. We note that Medezinski et al. (2007) used a slightly different
analysis pipeline so as to include the shape measurements of bright
cluster members and noisier objects (as well as better resolved red
background galaxies), optimizing for the weak lensing dilution analysis
including the measurements of cluster light and cluster luminosity
function where the completeness is crucial; Medezinski et al. (2007)
included a blue background population in the weak lensing shape
analysis, in addition to the red background population which B05a and
this work are based on; The magnification information, on the other
hand,  was not taken into account in Medezinski et al. (2007).

In B05a we simply assumed that the mean redshift of the red background
galaxies is $z_s=1$.
% based on deep photo-$z$ estimation for deep field
%data ($\bar z_s\simeq 1\pm 0.1$, Benitez et al. 2002).
In the present work we improve the accuracy of determination of the
cluster mass and concentration parameters, by taking into account the
redshift distribution of 
red background galaxies 
examined by Medezinski et al. (2007)
based on the multicolor photometry of Capak et al. (2004) in
the HDF-N (see \S \ref{subsec:red}).

\subsection{Subaru Data and Photometry}
\label{subsec:data}

For our weak lensing analysis
we used Subaru/Suprime-Cam imaging data of A1689 in $V$ 
%(1,920s) 
and SDSS $i'$ 
%(2,640s) 
retrieved from the Subaru archive,
SMOKA (see B05a and Medezinski et al. 2007 for more details).
The FWHM in the co-added mosaic image is 
$0\arcsec\!\!.82$ in $V$ and 
$0\arcsec\!\!.88$ in $i'$ with $0\arcsec\!\!.202$
pix$^{-1}$, covering a field of $\approx 30'\times 25'$.
%%%%

Photometry is based on a combined $V+i'$ image with SExtractor
(Bertin \& Arnouts 1996), where the $i'$ image is used as
the source detection image.
We adopt the following key configuration parameters 
in SExtractor:
${\it DETECT\_MINAREA} = 5$,
${\it DETECT\_THRESH}={\it ANALYSIS\_THRESH}=3$.
The limiting magnitudes are obtained as
$V=26.5$ and $i'=26.0$ for a $3\sigma$ detection within a $2$\arcsec
aperture.
A careful background selection is critical for a weak lensing analysis 
so that unlensed cluster members and
foreground galaxies to not dilute the true lensing signal of the background
(see B05a and Medezinski et al. 2007).
%%%
We identify an E/S0 sequence of cluster galaxies in the 
%%% @@ edited by KU (2008/02/18)
%% color-magnitude (CM) diagram, which can be defined by the linear
color-magnitude (CM) diagram, which can be defined by the linear
CM relation:
$(V-i')_{E/S}=-0.02094i'+1.255$ (B05a;
see Figure 1 of Medezinski et al. 2007 for the CM diagram).
%% We note that Medezinski et al. (2007) 
%% used slightly different values of the CM parameters for A1689.
For the number counts to measure lensing magnification, 
we define a sample of galaxies that are redder than the
cluster sequence by 
$(V-i')>1.0$ and $20<i'<25.5$,
which yields a total of $N_\mu= 8,907$ galaxies,
or the mean surface number density of $\bar{n}_{\mu}=12.0$ arcmin$^{-2}$.
%%%
For the magnification bias analysis, a conservative
magnitude limit of $i'<25.5$
is adopted to avoid incompleteness.

%%%%%%%%%%%%%%%%%%%%%%%%%%%%%%%%%%%%%%%%%%%%%%%%%%%%%%%%%%%%%%%%%%%%%%%%%

\subsection{Weak Lensing Distortion Analysis}
\label{subsec:shear}

We use the IMCAT package developed by N. Kaiser
\footnote{http://www.ifa.hawaii/kaiser/IMCAT} to perform object
detection, photometry and shape measurements, following the formalism
outlined in  Kaiser, Squires, \& Broadhurst (1995; hereafter KSB).
We have modified the method somewhat following the procedures
described in Erben et al. (2001). We used the same analysis pipeline
as in 
    B05a,
    Umetsu et al. (2007), 
and Okabe \& Umetsu (2008).

\subsubsection{Object Detection}
\label{subsubsec:detection}

Objects are first detected as local peaks in the image by 
using the IMCAT hierarchical peak-finding
algorithm {\it hfindpeaks}
which for each object yields object parameters such as a peak position,
an estimate of the object size ($r_g$), the significance of
the peak detection ($\nu$).
%%%
The local sky level and its gradient are 
measured around each  object from the mode of pixel values 
on a circular annulus defined by
inner and outer radii of 
$16\times r_g$ and $32\times r_g$ (see Clowe et al. 2000).
In order to avoid contamination in the background estimation
by bright neighboring stars and/or
foreground galaxies,
all pixels within $3\times r_g$ of another object are excluded from the
mode calculation. Total fluxes and half-light radii ($r_h$) are then
 measured on sky-subtracted images using a circular aperture of
 radius $3\sqrt{2}\times r_g$ from the object center. Any pixels within
 $2.5\times r_g$ of another object are excluded from the aperture.
The aperture $i'$-magnitude is then calculated from the measured total
flux and a zero-point magnitude.
Any objects with positional  differences between the
peak location and the weighted-centroid 
greater than $d=0.4$ pixels are excluded from
the catalog.

Finally, bad objects such as spikes, saturated stars, and noisy
detections 
%and groups
%of objects detested as a single object 
must be removed from the weak lensing object
catalog. 
%%%
We removed from our object catalog 
(1) extremely large objects with $r_g>10$ pixels, 
(2) objects with low detection significance, $\nu<7$,
(3) objects with large raw ellipticities, $|e|>0.5$, 
(4) noisy detections with unphysical negative fluxes, and
(5) objects containing more than $10$ bad pixels, ${\it nbad}>10$.
%%%%
This selection procedure yields an object catalog with
$N=62,384$ ($82.6$ arcmin$^-2$).

\subsubsection{Weak Lensing Distortion Measurements}
\label{subsubsec:shape}

To obtain an estimate of the reduced shear,
$g_{\alpha}=\gamma_{\alpha}/(1-\kappa)$ ($\alpha=1,2$), 
we measure the image ellipticity 
$e_{\alpha} = \left\{Q_{11}-Q_{22}, Q_{12} \right\}/(Q_{11}+Q_{22})$ 
from the weighted quadrupole moments of the surface brightness of
individual galaxies defined in the above catalog,
\begin{equation}
Q_{\alpha\beta} = \int\!d^2\theta\,
 W({\theta})\theta_{\alpha}\theta_{\beta} 
I({\btheta})
\ \ \ (\alpha,\beta=1,2)
\end{equation} 
where $I(\btheta)$ is the surface brightness distribution of an object,
$W(\theta)$ is a Gaussian window function matched to the size of the
object.

\begin{figure*}[!htb]
 \begin{center}
  \includegraphics[width=60mm, angle=270]{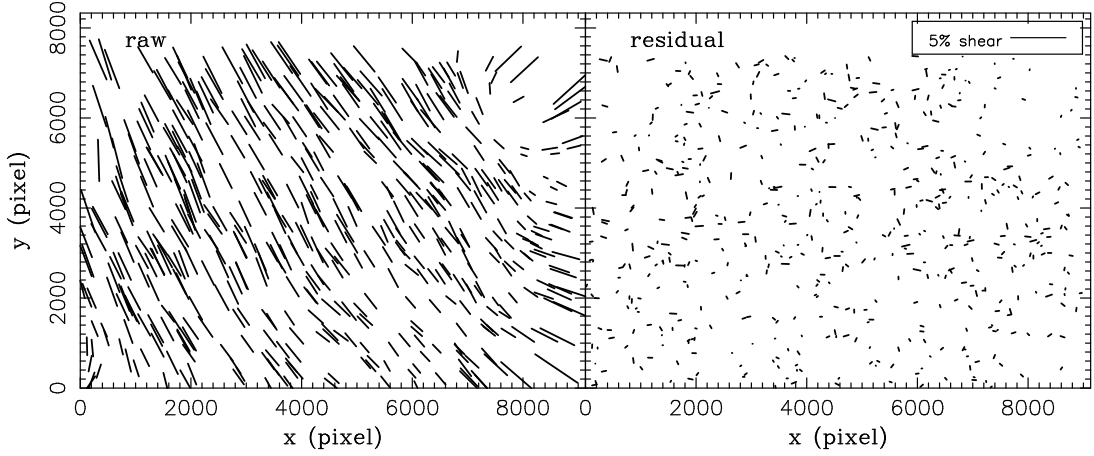}
 \end{center}
\caption{
The quadrupole PSF anisotropy field as measured from
stellar ellipticities before and after the PSF anisotropy correction.
The left panel shows the raw ellipticity field of stellar objects,
and the right panel shows the residual ellipticity field after
the PSF anisotropy correction.
The orientation of the sticks indicates the position angle of
the major axis of stellar ellipticity, whereas the length is
 proportional to the modulus of stellar ellipticity. A stick with the
 length of $5\%$ ellipticity is indicated in the top right of the right
 panel. 
}
\label{fig:anisopsf1}
\end{figure*}

\begin{figure*}[!htb]
 \begin{center}
  \includegraphics[width=60mm, angle=270]{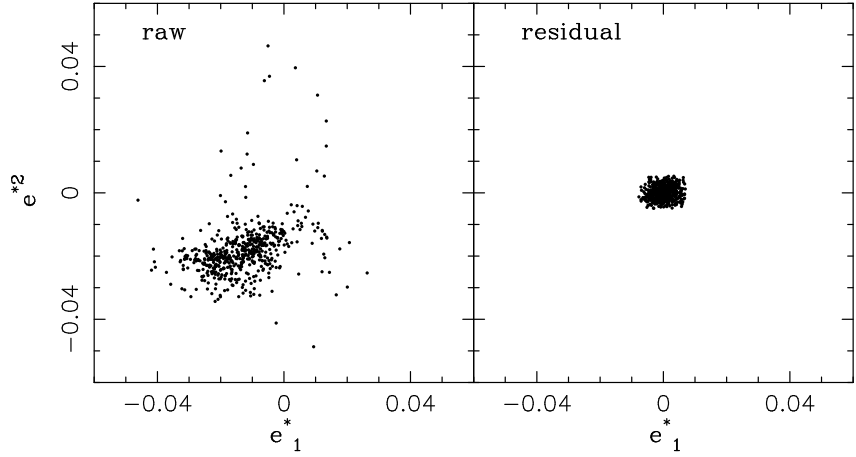}
 \end{center}
\caption{
Stellar ellipticity distributions before and after the PSF anisotropy 
correction. The left panel shows the raw ellipticity components 
$(e_1^*,e_2^*)$ of stellar objects, and the right panel shows
the residual ellipticity components $(\delta e_1^*, \delta e_2^*)$
after the PSF anisotropy correction.
}
\label{fig:anisopsf2}
\end{figure*}

Firstly the PSF anisotropy needs to be corrected using the star images
as references:
\begin{equation}
e'_{\alpha} = e_{\alpha} - P_{sm}^{\alpha \beta} q^*_{\beta} 
\label{eq:qstar}
\end{equation}
where $P_{sm}$ is the {\it smear polarizability} tensor  
being close to diagonal, and
$q^*_{\alpha} = (P_{sm}^*)^{-1}_{\alpha \beta}e_*^{\beta}$ 
is the stellar anisotropy kernel.
We select bright, unsaturated foreground stars 
of $20\simlt i' \simlt 22.5$
identified in a branch
of the half-light radius ($r_h$) vs. magnitude ($i'$) diagram
%($20<i'<22.5$, $\left<r_h\right>_{\rm median}=2.38$ pixels) 
to measure $q^*_{\alpha}$.
%%% 
In order to obtain a smooth map of $q^*_{\alpha}$ which is used in
equation (\ref{eq:qstar}), we divided the co-added mosaic image of
$9{\rm K}\times 7.4{\rm K}$ pixels into $5\times 4$ blocks,
each with $1.8{\rm K}\times 1.85{\rm K}$
pixels. The block length is based on the typical coherent
scale of PSF anisotropy patterns.  In this way the PSF anisotropy in
individual blocks can be well described by fairly low-order
polynomials.
We then fitted the $q^*$ in each block independently with
second-order bi-polynomials, $q_*^{\alpha}(\btheta)$, in
conjunction with iterative $\sigma$-clipping rejection on each
component of the residual:
$\delta e^* = e^*_{\alpha}-(P_{sm}^*)^{\alpha\beta}q^*_{\beta}(\btheta)$.  
%%%%
The final stellar sample consists of 540 stars
(i.e., $N_* \sim 30$ stars per block), 
or the mean surface number
density of $\bar{n}_*=0.72$ arcmin$^{-2}$. 
%%%%%
It is worth
noting that the mean stellar ellipticity before correction is
$(\bar{e_1}^*, \bar{e_2}^*) \simeq (-0.013, -0.018)$ 
over the data field, while the residual
$e^*_{\alpha}$ after correction
is reduced to 
$ {\bar{e}^{*{\rm res}}_1} = (0.47\pm 1.32)\times 10^{-4}$, 
$ {\bar{e}^{*{\rm res}}_2} = (0.54\pm 0.94)\times 10^{-4}$;
The mean offset from the null expectation is reduced down to 
$|\bar{e}^{* \rm res}| = (0.71\pm 1.12) \times 10^{-4}$.
%%%%%%
On the other hand, the rms value of stellar ellipticities,
$\sigma_{e*}\equiv \sqrt{\left<|e^*|^2\right>}$, is reduced from $2.64\%$ to
$0.38\%$ when applying the anisotropic PSF correction.
%%%%%
We show in Figure \ref{fig:anisopsf1}
the quadrupole PSF anisotropy field
as measured from stellar ellipticities before and after the anisotropic
PSF correction.
Figure \ref{fig:anisopsf2} shows
the distribution of stellar ellipticity
components before and after the PSF anisotropy correction.
%%%%
From the rest of the object
catalog, we select objects with 
$\bar{r}_{h*} \lesssim r_h \lesssim 10$ pixels
as an $i'$-selected weak lensing galaxy sample, 
where $\bar{r}_{h*}\approx 2.4$ pixels
is the median value of stellar half-light radii,
corresponding to half the median width of circularized PSF
over the data field.
An apparent magnitude cutoff 
of $20 \simlt i' \simlt 26$
is also made to remove from
the weak lensing galaxy sample bright foreground/cluster galaxies
and very faint galaxies with noisy shape measurements.
%%%%

%%%
Second, we need to correct image ellipticities for
the isotropic smearing effect 
caused by atmospheric seeing and the window function used for the shape
measurements. The pre-seeing reduced shear $g_\alpha$ can be
estimated from 
\begin{equation}
\label{eq:raw_g}
g_{\alpha} =(P_g^{-1})_{\alpha\beta} e'_{\beta}
\end{equation}
with the {\it pre-seeing shear polarizability} tensor
$P^g_{\alpha\beta}$.
We follow the procedure described in Erben et al. (2000)
to measure
$P^g$.
%%%
We adopt the scalar correction scheme, namely
\begin{equation}
(P_g)_{\alpha\beta}
=\frac{1}{2}{\rm tr}[P_g]\delta_{\alpha\beta}\equiv
P^{\rm s}_{g}\delta_{\alpha\beta}
\end{equation}
(Erben et al. 2000; 
 Hoekstra et al. 1998; 
 Hudson et al. 1998; 
 Okabe \& Umetsu 2008).

\begin{figure*}[!htb]
 \begin{center}
  \includegraphics[width=80mm]{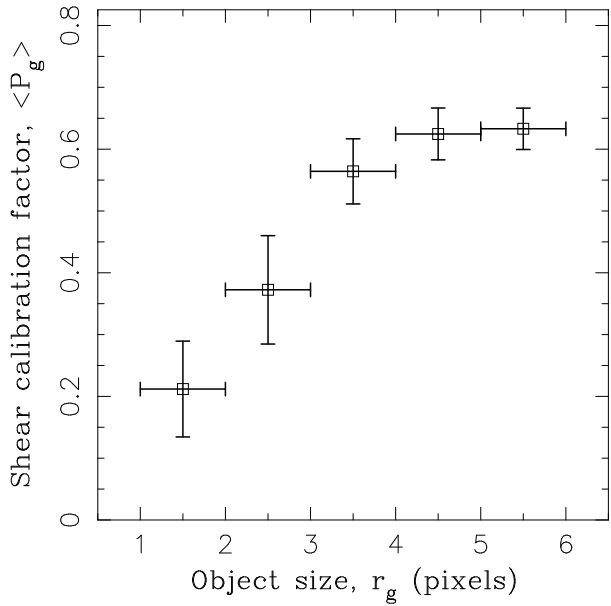}
 \end{center}
\caption{
Averaged
shear correction factor, $\langle P_g^{\rm s} \rangle$,
as a function of object size, $r_g$.
The horizontal error bar represents the size of the bin ($\Delta r_g=1$ 
 pixel), 
and the vertical error bar represents the {\it rms} scatter in 
the smoothed scalar correction factor, $\langle P_g\rangle$.
}
\label{fig:Pg}
\end{figure*}

In order to suppress artificial effects due to the noisy
$P_g^{\rm s}$ estimated for individual galaxies, we apply filtering 
to raw $P_g^{\rm s}$ measurements. Firstly, we discard those noisy
objects which have negative raw $P_g^{\rm s}$ values.
%%%%%%%%%%%%%%%%%%%
Secondly, 
we compute for each object 
a median value
of $P_g^{\rm s}$ among $N$-neighbors in the size $r_g$ and magnitude
$i'$ plane to define object parameter space: for each object,
$N$-neighbors are identified in the size ($r_g$) and magnitude ($i'$)
plane;
the median value of $P_g^{\rm s}$ is used as the smoothed
$P_g^{\rm s}$ for the object, $\langle P_g^{\rm s}\rangle$,
and the variance $\sigma^2_{g}$ 
of $g=g_1+ig_2$ is calculated using equation (\ref{eq:raw_g}).
%%%
The dispersion $\sigma_g$ is used as an rms error of the shear estimate
for individual galaxies.
We take $N=30$.
After filtering noisy $P_g^{\rm s}$ measurements,
the minimum value of $\langle P_g^{\rm s}\rangle$ is $\approx 0.035$.
%%%%%
Figure \ref{fig:Pg} shows the averaged $\langle P_g^{\rm s}\rangle$
as a function of object size $r_g$ for the $i'$-selected
weak lensing galaxy sample.
The mean of 
$\langle P_g^{\rm s}\rangle$ over all galaxies in the sample is obtained
as $0.307$, mostly weighted by galaxies with $r_g = 2-3$ pixels.
%%%
The mean variance $\overline{\sigma_g^2}$ 
over the galaxy sample is obtained as
$\simeq 0.152$, or $\sqrt{\overline{\sigma_g^2}}\approx 0.39$. 
%%%
Finally, we use the following estimator for 
the reduced shear:
$g_{\alpha} = e'_{\alpha}/\left< P_g^{\rm s}\right>$.
%%%
The final $i'$-selected galaxy sample 
contains $30,369$
galaxies or $\bar{n}_g\simeq 40.3$ arcmin$^{-2}$.

\subsection{Red Background Selection}
\label{subsec:red}

As demonstrated by B05a
and Medezinski et al. (2007), it is crucial 
in the weak lensing analysis
to make a secure selection of background galaxies in order
to minimize contamination by 
cluster/foreground galaxies and hence to make an accurate determination 
of the cluster mass,
otherwise dilution of the distortion signal results from
the inclusion of unlensed cluster galaxies, particularly at small radius
where the cluster is relatively dense.
%%%
This dilution effect is 
simply to reduce the strength of
the lensing signal when averaged over a local ensemble of 
galaxies, 
in proportion to the fraction of unlensed cluster/foreground galaxies
%cluster/foreground galaxies
whose orientations are randomly distributed, thus diluting the lensing
signal relative to the reference background level derived from
the background population (Medezinski et al. 2007).
%%%%%
With a pure red background sample (B) as a reference,
one can quantify the {\it degree of dilution} for a galaxy sample (G)
containing $N_{\rm CL}$ cluster galaxies and $N_{\rm BG}$
background galaxies
in terms of the strengths of the averaged tangential shear signal
$\langle g_+(\theta)\rangle$
as (Medezinski et al. 2007) 
\begin{equation}
1+\delta_d(\theta) \equiv \frac{N_{\rm BG}+N_{\rm CL}}{N_{\rm BG}}
=\frac{\langle g_+^{(B)}(\theta)\rangle}
      {\langle g_+^{(G)}(\theta)\rangle}
 \frac{\langle D_{ds}/D_s\rangle^{(G)}_{z_s>z_d}}
      {\langle D_{ds}/D_s\rangle^{(B)}_{z_s>z_d}},
\end{equation}
where  $\langle D_{ds}/D_s\rangle_{z_s>z_d}$'s are 
averaged distance ratios 
for respective background populations;
if the two samples contain the same
background population, then 
$\delta_d = 
\langle g_+^{(B)}\rangle/\langle g_+^{(G)}\rangle-1$.
%%%
The degree of dilution thus varies depending on the radius from
the cluster center,  increasing towards the cluster center.
%%%%%%%%%%%%%%%%%%%%%%%%%
Medezinski et al. (2007) found 
for their {\it green} galaxy sample
 ($[V-i']_{E/S0-0.3}^{+0.1}$) containing
the cluster sequence galaxies in A1689
that the fraction of cluster membership,
$N_{\rm CL}/(N_{\rm BG}+N_{\rm CL})$,
tends $\sim 100\%$ within $\theta \simlt 2'$.

\begin{figure*}[!htb]
 \begin{center}
  \includegraphics[width=70mm,angle=270]{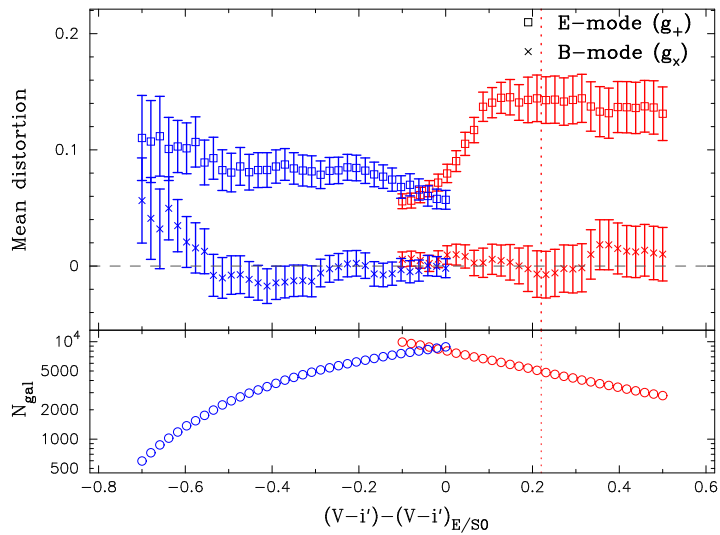}
 \end{center}
\caption{
\label{fig:dilution}
Top panel:
mean distortion strength averaged over a wide radial range of $1\arcmin
 < \theta < 18\arcmin$,
done separately for the blue and red galaxy samples.
No area weighting is used here to enhance the effect of dilution in the
 central cluster region.
%in order to
%establish the boundaries of the color distribution free of cluster members. 
Shown are the measurements of the tangential component ($g_+$)
with open squares, and those of the $45{\rm deg}$-rotated component
 ($g_\times$).  
%%%
On the right ({\it red}), the square symbols show
that $g_+$ drops rapidly when the bluer limit of the entire red sample
is decreased below a color indicated by the vertical dashed line
which lies $+0.22$ mag  redward of the cluster sequence.
%%%%
This sharp decline marks the point at which the red sample encroaches on
the E/S0 sequence of the cluster. 
For galaxies with colors bluer than the cluster seqeuence, cluster
members are present along with  background galaxies. Consequently, the
 mean lensing strengh of the blue sample, as shown on the left ({\it
 blue}), is systematically lower than that of the red sample.
Bottom panel: the respective numbers of galaxies as a function of color
 limit, contained in the range
 $1\arcmin <\theta < 18\arcmin$ in the red  ({\it right}) and the blue
 ({\it left}) samples.
} 
\end{figure*}

%%%%%
For our weak lensing distortion analysis
we define a sample of {\it red background galaxies}
whose colors are redder due to large $k$-corrections than
%% @@ edited by KU (2008/02/18)
%% the color-magnitude (CM) sequence of cluster member galaxies.
the CM relation, or red sequence, of cluster member galaxies.
These red background galaxies are largely composed of early to mid-type
galaxies at moderate redshifts (Medezinski et al. 2007). 
%%%%
Cluster member
galaxies are not expected to extend to these colors in any significant
numbers because the intrinsically reddest class of cluster galaxies,
%% @@ edited by KU (2008/02/18)
%% i.e. E/S0 galaxies, are defined by the CM sequence and lie blueward
%% of
i.e. E/S0 galaxies, are defined by the red sequence and lie blueward of
chosen sample limit, so that even large photometric errors will not 
carry them into our red sample.
%%%% @@ KU (2008/03/29)
This can be demonstrated readily, as shown in Figure \ref{fig:dilution},  
where we plot the mean tangential shear strength $\langle g_+ \rangle$,
averaged over a wide radial range of $1\arcmin < \theta < 18\arcmin$,
%%the central $10\arcmin$ region ($1\arcmin < \theta < 10\arcmin$) as
as a function of color limit
by changing the lower color limit progressively blueward.
Here we do not apply area weighting to enhance the effect of dilution in
the central region.
%, finding a sharp drop in the lensing signal 
%at our limit, $(V-i')<(V-i')_{E/S0}+\Delta(V-i')$
%when the cluster red sequence starts to
%contribute significantly, thereby reducing the mean lensing signal.
%%%
We take the lower (bluer) color limit of $+0.22$ mag 
where the cluster contribution is negligible, 
%and the upper color limit of $6$ to include the
%majority of the background red population,
defining the red background sample by
$\Delta(V-i')\equiv (V-i')-(V-i')_{E/S0}  > 0.22$, 
as adopted by B05a.
Figure \ref{fig:dilution} shows a sharp drop in the
lensing signal at $\Delta(V-i') \simlt 0.1$,
when the cluster red sequence starts to contribute significantly,
thereby reducing the mean lensing signal.
At $\Delta(V-i')\simgt 0.1$,
the mean lensing signal of the red background
stays fairly constant, $\langle g_+ \rangle \simeq 0.143$,
ensuring that our weak lensing measurements 
are not sensitive to this particular choice of the color limit
(see \S \ref{subsec:sys}).
%%%%
Similarly, on the left of Figure \ref{fig:dilution}, we show the mean
distortion strength for our blue galaxy sample with the magnitude
limit in the interval $23 < i' < 25.5$, so as to take only faint blue
galaxies. The lower color limit is set to $\Delta(V-i')>-1.5$.
For galaxies with colors bluer than the
cluster sequence,  cluster galaxies are present along
with background 
galaxies, since the cluster population extends to bluer colors
of the later type members.
Consequently, the mean lensing signal of the blue sample
is systematically lower than that of the red sample, unless we take the
upper (redder) color limit around $\Delta(V-i')\sim 0.7$ where, however,
the distortion measurement is quite noisy (Figure \ref{fig:dilution}).
We therefore exclude blue galaxies from our weak lensing analysis.
Note that 
the background populations do not need to be complete
in any sense but should simply be well defined and contain only
background.
%%%
This color-magnitude selection criteria yielded a total of 
$N_g= 5728$ galaxies, or the mean surface number density of
$\bar{n}_g=7.6$ galaxies arcmin$^{-2}$.
For the red background sample, we found
%the mean variance $\bar{\sigma}_g^2$ 
%over the red galaxy sample is obtained as
$\overline{\sigma_g^2}\simeq 0.133$, or 
$\sqrt{\overline{\sigma_g^2}}\approx 0.36$, which is slightly 
smaller than the rms dispersion for the $i'$-selected galaxy sample.

We need to estimate the depths of our color-magnitude selected 
red samples when measuring the cluster mass profile, 
because the lensing signal depends on the source redshifts in proportion
to $D_{ds}/D_s$.
%%%%%
Medezinski et al. (2007) utilized the multicolor 
photometry of Capak et al. (2004)
based on Subaru $UBVRIZ$ imaging covering $0.2$ deg$^2$, and 
estimated
photometric redshifts for color-magnitude 
selected background galaxy samples.
%%%%
Using the Capak et al.'s photometric redshift distributions,
Medezinski et al. (2007) estimated a mean redshift for the red
background with $i'>18$ mag to be 
$\bar{z}_s \approx 0.87$ at fainter magnitude limits of $i'=26$--$27$,
and also calculated weighted mean lensing depths $\langle
D_{ds}/D_s\rangle$ for respective background populations as a function
of apparent $i'$ limiting magnitude, and found that the averaged
distance ratio $\langle D_{ds}/D_s\rangle$ grows only slowly with
increasing apparent magnitude limit: 
$\langle D_{ds}/D_s \rangle =  0.693 \pm 0.02$ for 
the red sample with $i'_{\rm cut}=25.5$--$26.5$
(see Figure 9 of Medezinski et al. 2007),
and the corresponding redshift equivalent to this mean distance is
$z_{s,D}=0.68 \pm 0.05$, where the distance-equivalent redshift $z_{s,D}$ is
defined by the following equation:
\begin{equation}
\label{eq:zD}
\bigg\langle \frac{D_{ds}}{D_s} \bigg\rangle_{z_s} =
 \frac{D_{ds}}{D_s}\bigg|_{z_s=z_{s,D}}.
\end{equation}
%%%
Therefore, we can safely assume that our two different red samples for
the distortion and magnification measurements (with $i'_{\rm cut}=26.0$
and $25.5$, respectively) have nearly the same depths. 
%%%
We note that B05a assumed that all of the background galaxies are
located at a single redshift of $z_s=1$, corresponding to
$D_{ds}/D_s=0.772$, which underestimates $\Sigma_{\rm crit}$ by
$\sim 11\%$ 
%%%($0.772/0.693=1.11$),
and hence underestimates the cluster mass accordingly.
%%%%
We will come back to this issue in \S \ref{subsec:sys}.

\section{Weak Lensing Map-making}
\label{sec:map}

This section describes a maximum entropy method (MEM) for reconstructing  
the two-dimensional cluster mass distribution from combined shape
distortion and magnification bias observations. 

As described above (see \S \ref{subsec:red}), only the red selected
background is used for the measurement of the lens distortion and
magnification, in order to minimize the effect of dilution.

\subsection{Shape Distortion Data}
\label{subsec:distortion}

For map-making,
we pixelize the distortion data into a regular grid
of $N_{\rm data} = 21\times 17 = 357$ independent pixels
covering a field of $\approx 30'\times 24'$.
%%%
The pixel size is set to 
$\Delta_{\rm pix}=1\farcm 4$, 
and the mean
galaxy counts per pixel is $\bar{n}_g\Delta\Omega_{\rm pix} \sim  15$
with $\Delta\Omega_{\rm pix}=\Delta_{\rm pix}^2$ being
the solid angle per pixel.

We compute an estimate $\tilde g_{\alpha}(\btheta_i)$ for the reduced
shear $g_{\alpha}(\btheta_i)$ on a regular grid of cells
($i=1,2,...,N_{\rm pix}$) as
\begin{equation}
\label{eq:bin_shear} 
\tilde g_{\alpha}(\btheta_i)
=
\sum_{k\in {\rm cell} i} u_k g_{\alpha,k}
\Big/
\sum_{k\in {\rm cell} i} u_k
\equiv \tilde{g}_{\alpha,i},
\end{equation}
where  
$g_{\alpha,k}$ is a noisy estimate of the $\alpha$th component of the
reduced shear for the $k$th galaxy, and 
$u_k=1/\sigma_{g,k}^2$ is its inverse-variance weight
(see \S \ref{subsec:shear}).
%%%
The error covariance matrix
for the pixelized reduced shear
$\tilde{g}=\tilde{g}_1+i\tilde{g}_2$ 
(\ref{eq:bin_shear}) is then 
diagonal, and
given by
\begin{equation}
\langle 
\Delta\tilde{g}_{\alpha,i}\Delta\tilde{g}_{\beta,j}
\rangle
=
\frac{1}{2}
\sigma^2_{\tilde{g},i}
\delta_{\alpha\beta}
\delta_{ij},
\end{equation}
%where $\delta_{ij}$ is the Kronecker's delta, and
where $\sigma^2_{\tilde{g}, j}$ is the error variance for the $j$th
pixel defined as
\begin{equation}
\label{eq:bin_shearvar}
\sigma^2_{\tilde{g}}(\btheta_j) = 
%\frac{\sum_{k\in {\rm cell} j} u_{k}^2 \sigma^2_{g,k}}
\frac{1}
%{ \left( 
      {\sum_{k\in {\rm cell} j} u_{k}}.
% \right)^2}.
\end{equation} 
Here we have used $\langle \Delta g_{\alpha,k}\Delta g_{\beta,l}
\rangle = (1/2)\sigma^2_{g,k} \delta_{\alpha\beta}\delta_{kl}$
for individual shear estimates $g_{\alpha,k}$.
%%%%
The per-pixel rms dispersion for $\tilde{g}(\btheta)$
is then reduced down to $\sqrt{\overline{\sigma^2_g}/N}
\sim 0.36/\sqrt{15}\sim 0.092$.

\begin{figure*}[!htb]
 \begin{center}
  \includegraphics[width=70mm, angle=270]{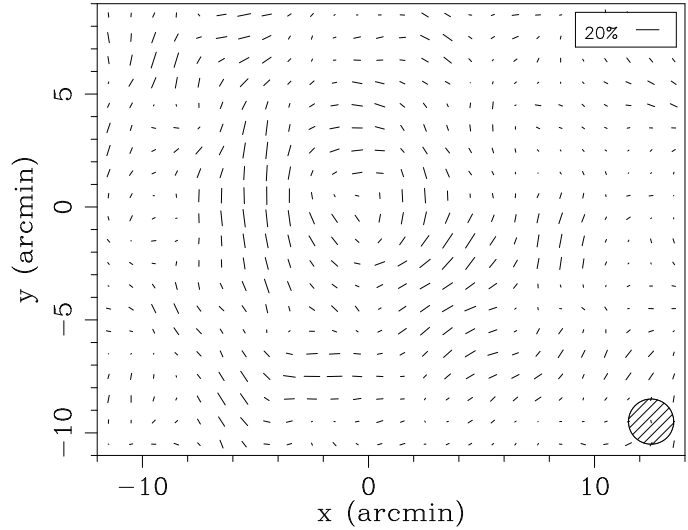}
 \end{center}
\caption{Gravitational reduced-shear field in A1689 obtained from shape
 distortions of the red background galaxies, smoothed with a
 Gaussian with ${\rm FWHM} = 2'$ for visualization purposes.
A stick with the length of $20\%$ ellipticity is indicated in the top
 right corner. 
The shaded circle indicates the FWHM of the Gaussian.
The coordinate origin is at the optical center (see Figure
 \ref{fig:image}). 
}
\label{fig:shear}
\end{figure*}

%%%%
In Figure \ref{fig:shear} we show the reduced-shear field 
obtained from the red galaxy sample, where for visualization purposes
the $\tilde{g}_{\alpha}(\btheta)$ is resampled on to a finer grid and
smoothed with a Gaussian with ${\rm FWHM}=2'$.
A coherent tangential shear pattern is clearly seen 
in Figure \ref{fig:shear} around the cluster center.

\subsection{Magnification Bias Data}
\label{subsec:mag}

%The magnification bias arises because gravitational lensing changes the
%apparent solid angle of the background but conserves the surface
%brightness, which influences the observed surface density of
%background sources, 
Lensing magnification, $\mu(\btheta)$, influences the observed
surface density of background sources, expanding the area of
sky, and enhancing the observed flux of background sources.
In the subcritical regime 
(Broadhurst, Taylor \& Peacock 1995; Umetsu et al. 1999),
the magnification $\mu(\btheta)$ is given by equation (\ref{eq:mu}).
%%%%%%%%%%%%%%%%%%%%%%%%%%%%%
The count-in-cell statistic is measured from 
the flux-limited red galaxy sample 
(see \S \ref{subsec:data}) on the same grid as the distortion data:
\begin{equation}
\label{eq:cic}
\tilde{N}(<m_{\rm cut}; \btheta_i) = \sum_{k\in {\rm cell} i}1
\equiv \tilde{N}_i
\end{equation}
with $m_{\rm cut}$ being the magnitude cutoff corresponding to the
flux-limit ($i'_{\rm cut}=25.5$).
%%%
The normalization and slope of the unlensed number counts
$N_0(<m_{\rm cut})$ for our red galaxy sample are reliably estimated as 
$n_{\mu,0}=12.6\pm 0.23$ arcmin$^{-2}$ and 
$s\equiv d\log{N_0(m)}/dm=0.22\pm 0.03$
from the outer region $\ge 10'$ (B05a).
%%%
The mean galaxy counts per pixel is thus 
$N_0 = n_{\mu,0}\Delta\Omega_{\rm pix}\sim 25$.
%%%
The magnification bias at the $i$th cell is then estimated as
$1+\tilde\delta_{\mu,i} = 
\tilde{n}_{\mu}/n_{\mu,0}
=
\tilde{N}_i/N_0$,
with $\tilde n_{\mu,i}=\tilde N_{\mu,i} / \Delta\Omega_{\rm pix}$,
where the dilution effect $\delta_d$ is negligible for our
red-background
sample; otherwise $\tilde N_i/N_0 = 1+\tilde{\delta}_\mu+\tilde{\delta}_d$.
%%%
The slope is less than the lensing invariant slope, $s=0.4$, 
and hence a net deficit of background galaxies is expected:
\begin{equation} 
\label{eq:magbias}
\delta_{\mu}(\btheta) \equiv \langle \tilde\delta_{\mu}(\btheta)\rangle 
= \mu_i^{2.5s-1}(\btheta)-1
\end{equation}
(Broadhurst et al. 1995; B05a).
%%%%
In the limit of weak lensing 
where $\kappa,|\gamma|\ll 1$, 
$\delta_{\mu}(\btheta) \simeq (5s-2) \kappa(\btheta)
\simeq -0.9\kappa(\btheta)$ with $s=0.22$.  
%%%
%%%

The masking effect due to bright cluster galaxies and bright
foreground objects is properly taken into
account and corrected for (B05a).
%%%%
We conservatively account for the masking of observed sky
by excluding a generous area $\pi a b$ around each masking object, where
$a$ and $b$ are defined as $\nu_{\rm mask}\equiv 3$ times 
the major ({\tt A\_IMAGE}) and minor axes ({\tt B\_IMAGE}) computed
from SExtractor, corresponding roughly
to the isophotal detection limit in our configuration
(see \S \ref{subsec:data}). 
%%%
We calculate the correction factor for this masking effect
as a function of radius from the cluster center, 
and renormalize the number density of each cell accordingly.
%%%%
The masking area is negligible at large radii, and increases up to
$\sim 20\%$ of the sky close to the cluster center, $\theta\simlt 3'$.
%%%
B05a showed that
the magnification bias measurements with and without the masking
correction are roughly consistent with each other, and with the NFW
prediction from  the ACS strong lensing observations (B05b)
and the Subaru distortion profile (see Figure 2 of B05a), 
even though all the measurements have different
systematics. 
Note that if we use the masking factor $\nu_{\rm mask}$ of $2$ or $4$,
instead of 3, 
the results shown below remain almost unchanged (see \S \ref{subsec:sys}
for more details). 
%%%%
%We found that 
%the results are little
%changed when the masking factor of 2 or 4 are adopted instead of 3.
%The magnificaiton-bias map is displayed in Figure \ref{fig:magbias}.
%%%%

\begin{figure*}[!htb]
 \begin{center}
   \includegraphics[width=60mm, angle=270]{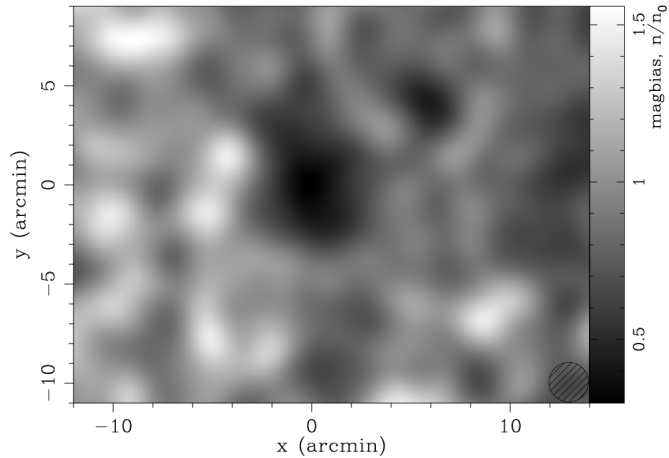}
 \end{center}
\caption{
Distribution of the lensing magnification bias  $n/n_0$
measured from a red-galaxy sample in the background of A1689, 
smoothed with a Gaussian 
 with ${\rm FWHM} = 2'$ for visualization purposes.
The shaded circle indicates the FWHM of the Gaussian.}
\label{fig:magbias}
\end{figure*}

Figure \ref{fig:magbias} shows the resulting magnification-bias
distribution
derived from the red galaxy sample based on the SExtractor
photometry (\S \ref{subsec:data}). 
%%%
A clear depletion of the red galaxy
counts is visible in the central, high-density region of the cluster.
%%% :: Keiichi's original ::
%A clear depletion of the red galaxy
%counts is seen in the central, high-density region of the cluster.
%%%
On the other hand, 
it has been argued that estimates of the lensing magnification 
based on number counts suffer from noise 
arising from the intrinsic clustering of the source galaxies
(e.g., Zhang \& Pen 2006).
Indeed,
some
variance is apparent in the spatial distribution of red galaxies.
In particular, a local enhancement 
%$n_{\mu}/n_0\sim 1.5$
of red background galaxies
can be seen in Figure \ref{fig:magbias} around $\btheta \sim (0',5')$,
having an overdensity of $\tilde n_{\mu}/n_0 \sim 1.5$.
This may explain the discrepant point at $\theta\sim 5'$ in the
magnification bias profile of B05a.
%Estimates of the lensing magnification 
%based on number counts suffer from noise 
%arising from the intrinsic clustering of the source galaxies,
%To overcome this intrinsic clustering problem
%we wil
%intrinsic clustering signal is ..
%Such an intrinsic clustering signal of red galaxies ...
%must be unweighted
We will come back to this issue later in \S \ref{subsec:declustering}.
%%%%
%Indeed, in the one dimensional analysis of 
%Broadhurst et al. (2005b), we found a discrepant point at $\theta\sim
%5'$ in the magnification-bias profile (Figure 2 of Broadhurst et
%al. 2005b). 

\subsection{Maximum Entropy Mass Reconstruction Method}
\label{subsec:massrec}
 
The relation between distortion and convergence is non-local, and
the convergence $\kappa$ derived from distortion data alone suffers from
a mass sheet degeneracy (see \S \ref{sec:wl}). 
However, by combining the distortion and magnification measurements the
convergence can be obtained unambiguously with the correct mass
normalization.  
%%%%

Here we combine pixelized distortion and depletion data of red
background galaxies in a maximum likelihood sense, and reconstruct the
two-dimensional distribution of the lensing convergence field,
$\kappa(\btheta)$. 
%%%
Several authors have proposed maximum likelihood and maximum entropy
methods for reconstructing the projected mass distribution from 
joint weak lensing observations of the shape distortion and
magnification bias effects (Bartelmann 1995; Seitz, Schneider, \&
Bartelmann 1998; Bridle et al. 1998; Umetsu et al. 2007). 
%%%
In the present study, we utilize a maximum entropy method 
extended to account for positive/negative distributions of the
underlying field starting with the method proposed in the context of 
interferometric observations of the Cosmic Microwave Background (CMB)
radiation by Maisinger, Hobson, \& Lasenby (1997) and Hobson \& Lasenby
(1998).

Bayes' theorem states that given a hypothesis $H$ and some data $D$ the 
posterior probability ${\rm Pr}(H|D)$ is the product of of the
likelihood ${\rm Pr}(D|H)$ and the prior probability  ${\rm Pr}(H)$,
with a proper normalization by the evidence ${\rm Pr}(D)$ (Maisinger et
al. 1997):
\begin{equation}
\label{eq:bayes}
{\rm Pr}(H|D) = \frac{{\rm Pr}(H)\, {\rm Pr}(D|H)}{{\rm Pr}(D)}.
\end{equation}
In the context of map-making, the hypothesis $H$ is taken as the
pixelized image  ${\bf p}=\{p_i\}$ ($i=1,2,...,N_{\rm pix}$).
Since the evidence in Bayes' theorem is a normalization constant for a
given dataset $D$, we maximize ${\rm Pr}({\bf p}|D)\propto {\rm
Pr}({\bf p})\, 
{\rm Pr}(D|{\bf p})$ with respect to ${\bf p}$.
We assume that the errors on the data follow a Gaussian distribution, so
that the likelihood is given as ${\rm Pr}(D|{\bf p})\propto
\exp\left[-\chi^2({\bf p})/2\right]$, with $\chi^2$ being the standard
misfit statistic. Then, for a given form of the entropy function $S({\bf
p}, {\bf m})$, the entropic prior is written as
\begin{equation}
\label{eq:entropy}
{\rm Pr}({\bf p}) \propto 
\exp\left[\alpha S({\bf p},{\bf m})\right],
\end{equation}
where ${\bf m} = \{m_i\}$ ($i=1,2,...,N_{\rm pix}$)
is a set of model parameters for the pixelized image, ${\bf p}$.
We follow the prescription given in Maisinger et al. (1997) and Hobson \&
Lasenby (1998), and define the cross-entropy function for ${\bf p}$ as
%the cross-entropy function for a pixelized {\it image} ${\bf p}=\{p_i\}$
%($i=1,2,...,N_{\rm pix}$) is written in discretized form as (Maisinger
%et al. 1997; Hobson \& Lasenby 1998)
\begin{equation}
S({\bf p}, {\bf m}) = 
\sum_{i=1}^{N_{\rm pix}}\left[
\psi_i -2m_i-p_i\ln\left(
\frac{\psi_i+p_i}{2m_i}
\right)
\right],
\end{equation}
where $\psi_i\equiv \sqrt{p_i^2+4m_i^2}$.
%and ${\bf m}=\{m_i\}$  ($i=1,2,...,N_{\rm pix}$)
%is a set of model parameters for ${\bf p}$.
%%%
We take $p_i = \kappa_i\equiv \kappa(\btheta_i)$ 
as the image to be reconstructed,
and express a set of discretized $\kappa$-values as 
${\bf p}=\{\kappa_i\}$ $(i=1,2,...,N_{\rm pix})$.
%%%%
In general, an entropy regularization 
helps to reduce the sensitivity of the least $\chi^2$
solutions to small-scale noise in data, by imposing 
smoothness constraints on the solutions.
%%%%
This MEM prior ensures that
$\kappa(\btheta) \to 0$ in the noise-dominated regime,
or low-density regions in the outskirts of the cluster.
%%%%%%%%%%%%%%%%%%%%%%%%%%%
Note that unlike conventional MEM functions, this MEM prior is free from
the ``positive bias''  in the reconstructed image ($= {\rm signal} +
{\rm residual \ noise}$), and this bias is more significant in the low
signal-to-noise ratio, or low density regions ($r\to r_{\rm vir}$) where
we are interested in measuring $\kappa$.

%%%%
We take into account the non-linear, but subcritical, regime of the
lensing properties, $\kappa$ and $\gamma_{\alpha}$,
for a MEM mass reconstruction (see Bridle et al. 1998).
We use equations (\ref{eq:gamma2kappa}) and (\ref{eq:kappa2gamma})
to relate the gravitational shear and the convergence fields, 
$\gamma$ and $\kappa$.
%%%%
%%%%
The log posterior probability function,
$F({\bf p}) = -\ln{{\rm Pr}({\bf p}|D)}$, is then expressed as a linear
sum of the shear/magnification data log-likelihoods (Schneider, King, \&
Erben 2000) and the entropy term (Maisinger et al. 1997): 
\begin{eqnarray}
\label{eq:loglikelihood}
F({\bf p})& = &
\frac{1}{2} \chi^2({\bf p})
%\left[
%\chi^2_g({\bf p}) 
%+
%\chi^2_\mu({\bf p})
%\right]
- \alpha S({\bf p}, {\bf m}),\\
\chi^2({\bf p}) &=& \chi^2_g({\bf p}) + \chi^2_\mu({\bf p})\\
\chi^2_g &\equiv& 
\sum_{i=1}^{N_{\rm pix}}
\sum_{\alpha=1}^{2} 
\frac{ [\tilde{g}_{\alpha,i}-g_{\alpha,i}({\bf p}) ]^2 } {
\sigma_{\tilde{g}_\alpha,i}^2 },\\
\chi^2_\mu &\equiv& 
\sum_{i=1}^{N_{\rm pix}}
\frac{ [\tilde{N}_{i}-N_{i}({\bf p})]^2} {\tilde N_i},
\end{eqnarray}
where $\alpha (>0)$ is the dimensionless regularization constant, and
$\sigma_{\tilde{g}_{\alpha,i}}$ is the per-component rms error
for the pixelized distortion measurement ($\alpha=1,2$),
$\sigma_{\tilde{g}_{1,i}} =\sigma_{\tilde{g}_{2,i}} =
\sigma_{\tilde{g},i}/\sqrt{2}$;   
%%%%%
$g_{\alpha,i}({\bf p})$ (equation [\ref{eq:redshear}])
and  $N_i({\bf p}) = \mu_i({\bf p})^{2.5s-1} N_0$
%n_{\mu,0}\Delta\Omega_{{\rm pix},i}$
are the theoretical expectations for 
$\tilde{g}_{\alpha,i}$ and $\tilde{N}_i$, respectively,
where $\Delta\Omega_{{\rm pix},i}$ is the effective observed area
of the $i$th pixel excluding the masking area by the cluster members. 
%%%%%%%
%In defining the magnification bias likelihood function $l_\mu$,
In defining $\chi^2_\mu$,
we have used the Gaussian approximation for the 
Poisson distribution 
of count-in-cell statistics 
($N_0 \sim 25$),
as done in the one-dimensional analysis by B05a.
%%%
We also note that the dispersion for $\tilde{g}$
is modified as $\sigma[\tilde g(\btheta)] \approx \sigma_{\tilde
g}(1-|g({\bf p})|^2)$ in the subcritical, non-linear regime (Schneider
et al. 2000);  however, we neglect this non-linear correction for the
dispersion $\sigma_{\tilde g}$ to simplify various calculations, as in
B05a.  
%%%% 
We found these are indeed good approximations for our combined
distortion and depletion datasets, and the results presented here
are little changed when the full likelihood function is used.

The maximum likelihood solution, $\hat{\bf p}$,
is obtained by minimizing
the function $F({\bf p})$ with respect to ${\bf p}$
for given $\alpha$ and ${\bf m}$. 
%%%%%
To do this,
we compute numerically the
derivatives $\partial F({\bf{p}})/\partial p_i$ using a
conjugate-gradient algorithm (Press et al. 1992).
%%%
%We take $m_i={\rm const}\equiv m$, 
We determine by iteration
the Bayesian value of $\alpha=\hat{\alpha}(\hat{\bf p}, {\bf m})$ 
by the following equation
(Bridle et al. 1998):
%for a given value of $m$. 
\begin{equation}
\label{eq:alpha}
-2\hat\alpha S(\hat{\bf p},{\bf m}) =  N_{\rm pix} - \hat\alpha {\rm Tr}({\bf
 M}^{-1}) \, \equiv N_{\rm good},
\end{equation}
where ${\bf M}$ is a $N_{\rm pix}\times N_{\rm pix}$ matrix defined by
\begin{equation}
{\bf M} = {\bf G}^{-1/2} {\bf H} {\bf G}^{-1/2},
\end{equation}
with
\begin{eqnarray}
H_{ij}(\hat{\bf p})&=&\frac{\partial^2 F({\bf p})}
{\partial p_i \partial p_j}\bigg|_{{\bf p}=\hat{\bf p}},\\
G_{ij}(\hat{\bf p}) &=& \frac{\partial^2 S({\bf p})}
{\partial p_i \partial p_j}\bigg|_{{\bf p}=\hat{\bf p}},
\end{eqnarray}
evaluated at ${\bf p}=\hat{\bf p}$; 
$N_{\rm good}$ is a measure of the
effective number of parameters (Suyu et al. 2006).
With the optimal $\alpha$, we thus expect that 
the final value of the misfit statistic $\chi^2$ is close to the {\it
classical number of degrees of freedom} (hereafter NDF, see Suyu et
al. 2006), ${\rm NDF} \equiv N_{\rm data} - N_{\rm good}$ (classic MEM).  
We take $m_i={\rm const}\equiv m$, 
where
$m$ can be regarded as a characteristic amplitude 
of the image (Maisinger et al. 1997).
%%%%
%%%%
We find that 
the maximum-likelihood solution $\hat{\bf p}$
for the Bayesian $\hat\alpha$ 
is insensitive to the choice of $m$ (see \S \ref{subsec:sys}). 
In the following we set $m$ to be $0.5$. 
%%%
In order to be able to 
quantify the errors on the mass reconstruction we
evaluate the Hessian matrix ${\bf H}$ of the function 
$F({\bf p})$ at ${\bf p}=\hat{\bf p}$,
%\begin{equation}
%\label{eq:hess}
%H_{ij}(\hat{\bf p})=\frac{\partial^2 F({\bf p})}
%{\partial p_i \partial p_j}\bigg|_{{\bf p}=\hat{\bf p}},
%\end{equation} 
from which the covariance matrix 
of the reconstructed $\kappa$ map is given by 
%parameters ${\bf p}$ is given by
\begin{equation}
 C_{ij}
\equiv \langle 
\delta\kappa_i
\delta\kappa_j 
%(\kappa-\hat\kappa)_i
%(\kappa-\hat\kappa)_j
\rangle=
(H)^{-1}_{ij}(\hat{\bf p}).
\end{equation}
%%%%

\subsection{Strong Lensing Constraints}
\label{subsec:acs}

The map-making method described in \S \ref{subsec:massrec} 
is only applicable to subcritical regions 
lying outside of the critical curves (see Bartelmann \& Schneider 2001).
Therefore, our Subaru distortion and depletion datasets cannot be used
to constrain the cluster mass distribution in the strong lensing region
within the tangential critical curve (see B05b).
%%%%
Instead, we can 
pause quadratic constraints on such pixels
to the total log-likelihood function
based on strong lensing models as:
\begin{equation}
 F({\bf p})
\to F({\bf p}) + \frac{\beta}{2}\sum_{i=1}^{N_c} 
\left(\kappa-\kappa_{\rm c}\right)_i^2,
\end{equation}
where 
$\kappa_{{\rm c}.i}$ is the constraint on $\kappa_i$ by strong lensing,
$N_{\rm c}$ is the number of constraints, and $\beta$ is a large number
without causing this external term to become singular to working precision.

In the present pixelization scheme, there is one such pixel ($N_{\rm
c}=1$) containing the strong lensing region for which the Einstein
radius of $\theta_{\rm E}\simeq 43''$ corresponds to a mean redshift of
$\bar z_s=0.87$ for our red background population (see \S
\ref{subsec:red}). 
To constrain the central $\kappa$ pixel, we utilized a mass model of
A1689, which is well constrained by ACS strong lensing observations
restricted to the central region $\simlt 2'$ (B05b).
The central ACS-derived mass profile of B05b is best described by an NFW
model with a virial mass $M_{\rm vir}=2.6\times 10^{15}M_{\odot}/h$ and a
concentration $c_{\rm vir}=8.2$  having a scale radius of $r_s \equiv
r_{\rm vir}/c_{\rm vir}=310 {\rm kpc}/h$ (B05b).
With this NFW model, we find $\kappa_{\rm c} \approx 0.781 \pm 0.1$ at
the central pixel of $\btheta\sim (0',0')$ for a reference source
redshift of $z_s=1$, corresponding to $\kappa_{\rm c}\approx 0.700$ for
our red background sample (\S \ref{subsec:red}). Hence, assuming $z_s=1$
for the source redshift will overpredict the central density by a
factor of $0.772/0.693\simeq 1.113$, which will affect slightly the
overall amplitude of the reconstructed $\kappa$ map (see \S
\ref{subsec:sys}).

\subsection{Downweighting Intrinsic Clustering Contributions and Noisy
  Measurements}
\label{subsec:declustering}

In contrast to the shearing effect, the magnification bias
is a local, direct measure of the lensing convergence,
free from the mass-sheet degeneracy.
However, a practical difficulty of the magnification bias measurement 
is the intrinsic clustering contribution which locally can be larger 
than the lensing induced signal in a given pixel.
%% :: Keiichi's original ::
%However, the practical difficulty of 
%the magnification bias measurement is the intrinsic clustering
%contribution which is usually larger than the lensing induced signal.
In order to obtain a clean measure of the lensing signal,
such intrinsic clustering needs to be eliminated.
%%%%%%%%%%%%%%%%%
%% Zhang \& Pen (2005) 
Broadhurst et al. (1995) proposed an active declustering method 
based on the facts that 
the magnification bias depends strongly on the shape of the luminosity
function, whereas intrinsic clustering depends weakly on the
intrinsic luminosity function,
and that they have different redshift dependence 
(see also Zhang \& Pen 2005). 
%%%%
This method however requires
the addition of redshift information for
%%% :: Keiichi's original ::
%information of photometric redshifts as well as fluxes of 
individual background galaxies. 
Furthermore, the shape information of 
luminosity functions for respective background populations must be
provided in order to convert a density depletion or enhancement
into the lens magnification, $\mu$.
%Therefore, 
%which enables us to 

Alternatively, one may employ prior information from, for example,
the surface luminosity density of cluster member galaxies
to predict the projected mass distribution,
which is then used to 
downweight intrinsic clustering contributions. 
In the present study, we have adopted 
an objective rejection scheme 
based on the gravitational shear predictions, summarized in
the following steps:
 (1) Using shape distortion data alone
  we derive an entropy-regularized
  maximum likelihood solution $\bf{p}$ for the mass distribution.
 (2) Then, the shear-based $\kappa$ map is used to predict
  the magnification bias on the same grid of pixels.
 (3) Finally, we make a pixel-to-pixel comparison between
  the observed and predicted galaxy counts, and
  reject those magnification measurements $\tilde N_i$
  which are in conflict with
  the shear-based predictions $N_i(\bf{p})$ as:
\begin{equation}
|\tilde N_i - N_i({\bf p})| > \nu_{\rm clust} \sqrt{N_0}
\end{equation}
where $N_0\sim 25$ is the unlensed, mean number counts of red galaxies,
and $\nu_{\rm clust}$ is a rejection threshold in units of $\sigma$. We
set a rejection threshold at $4\sigma$, i.e., $\nu_{\rm clust}=4$.
%%%
Changing this threshold will affect the details of the 
reconstruction especially in the outer regions ($\theta \simgt
7\arcmin$).  
However, we confirmed that changing the threshold $\nu_{\rm clust}$ by
$\pm 1$ leaves our results unchanged within our statistical
uncertainties (see 
\S \ref{subsec:sys}), and that the resulting mass map is fairly
consistent with the surface luminosity and density distributions of
cluster member galaxies, as described in detail below.

In addition to the above,
in order to exclude unreliable measurements,
we have assigned zero weight to those pixels which 
satisfy either of the following rejection criteria:
\begin{enumerate}
\item Measurement pixel with no usable galaxy,
\item Measurement pixel in the strong lensing region (\S \ref{subsec:acs}),
%\item Magnification measurement rejected according to the shear-based
%      prediction (\S \ref{subsec:declustering}),
\item Magnification measurement near boundary regions 
      with low
      completeness at fainter magnitudes.
\end{enumerate}

The last rejection criterion is 
required since the slope of the unlensed number counts,
$s = d\log{N_0(m)}/dm$, could depend on the selection completeness;
such an apparent variation of the slope $s$ could give rise
to additional errors in the mass reconstruction.

\subsection{Map-making of A1689}
\label{subsec:a1689map}

We have applied our MEM method to the
joint measurements of the shape distortion and magnification bias
effects of red background galaxies 
%combined shape distortion and magnification bias measurements
in order to reconstruct the two-dimensional mass distribution of A1689.
We utilized the FFTW implementation 
%(Frigo \& Johnson 1998)
of 
Fast Fourier Transforms (FFTs) to convert between
the lensing convergence and the gravitational shear fields
using equations (\ref{eq:gamma2kappa}) and (\ref{eq:kappa2gamma}).
The FFT however implies a periodicity in 
both horizontal and vertical directions, 
producing aliasing effects at the borders of the computational domain. 
In order to avoid such aliasing effects, 
we used large arrays of $31 \times 27$ pixels
with $N_{\rm pix}=31\times 27=837$, 
covering a field of $43\farcm 4\times 37\farcm 8$. 
%%%
On the other hand,
the distortion and depletion measurements 
are limited to the central $30'\times 24'$ region
($21\times 17=357$ pixels);
the wide-field Subaru data thus allow us to probe the 
projected mass distribution on scales ranging from
$\theta\sim 1'$ up to $\theta\sim 20'$.
%%%
To minimize spurious effects from the periodic boundary
condition,
pixelized maps are further zero padded 
by a factor of $2$ in each dimension 
(Seljak 1998; Sato et al. 2003).
%%% Seljak 1998, ApJ, 506, 64
%%%%

For the spin-2 distortion measurements, we have in total $N_{{\rm
data},g}=2 \times 355 = 710$ usable (real) observations.
%%%%%%
For the magnification measurements,
we have $N_{{\rm data},\mu}=302$ usable observations.
%we discarded depletion measurements outside
%the central $24'\times 24'$ region
%we only used depletion data in the
%central $24'\times 24'$ region, 
%yielding a net number of $N_{{\rm data},\mu}=270$.
%%%%%
%The first reason for this masking is 
%because, unlike the distortion signal,
%the depletion signal measures the local surface mass density $\kappa$,
%and hence falls rapidly with increasing radius.
%Furthermore, 
%Secondly,
%local overdensities of  
%red galaxies are apparent in the outer region
%(Figure \ref{fig:magbias})
%where the distortion signal is relatively weak; thus, this intrinsic
%clustering of the red background could dominate over the lensing singal 
%in such low density regions, and hence strongly affect the mass
%reconstruction.  
%%%%
The total number of constraints is thus $N_{\rm data}=N_{{\rm
data},g}+N_{{\rm data},\mu}=1012$, yielding $N_{\rm data}-N_{\rm
pix}=175$ degrees of freedom (dof). 
The Bayesian value of the regularization parameter $\hat{\alpha}$
that satisfies equation (\ref{eq:alpha}) was obtained as 
$\hat{\alpha}=96.6$, when $-2\hat\alpha S \approx N_{\rm pix} -
\hat\alpha {\rm Tr}({\bf  M}^{-1}) \approx 213.1 (=N_{\rm good})$. 
The minimum function value of $F$ was found to be 
$F_{\rm min}=F(\hat{\bf p})=
(1/2)\chi^2(\hat{\bf p})-\hat{\alpha}S(\hat{\bf p})
= 651.2$, and $\chi^2(\hat{\bf p})= 1089.3$, 
corresponding to ${\rm NDF} = N_{\rm data} - N_{\rm good} = 798.9$,
i.e., $\chi^2(\hat{\bf p})/{\rm NDF} = 1.36$.
Hence, the resulting $\chi^2$ is somewhat large, and is rather close to
historic MEM, i.e., $\chi^2(\hat {\bf p}) \approx N_{\rm data}$.
However, we note that our MEM mass reconstructions without the magnification
data yield $\chi^2(\hat {\bf p})/{\rm NDF} \approx 1$ (see Table
\ref{tab:memmethod}) as expected for classic MEM. Therefore, slightly
large values of this misfit statistic using the magnification data could be
attributed to the fact that the intrinsic clustering noise in the
magnification measurements is not included in the likelihood
calculation, underestimating the errors for the magnification measurements.

\begin{figure*}[!htb]
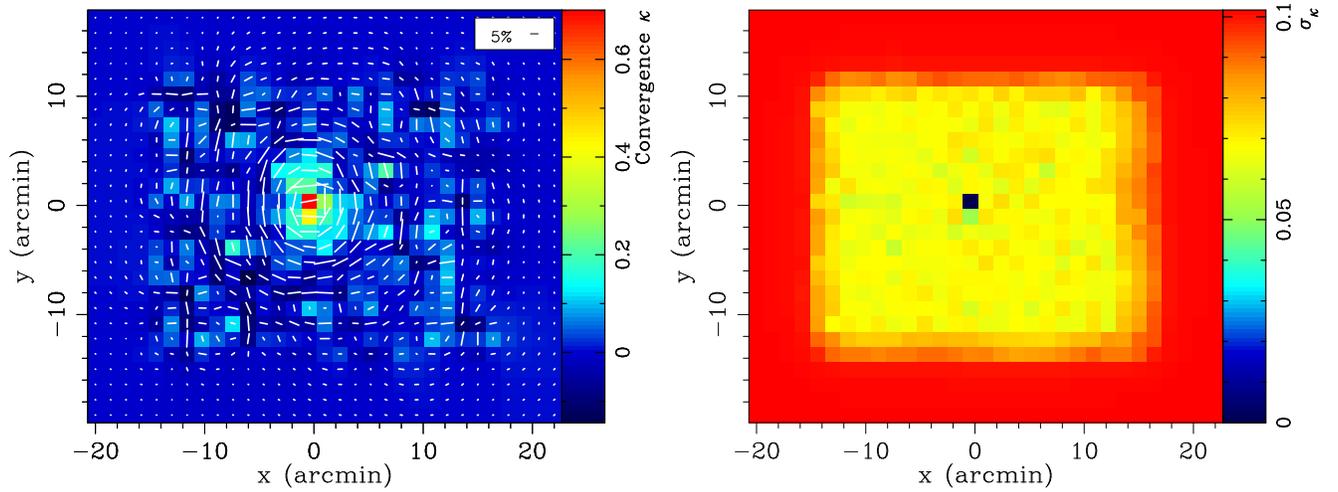

 \begin{center}
   \begin{tabular}{cc}
   \includegraphics[width=65mm,angle=270]{f7a.ps} &
   \includegraphics[width=65mm,angle=270]{f7b.ps}\\
  \end{tabular}
 \end{center} 
\caption{
Left panel:
The projected mass distribution $\kappa$ of A1689 
on a grid of $31\times 27$ pixels
reconstructed 
from an entropy-regularized, maximum-likelihood combination of
%from a maximum entropy combination of 
 Subaru shape distortion and
number-count depletion
data of red background galaxies. 
ACS strong lensing constraints are used to determine the
convergence value 
at the central pixel 
that falls in the strong
lensing region.
%%%%
The pixel width is $1.4$ arcmin.
The observing region is limited to the central $30'\times 24'$,
covering a projected area of $3.9 \times 3.1$ ${\rm Mpc}^2/h^2$,
while the field size for the reconstruction is $43\farcm 4\times 37\farcm 8$.
%%%
Overlayed up on the image is the reconstructed spin-2 gravitational
 shear field. A stick with the length of $5\%$ shear is 
indicated in the top right corner.
The north is to the top, and the east is to the left.
Right panel: 
distribution of the reconstruction errors for the 
convergence $\kappa$ 
calculated using the maximum entropy method.
The uncertainties are given by the square root of the diagonal part 
of the covariance matrix.
\label{fig:rawkappa}
}
\end{figure*}

\begin{figure*}[!htb]
 \begin{center}
   \includegraphics[width=120mm]{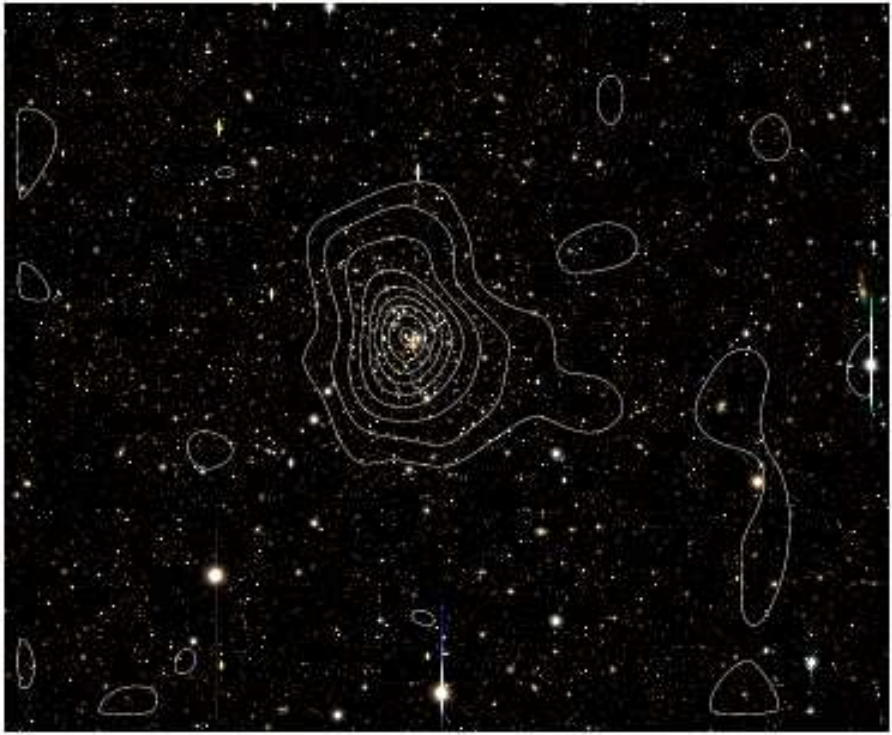}
 \end{center} 
\caption{
Contours ({\it thick}) of the  
dimensionless surface mass density $\kappa$ 
smoothed with a Gaussian of ${\rm FWHM=1\farcm 4'}$,
superposed on the
$V+i'$ pseudo-color image of A1689.
The image size is $\approx 30'\times 25'$,
covering a projected area of $3.9\times 3.2$
${\rm Mpc}^2/h^2$ at $z=0.183$.
The lowest contour
 and the contour interval are $0.05$.
 North is to the top and east is to the left.
}
\label{fig:image}
\end{figure*}

\begin{figure*}[!htb]
 \begin{center}
   \includegraphics[width=150mm]{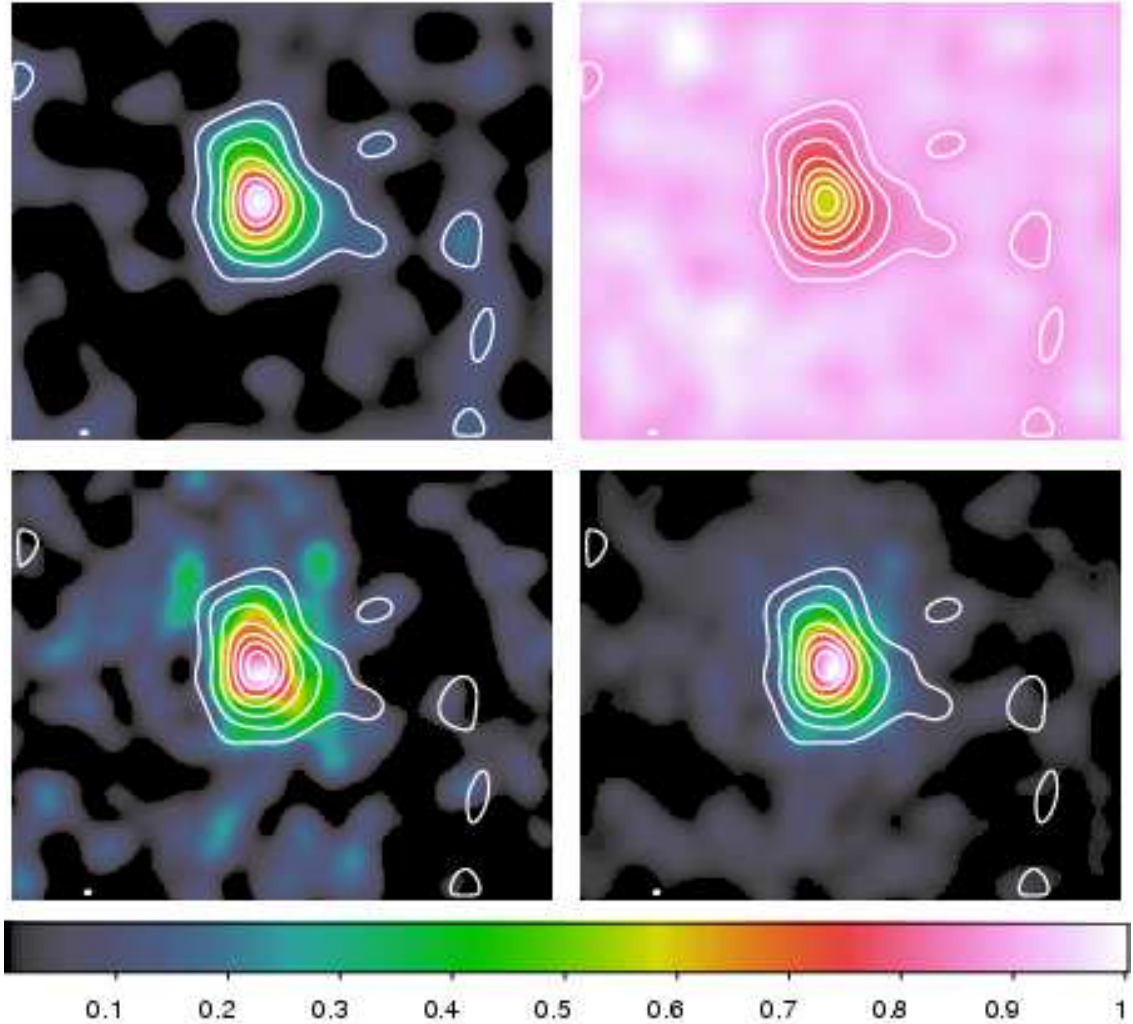}
 \end{center}  
\caption{
Comparison of the reconstructed lensing fields and the
cluster galaxy distributions in A1689.
Top: reconstructed convergence $
\kappa$ ({\it left}) and magnification bias
 $\mu^{2.5s-1}$ ({\it right}) fields.
%smoothed with a Gaussian of ${\rm FWHM}=2'$.
Bottom:
%distributions of 
observed $i'$-band surface luminosity ({\it left}) and number ({\it right}) 
density distributions of  $(V-i')$-selected 
member galaxies with $i'<23$ ABmag.
% in A1689,
%smoothed with a Gaussian of ${\rm FWHM}=2'$.
All images are smoothed with a Gaussian of ${\rm FWHM}=2'$.
The field size is $30'\times 24'$. North is to the top,
east to the left.
%%%
The contours show the 
Gaussian smoothed convergence distribution
shown  in the top-left panel.
%as the  surface luminosity and number density maps. 
%smoothed with a Gaussian of ${\rm FWHM}=2'$.
The lowest contour level and the contour interval are $0.05$.
%$\Delta\Sigma_m \approx 2.4\times h 10^{14} M_{\odot} {\rm Mpc}^{-2}$.
%with contours levels
For each panel the color scale is linear, and 
ranges from $0\%$--$100\%$ of the peak value.
}
\label{fig:lmap}
\end{figure*}

In the left panel of 
Figure \ref{fig:rawkappa} we show the resulting 
$\kappa$ map
on a grid of $31\times 27$ pixels reconstructed from 
the combined distortion and depletion data with the ACS constraint
on the mean value of $\kappa$ for the central pixel.
%%%%%%%%%%%%%%%%%%%%%%%%%%%%%%%%%%%%%%%%%%%%%%%%%%%%%%%%
The reconstructed spin-2 shear field ($\gamma[\hat{\bf p}]$)
is overlayed up on the $\kappa$ map.
The right panel in Figure \ref{fig:rawkappa} 
shows the two-dimensional distribution of the   
rms reconstruction error for $\kappa_i$, estimated from the diagonal
part of the pixel-pixel covariance matrix of errors:
$\sigma(\kappa_i) = C^{1/2}_{ii}({\bf \hat p})$. 
%%%
Figure \ref{fig:image} presents the contours of the reconstructed
$\kappa$ map 
superposed on the $V+i'$ pseudo-color image covering a field of
$30'\times 25'$ around the cluster.
Here,
for visualization purposes,
the $\kappa$ map is resampled on to a finer grid 
and convolved with a Gaussian of ${\rm FWHM}=1\farcm 4$,
corresponding to a physical resolution of ${\rm FWHM}\sim 180 {\rm
kpc}/h$ at the cluster redshift.  
%%%
The projected mass distribution of the cluster is 
smoothly varying and symmetric, with no significant substructure at 
$r\simgt 100 {\rm kpc}/h$.
%showing that mass and
%light in A1689 are similarly distributed 
%with a fairly round shape,
%and concentrated around 
%the cD galaxy.
%We note that brightest cluster galxies tend to deviate from the
%linear CM relation, so that we visually checked if such brightest
%cluster
%galaxies were properly included in our cluster member sample, and
%included them if they were missing.
%A field correction is applied to the measured luminosity and number
%densities of cluster galaxies to account for contamination by field
%galaxies.  We estimate the background luminosity/number 
%density level from an annular
%region outside of the cluster region ($r\simgt r_{\rm vir}$), and then
%subtract this background contribution from the observed
%luminosity/number density of the cluster.
%%
Figure \ref{fig:lmap} compares the reconstructed lensing fields 
and the member galaxy distributions in A1689, Gaussian smoothed
to  a resolution of ${\rm FWHM}=2'$.
%%%%
The top panels show the reconstructed $\kappa$ 
and the magnification-bias $\delta_\mu=\mu^{2.5s-1}-1$ 
fields in the left and right panels, respectively.
%%%
The bottom left and right panels display the field-corrected
$i'$-band luminosity and number density maps, respectively,
of bright red-sequence galaxies with $i'<23$ mag.
%where the background density levels are estimated from an
%annular region 
%$15'\simlt \theta \simlt 20'$
%outside of the cluster region;
%Overlayed up on the images are the mass contours shwon in the top-left
%panel. 
For each panel the color scale is linear, and 
ranges from $0\%$--$100\%$ of the peak value.
%A field correction is applied to the measured luminosity and number
%densities of cluster galaxies to account for contamination by field
%galaixes. 
It is clear from Figure \ref{fig:lmap}
that mass and light in A1689 are similarly distributed
with a fairly round shape, and well centered on the main cD galaxy.

\subsection{Mass Maps from Different Datasets and Boundary Conditions}
\label{subsec:3mem}

Any mass reconstruction technique based on the shear information
involves a non local process, meaning that one has to assume
certain boundary conditions to convert the gravitational shear field,
$\gamma(\btheta)$,into the lensing convergence field, $\kappa(\btheta)$
(Bartelmann \& Schneider 2001; Umetsu et al. 1999).
%%%
If the data field is sufficiently large so as to ensure that
projected mass fluctuations over the field average out,
or that the mean convergence over the field is zero, then 
one may simply use equation (\ref{eq:gamma2kappa}) to invert
the observed shear field into the convergence field (Kaiser \& Squires
1993). 
Or, if magnification data are available, then
a combination of complementary shearing and magnification 
information can be used to break the degeneracy between
the observables and the underlying gravitational potential field 
(Bartelmann 1995; Seitz et al. 1998).
Besides, strong lensing observations,
if available, will place additional, tight constraints on the
central mass distribution of the cluster where weak lensing
alone cannot constrain (e.g., mass reconstruction of A370 in 
Umetsu et al. 1999).

Here we consider three sets of combinations of datasets and boundary
conditions 
as summarized in 
Table \ref{tab:memmethod}: 
(i) {\it 2D MEM+} method using 
shear and magnification data with the central ACS constraint,
(ii) {\it 2D MEM} method using shear and magnification data without the ACS
constraint,
(iii) {\it 2D MEM-S} method using shear data with the central ACS constraint.
For each MEM reconstruction, we derive an
entropy-regularized maximum likelihood solution for $\kappa$
with the Bayesian value of $\alpha$.
The resulting Bayesian value of $\alpha$ and the minimized 
functional values of
$F$ and $\chi^2$ are also listed in 
Table \ref{tab:memmethod}.
%%%%

In order to quantify the significance of the reconstruction, 
we define an estimator for the detection signal-to-noise
ratio (S/N) by the following equation:
\begin{equation}
\label{eq:sn}
\left({\rm S/N}\right)^2 \equiv
\sum_{\kappa_i>0}\sum_{\kappa_j>0}
 \kappa_i \kappa_j C^{-1}_{ij},
\end{equation}
where the indices $i$ and $j$ run over all pixels 
except those with negative values of $\kappa$ 
and those in the strong lensing region
(see Table \ref{tab:memmethod}).
The reconstructions based on 
both the distortion and depletion data yield a similar
S/N  of $\sim 18$--$19$, whereas  the reconstruction
from the distortion data alone gives a slightly lower S/N of
$\sim 15$ (see Table \ref{tab:memmethod}). 
For comparison we quote the detection significance
from the 1D Subaru analysis
by B05a based on the same red background catalogs for
the weak lensing distortion and depletion analyses:
%%%%%%%%%%%%%%%%
B05a measured the lens distortion and depletion profiles 
over a radial range of $1'\simlt \theta \simlt 18'$.
%%%%%%%%%%%%%%%%
We find ${\rm S/N}\simeq 14.2$ and ${\rm S/N}\simeq 9.2$ for the 
measurements of the 
distortion and depletion profiles,
respectively. 
%%%%%%%%%%%%%%%%
Since covariance between the distortion and magnification measurements
can be neglected,
the total S/N is simply obtained as
$\sqrt{14.2^2+9.2^2} \sim 17$.
%%%%%
These numbers are quite comparable to those as measured from the
reconstructed $\kappa$ field and its covariance matrix, 
indicating that the lensing information contained in the 
red catalogs is properly propagated into the $\kappa$-basis\footnote{
This is not trivial since 
the noise level in a MEM-reconstructed
map is affected by the smoothness constraint specified by
$\alpha$. Besides, 
MEM is non-linear, so that the resulting reconstruction errors depend
on the signal as well as noise properties.}.
We find it important to use the Bayesian value for $\alpha$ in
order to have an optimal smoothness for the mass reconstruction,
avoiding oversmoothing.
A more quantitative comparison between different reconstructions
will be given in \S \ref{sec:massprofile}.

%%%%%%%%%%%%%%%%%%%%%%%%%%%%%%%%%%%%%%%%%%%%%%%%%%%%%%%%%%%%%%%%%%%%%%
%%%%%%%%%%%%%%%%%%%%%%%%%%%%%%%%%%%%%%%%%%%%%%%%%%%%%%%%%%%%%%%%%%%%%%%%%
  
%%%%%%%%%%%%%%%%%%%%%%%%%%%%%%%%%%%%%%%%%%%%%%%%%%%%%%%%%%%%%%%%%%%%%%

\section{Mass Profile of A1689}\label{sec:massprofile}

In this section, we aim to quantify and characterize 
the projected mass distribution ($\kappa$) of A1689
reconstructed from Subaru weak lensing observations,
in order to derive quantitative constraints on the 
three-dimensional mass distribution.
Specifically,
we will adopt the NFW density profile 
$\rho(r)\propto r^{-1}(1+r/r_s)^{-2}$
(Navarro et al. 1997) to describe the cluster mass distribution,  
characterized by the virial mass $M_{\rm vir}$ and the concentration
parameter $c_{\rm vir}=r_{\rm vir}/r_s$,
 defined as the ratio of the
virial radius $r_{\rm vir}$ to the scale radius $r_s$.
The best-fitting NFW parameters $(M_{\rm vir},c_{\rm vir})$
can be then compared with $\Lambda$CDM predictions 
based on $N$-body simulations (Bullock et al. 2001; Neto et al. 2007).
%%%
As a test for the consistency and reliability, we will compare
the best-fitting NFW parameters derived from different combinations of  
datasets, boundary conditions, and weak lensing techniques.
Here we introduce three different methods to 
derive a convergence profile $\kappa(\theta)$ from weak lensing
observations.

\subsection{Method (I): 2D Convergence Map}

The first method makes a direct use of the 2D $\kappa$ map
reconstructed by the entropy-regularized maximum likelihood method
(\S \ref{sec:map}).
The $\kappa$ map is directly compared with the model convergence field
for an NFW spherical halo specified by two model parameters.
%%%%
We take 
the virial mass $M_{\rm vir}$ and the concentration parameter 
$c_{\rm vir}$
%$c_{\rm vir}=r_{\rm vir}/r_s$, the ratio of the virial radius
%$r_{\rm vir}$ to the scale radius $r_s$
%evaluated at the clutser redshift 
%($z_d=0.183$) 
for describing an NFW halo. 
We employ the radial dependence of the 
convergence profile for the NFW model given by Bartelmann (1996).
Note that the Bartelmann's
formulae for the NFW convergence and shear profiles are obtained
assuming the projection  integral to extend to infinity.
Alternatively, a truncated NFW profile can be used to model the 
convergence profile (Takada \& Jain 2003).
We have confirmed that 
the best-fitting NFW parameters obtained using 
the above two different models agree to within $1\%$ 
for the case of  A1689 lensing 
(for detailed discussions,
see Baltz, Marshall, \& Oguri 2007 and
Hennawi et al. 2007). We thus simply use 
Bartlemann's formulae to calculate the relevant
lensing fields for an NFW halo
as done in B05a and B05b.

We constrain the two NFW parameters from $\chi^2$ fitting to the 
2D convergence map $\kappa(\btheta)$ derived from 
Subaru weak lensing
observations. We adopt a flat prior of $c_{\rm vir}\le 30$ for the 
halo concentration because the NFW profiles with $c_{\rm vir} \simgt 20$
cannot be distinguished by the Subaru data alone due to lack of
information on the inner density profile (B05a).
The $\chi^2$-function for the Subaru weak lensing observations,
$\chi^2=\chi^2_{\rm WL}$,
 can be expressed as
(Oguri et al. 2005) 
\begin{equation} 
\label{eq:chi2_wl} 
\chi^2_{\rm WL} = \sum_{i,j}
 \left[\hat{\kappa}(\btheta_i)-\kappa(\btheta_i)\right]  
\left(C^{-1}\right)_{ij} 
 \left[\hat{\kappa}(\btheta_j)-\kappa(\btheta_j)\right],
\end{equation}  
where $\hat{\kappa}(\btheta_i)$ is the model prediction of the 
NFW halo for the $i$th bin ($i=1,2,...,N_{\rm SL}$), 
and $(C^{-1})_{ij}$ is the inverse of
the pixel-pixel covariance matrix; $N_{\rm pix}=31\times 27=837$.
In the model fitting 
we exclude the central pixel
in the strong lensing region
 (see \S \ref{subsec:acs}).
%%%%

We have derived sets of best-fitting NFW parameters for
the three different MEM reconstructions 
listed in Table \ref{tab:memmethod}.
%based on different combinations of datasets and boundary conditions.
%%%%
Table \ref{tab:nfwfit} shows a summary 
of the best-fitting NFW parameters $(M_{\rm vir},c_{\rm vir})$
%with $1\sigma$ uncertainties
 and the resulting
$\chi^2$ value for our Subaru weak lensing observations;
%%%%
the errors quote $68\%$ confidence intervals
estimated from $\Delta\chi^2\equiv
\chi^2-\chi^2_{\rm min}=1$ 
in the $(c_{\rm vir}, M_{\rm vir})$-plane.
%%%%
Combining the distortion and magnification measurements
with the central ACS constraint (MEM+)
yields 
the best-fitting NFW parameters,
$M_{\rm vir}=(1.97\pm 0.20)\times 10^{15}M_{\odot}$ and 
$c_{\rm
vir}= 13.4^{+5.4}_{-3.3}$,
with $\chi^2_{\rm min}/{\rm dof}=332/834 (421)$,
where the value in parentheses refers to an effective degrees of
freedom excluding upper limit bins with $\kappa<0$.
%%%
As a test for the consistency, we compare 
best-fit NFW parameters for the three MEM reconstructions 
%based on different combinations of
%datasets and boundary conditions
described in \S \ref{subsec:3mem}. Firstly, all of the three MEM
reconstructions yield 
consistent results on the concentration parameter in the range,
$c_{\rm vir}= 12.6-13.6$, but with rather large uncertainties
allowing a wide range of the concentration ($8\simlt c_{\rm vir} \simlt
20$). 
%%%
On the other hand, the best-fit values for $M_{\rm vir}$ range from
$1.47\times 10^{15}M_{\odot}$  for MEM-S, through $1.62\times
10^{15}M_{\odot}$ for MEM, to $1.97 \times 10^{15}M_{\odot}$ for MEM+,
while the $1\sigma$ error level for each is $\sigma(M_{\rm vir}) \sim
0.2\times 10^{15}M_{\odot}$.

The observed location of the Einstein radius can be used for a powerful,
model independent test of the $\kappa$ profile (B05b). This is based on
the fact that the enclosed mass interior to the Einstein radius is given
by fundamental constants and with knowledge of distances involved,
namely $\bar{\kappa}(\theta_{\rm E})=1$ or $M(<\theta_{\rm
E})=\pi(D_d\theta_{\rm E})^2\Sigma_{\rm crit}$,
provided that the critical curve is nearly circular; this is the case
for A1689 (B05b).
%%%%%
Although our weak lensing measurements 
do not resolve such strong lensing phenomena,
the observed Einstein radius of $\theta_{\rm E}=45''$ 
%from the ACS observations 
(for $z_s=1$)
%%%%%
may be used to test the derived NFW models.
%with the observed Einstein radius
%of $\theta_{\rm E}=45''$ for $z_s=1$ (B05b),
To do this we numerically solve the equation $1=\bar{\kappa}_{\rm
NFW}(\theta_{\rm E})$ for the Einstein radius $\theta_{\rm E}$ by the
Newton-Raphson method.
We found 
$\theta_{\rm E}=36.7^{+22}_{-18}, 36.7^{+22}_{-20}, 45.4^{+17}_{-15}$
for the MEM-S, MEM, and MEM+ reconstructions, respectively (Table \ref{tab:nfwfit}).
All of the MEM reconstructions are roughly consistent with the ACS
measurement of the Einstein radius; however,
the constraints on
the $\theta_{\rm E}$ 
placed by the Subaru data alone 
%from the Subaru observations alone 
are still rather weak, naturally due to the lack of central mass 
distribution.
%satisfy the Einstein radius
%constraint by the ACS observations

To make a direct comparison with model predictions,
we compute the convergence profile
$\kappa(\theta)$ 
from a weighted radial projection of the 2D $\kappa$ map as:
%Given a two-dimensional $\kappa(\btheta_i)$ map 
%and the covariance matrix $C_{ij}$ of the reconstruction errors
%($i,j=1,2,...,N_{\rm pix}$),
%we can calculate the convergence profile $\kappa(\theta)$ 
%of the cluster
%from a weighted radial projection of the pixelized
% $\kappa(\btheta)$ map:
\begin{equation}
\label{eq:binkappa}
\kappa_m\equiv
\kappa(\bar\theta_m) = \sum_{i\in {\rm Bin} m} W_{im} \kappa(\btheta_i)
\Big/
\sum_{i\in {\rm Bin} m} W_{im},
\end{equation}
where $W_{im}$ ($0\le W_{im}\le 1$) is the fraction of the area of the 
$i$th pixel lying within the $m$th annular bin. We use Monte Carlo
integration to calculate these area fractions for individual pixels
(Marshall et al. 2002), and $\bar\theta_m$ is the 
area-weighted center of the $m$th radial bin:
\begin{equation}
\theta_m = \sum_{i\in {\rm Bin} m} W_{im} |\btheta_i|
\Big/
\sum_{i\in {\rm Bin} m} W_{im}.
\end{equation}
 Since the $\kappa_m$ profile is
expressed as
a linear combination of $\kappa(\btheta_i)$ values, it is
straightforward to calculate the covariance matrix 
$C_{mn} \equiv \langle
\delta\kappa_m\delta\kappa_n
%%\delta\kappa_(\bar\theta_m)\delta\kappa(\bar\theta_n)
\rangle$
of the reconstruction errors:
\begin{eqnarray}
\label{eq:bin2bincovar}
C_{mn} 
&=&  \sum_{i\in {\rm Bin} m} \sum_{j\in {\rm Bin} m}
W_{im}W_{jn} C_{ij}
\Big/
\sum_{i\in {\rm Bin} m} W_{im}
\sum_{j\in {\rm Bin} n} W_{jn}\nonumber\\
&+&
\delta_{mn}
%\left(
\sum_{i\in {\rm Bin} m} 
W_{im}^2 [\kappa_m-\kappa(\btheta_i)]^2
%\right)
\Big/
\left(\sum_{i\in {\rm Bin} m} W_{im}
\right)^2,
\end{eqnarray}
where the first term represents the errors propagated from the 
pixel-pixel covariance matrix of the 2D $\kappa$ reconstruction, 
and the second term is responsible for variations of 
the $\kappa(\btheta)$-field along the azimuthal direction, i.e.,
departure from circular symmetry.

\begin{figure*}[!htb]
 \begin{center}
  \includegraphics[width=100mm, angle=270]{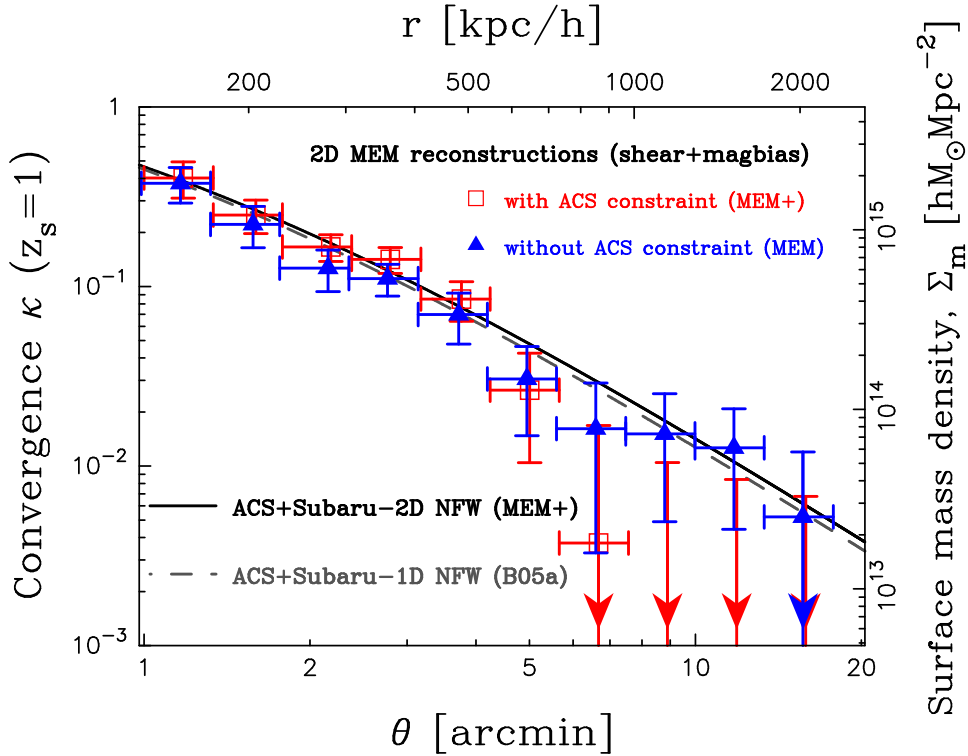}
 \end{center} 
\caption{
Mass profiles of A1689 
obtained by a radial projection of 
the 2D $\kappa$ map
(see Figure \ref{fig:rawkappa}) 
reconstructed from an entropy-regularized 
maximum likelihood combination of 
Subaru distortion and depletion measurements.
All of the profiles are scaled to a fiducial source redshift of 
$z_s=1$.
%%%%
The square and triangle symbols
represent the results with (2D MEM+) and without (2D MEM) 
ACS strong lensing constrains on the central mass density,
respectively.
The error bars are highly correlated in the different bins.
%Downwards-pointing arrows 
%are used  where the lower error bar drops below zero.
%%%
%When the ACS constraints are  not incorporated 
%in the mapmaking,
Without the central ACS constraint,
central $\kappa$-bins at $\theta\simlt 4'$ are 
slightly underestimated ($\sim 10\%$), but the two profiles are
overall in good agreement within the statistical uncertainties.
%%%
The solid curve shows the best-fitting NFW profile
for the 2D $\kappa$ map (2D MEM+) combined with the ACS-derived
inner mass profile.
%the mass profile with the ACS constraints (2D MEM+).
%%%%
For comparison
an NFW model based on the ACS+Subaru 1D analysis (B05a)
%on the 1D ACS data (B05b) and 
%combined Subaru distortion/depletion data (B05a)
is shown as a dashed curve.
%The dashed curve shows 
%the best-fitting NFW model 
%from the combined ACS+Subaru-1D analysis (B05a).
} 
\label{fig:kprof_mem}
\end{figure*}

\begin{figure*}[!htb]
 \begin{center}
  \includegraphics[width=100mm, angle=270]{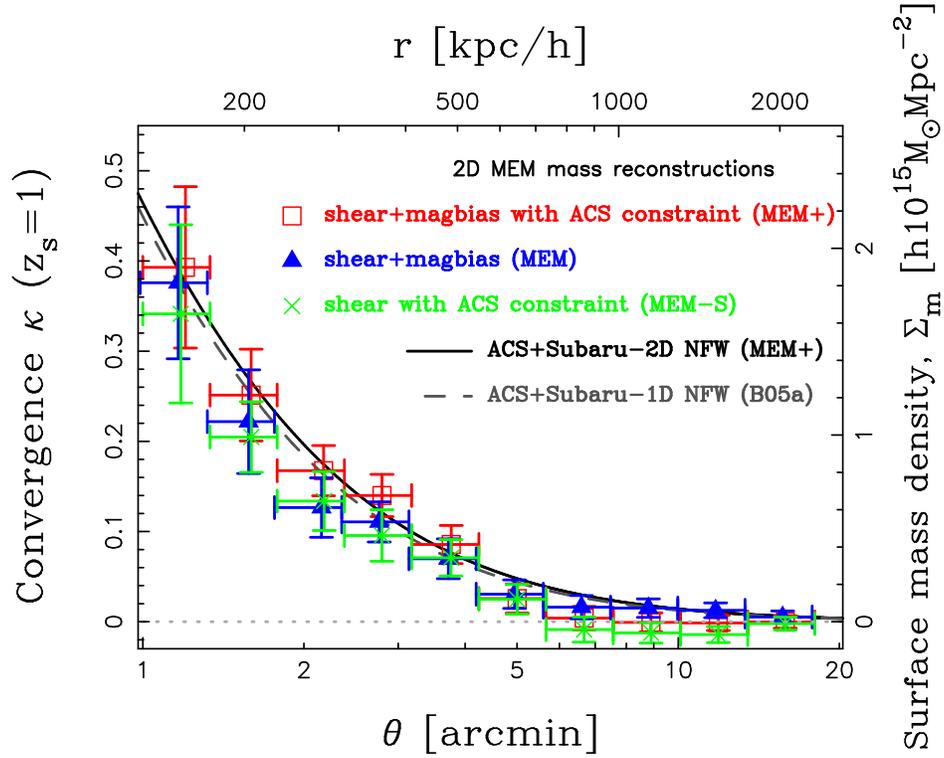}
 \end{center} 
\caption{
Comparison of 
mass profiles from MEM-reconstructed $\kappa$ maps
based on different combinations of Subaru datasets and boundary conditions.
%%%
All of the profiles are scaled to a fiducial source redshift of $z_s=1$.
The square and triangle  symbols represent the results from the
combined distortion and depletion measurements,
with (MEM+) and without (MEM) the 
ACS constraint on the mean surface mass density
in the central pixel, respectively.
%%% 
The crosses show the results from the distortion data with
the central ACS constraint (MEM-S).
%%%
The model curves are shown for comparison as in 
Figure \ref{fig:kprof_mem}.
The mass profile from the distortion data alone ({\it crosses})
shows a slight negative dip of $\kappa\sim -0.01$
at $6'\simlt \theta \simlt 10$ due to spurious boundary effects. 
%due to incorrect 
%boundary condition in the biased rich cluster field.
%The 1D $\zeta_{\rm c}$-based mass profile
%increases continously 
%from the outer boundary towards the center, but lies
%systematically below the MEM+ reconstruction with the
%ACS inner boundary condition.
} 
\label{fig:kprof_3mem}  
\end{figure*}

Here we derive 
for each of the 2D $\kappa$ maps
a discrete convergence profile 
(\ref{eq:binkappa})
in $N_{\rm bin}$
logarithmically spaced bins for $\theta = 
[\theta_{\rm min},\theta_{\rm max}]$; we adopt the same binning as in B05a:
$N_{\rm bin}=10$, $\theta_{\rm min}=1'$,
and $\theta_{\rm max}=18'$, as summarized in Table \ref{tab:mrec}.
In Figure \ref{fig:kprof_mem} we compare the convergence profiles from the
2D MEM+ and MEM reconstructions,
both of which are based on the 
lens distortion and magnification measurements,
 to clarify the effect of the ACS
constraint on the central surface mass density.
The error bars represent the $1\sigma$ uncertainties 
 from the diagonal
part ($C_{mm}$) of the bin-to-bin covariance matrix given by
equation (\ref{eq:bin2bincovar}), and hence are correlated between
the different bins.
One can see that overall the convergence profile obtained with the 
central ACS
constraint is slightly steeper than that without the ACS constraint,
and has a slightly higher overall normalization;
these features are well explained by
%as explained by 
slightly higher NFW mass ($M_{\rm vir}$) and concentration ($c_{\rm vir}$)
derived for the results with the central ACS constraint.
Nonetheless, the two convergence profiles 
are overall in good agreement within the
statistical uncertainties. 
Steep NFW profiles with 
$c_{\rm vir}\sim 13$ and 
$M_{\rm vir}\sim 2\times 10^{15}M_{\odot}$
are well fitted to the reconstructed convergence profiles 
(solid and dashed curves).
Figure \ref{fig:kprof_3mem} shows convergence profiles from the three
MEM reconstructions: MEM+ (squares), MEM (triangles), and MEM-S (crosses). 
Note that the vertical axis is linear here rather than logarithmic.
The convergence profile from the distortion data alone (MEM-S) shows a
slight negative dip of $\kappa\sim -0.01$ at $\theta'\simgt 6'$ due to
boundary effects.

\subsection{Method (II): 1D Maximum Likelihood Analysis}  

%upporting the assumption of. quasi-circular symmetry in the projected
%mass ... 2005, ApJ, 621, 53) and the Subaru weak lensing analysis with
%the 1D reconstruction method ...

In B05a we developed a
%model-independent 
method for reconstructing the discrete convergence profile $\kappa_m$ 
from a maximum likelihood combination of radial profiles of the lens
distortion and magnification effects on red background galaxies (see
Figure 3 of B05a).  
%%%
With the assumption of quasi-circular symmetry in the 
projected mass distribution (or the projected potential field),
one can express the lensing observables (i.e., tangential distortion and
magnification bias) in terms of the binned convergence profile (see also
\S \ref{subsec:zeta}). Owing to the nature of shear-based
reconstruction, 
boundary conditions need to be specified (as discussed in \S
\ref{subsec:3mem}). 
In B05a we set the inner boundary condition on the mass interior to 
$\theta_{\rm min}=1'$, which is readily obtained from the
well-constrained ACS mass model.
%this 1D method also requires to set inner or outer
%boundary conditions 

One of the advantages of such a 1D method is that one can improve up on 
the statistical significance of each measurement pixel by azimuthal
averaging, provided that the system is nearly symmetric. 
%%%
B05a measured from the red background sample
a tangential distortion 
and a radial depletion profile
in 10 discrete radial bins, 
with total significance of ${\rm S/N}\simeq 14.2$ 
and ${\rm S/N}=9.2$, respectively
(see \S \ref{subsec:3mem}). Thus, the per-pixel S/N of each measurement
is of the order of unity, which is optimal for an inversion problem.
Accordingly,
the 1D analysis does not require 
regularization techniques. 
The best-fitting NFW parameters for the Subaru 1D analysis 
are listed in 
Table \ref{tab:memmethod}.
%%%
%In \S \ref{subsec:comparison} we compare the convergence profile from
%the Subaru 1D analysis by B05a with the results from the 2D MEM
%reconstructions. 

\subsection{Method (III): Aperture Densitometry}
\label{subsec:zeta}

The spin-2 shape distortion of an object 
due to gravitational lensing
is described by
the complex reduced shear, $g=g_1+i g_2$ (see equation [\ref{eq:redshear}]),
 which is coordinate dependent.
For a given reference point on the sky, one can instead 
form coordinate-independent
quantities, 
the tangential distortion $g_+$ and the $45^\circ$ rotated component,
from linear combinations of the distortion coefficients
$g_1$ and $g_2$ as
\begin{equation}
g_+ = -(g_1 \cos 2\phi + g_2\sin 2\phi), \ \ 
g_{\times} = -(g_2 \cos 2\phi - g_1\sin 2\phi),
\end{equation}
%For a given point on the sky,
%the tangential component $g_+$ is used to obtain the azimuthally
%averaged distortion 
%due to lensing and computed from the distortion
%coefficients $g_1$ and $g_2$ as:
%\begin{equation}
%g_+= -(g_1 \cos 2\phi + g_2\sin 2\phi),
%\end{equation}
where $\phi$ is the position angle of an object with respect to
the reference position, and the uncertainty in the $g_+$ and
$g_{\times}$ 
measurement 
is $\sigma_+ = \sigma_{\times } = \sigma_{g}/\sqrt{2}\equiv \sigma$ 
in terms of the rms error $\sigma_{g}$ for the complex shear measurement.
In practice, the reference point is taken to be the cluster center,
which is well determined from symmetry of the strong lensing pattern.
%%%%%
To improve the statistical significance of the distortion measurement,
we calculate the weighted average of the $g_+$'s and its weighted error
as
\begin{eqnarray}
\label{eq:gt}
\langle g_+(\theta)\rangle &=& \frac{\sum  g_+/\sigma^2}{\sum
 1/\sigma^2},\\
\sigma_+(\theta) &=& \left(\sum 1/\sigma^2\right)^{-1/2}.
\end{eqnarray}

For a shear-based estimation of the cluster mass profile
we use a variant of the weak lensing aperture densitometry, or
so-called the $\zeta$-statistic (Fahlman et al. 1994; Clowe et
al. 2000), 
of the form:
\begin{eqnarray}
\label{eq:zeta}
\zeta_{\rm c}(\theta)
&\equiv &
2\int_{\theta}^{\theta_{\rm inn}}\!d\ln\theta' 
\gamma_+(\theta')\nonumber\\
&&+ \frac{2}{1-(\theta_{\rm inn}/\theta_{\rm out})^2}\int_{\theta_{\rm
inn}}^{\theta_{\rm out}}\! d\ln\theta'  
\gamma_+(\theta')
\nonumber\\
&=& 
\bar{\kappa}(\theta) - \bar{\kappa}
(\theta_{\rm inn}<\vartheta <\theta_{\rm out}),
\end{eqnarray}
where 
$\kappa(\theta)$ is the azimuthal average of 
the convergence field $\kappa(\btheta)$ at radius $\theta$,
$\bar{\kappa}(\theta)$ is the average convergence interior to 
 radius $\theta$,
 $\theta_{\rm inn}$ and $\theta_{\rm out}$ are the inner and
outer radii of the annular background region in which the mean
background contribution, 
$\bar{\kappa}_b\equiv
\bar{\kappa}(\theta_{\rm inn}<\vartheta <\theta_{\rm out})$,
is defined;
the $\gamma_+(\theta) = \bar{\kappa}(\theta)-\kappa(\theta)$ 
is an azimuthal average of the tangential
component of the gravitational shear at radius $\theta$
(Fahlman et al. 1994),
which is observable in the weak lensing limit: $\gamma_+(\theta)\approx
\langle g_+(\theta)\rangle$.
%%%%%
This cumulative mass estimator
subtracts  from the mean convergence 
$\bar{\kappa}( \theta)$
a constant 
$ \bar{\kappa}_{\rm bg}$
%=\bar{\kappa}(\theta_{\rm inn}<\vartheta <\theta_{\rm out})$ 
%for fixed inner and outer radii,
%$\theta_{\rm inn}$ and $\theta_{\rm out}$, 
for all apertures $\theta$ in the measurements, 
thus removing any DC component in the control
region $\theta = [\theta_{\rm inn}, \theta_{\rm out}]$.
We note that the $\bar{\kappa}_b$ is 
a non-observable free parameter, and we 
use this degree-of-freedom
to fix the outer boundary condition,
and hence to derive a convergence profile.
%%%%%
%%%%%

We compute the aperture densitometry profile
$\zeta_{\rm c}(\theta_m)$ 
in $N_{\zeta}=10$ logarithmically spaced bins
for $\theta = [\theta_{\rm min},\theta_{\rm max}]$,
which we set to 
$\theta_{\rm min}=1'$ and $\theta_{\rm max}=16'$;
the maximum radius $\theta_{\rm max}$ is comparable to 
the angular size of 
the cluster virial radius,
 $\theta_{\rm vir}=15\farcm 7$
($r_{\rm vir} \sim 2 {\rm Mpc}/h$), 
according to the ACS+Subaru-1D model (B05a).
%%%%
The inner and outer background radii 
%$(\theta_{\rm inn},\theta_{\rm out})$
are set to $\theta_{\rm inn}=\theta_{\rm max}=16'$ and $\theta_{\rm
out}=19'$, respectively.
%%%%
Using the ACS+Subaru-1D model by B05a,
the mean background level, $\bar\kappa_b$, is calculated to be 
$\bar\kappa_b\sim 4.0 \times 10^{-3}$
for our red background population with $\langle
D_{ds}/D_s\rangle=0.693$, at an effective source redshift of $z_{s,D}=0.68$ 
(see equation [\ref{eq:zD}]).
Note that the current 
$1\sigma$ upper limit is
$\kappa(\theta) \simlt 0.01$ at $\theta\simgt 8'$ (B05a).
%%%
For a given boundary condition $\bar{\kappa}_b$, the average
convergence $\bar{\kappa}(\theta_m)$ is estimated as
\begin{equation}
\label{eq:z2avk}
\bar{\kappa}(\theta_m) = \zeta_{\rm c}(\theta_m) + \bar{\kappa}_b.
\end{equation}
Then, we define a discretized estimator for $\kappa$ as
\begin{equation}
\label{eq:zeta2kappa}
\kappa_m\equiv
\kappa(\overline{\theta}_m)
= \alpha_2^m \zeta(\theta_{m+1})
- \alpha_1^m \zeta(\theta_{m})
+(\alpha_2^m-\alpha_1^m)\bar{\kappa}_b,
\end{equation}
where
\begin{equation}
\alpha_1^m = \frac{1}{2\Delta\ln\theta_m} 
\left( 
  \frac{\theta_{m}}{ \overline{\theta}_m }
\right)^2, \, \, 
\alpha_2^m = \frac{1}{2\Delta\ln\theta_m} 
\left(\frac{\theta_{m+1}}{ \overline{\theta}_m }\right)^2,
\end{equation}
and 
 $\bar\theta_m$ is the weighted center of the $m$th radial bin
($m=1,2,...,N_{\zeta}-1$; see Appendix \ref{app:kappad}).
%%%
It is worth noting that, unlike strong-lensing based boundary
conditions (e.g., B05a),
%where the location of Einstein rings or the model-predicted
%interior mass is used to normalize $\kappa$, 
this method utilizes an outer boundary condition on the mean background
density $\bar\kappa_b$ to derive a $\kappa$ profile (see Schneider \&
Seitz 1995 for an alternative method for a direct inversion of the mass
profile).   
The error covariance matrix $C_{mn}$ of $\kappa_m$ is expressed as
\begin{eqnarray}
C_{mn} \equiv \langle \delta\kappa_m \delta\kappa_n \rangle
&=& 
\alpha_2^m \alpha_2^n C^{\zeta}_{m+1,n+1}
+  
\alpha_1^m \alpha_1^n C^{\zeta}_{m,n}\nonumber\\
&-&
\alpha_1^m \alpha_2^n C^{\zeta}_{m,n+1}
-
\alpha_2^m \alpha_1^n C^{\zeta}_{m+1,n},
\end{eqnarray}
where $C^{\zeta}_{mn}\equiv \langle \delta\zeta_m\delta\zeta_n\rangle$
is the bin-to-bin error covariance matrix of the aperture densitometry
measurements which is calculated by propagating the rms errors
$\sigma_+(\theta_m)$ for the tangential shear measurement (Okabe \&
Umetsu 2008).

In the non-linear regime, however,
the $\gamma_+(\theta)$ is not a direct observable.
Therefore, non-linear corrections need to be taken into account 
in the mass reconstruction process.
%%%
In the subcritical regime (i.e., outside the critical curves),
the $\gamma_+(\theta)$
can be expressed in terms of the 
the averaged tangential reduced shear as
$\langle g_+(\theta) \rangle \approx
\gamma_+(\theta)/[1-\kappa(\theta)]$
assuming a quasi circular symmetry in the projected mass distribution
(B05a; Umetsu et al. 2007).
This non-linear equation (\ref{eq:zeta})
for $\zeta_{\rm c}(\theta)$ can be solved by an iterative procedure:
Since the weak lensing limit ($\kappa,|\gamma|,|g|\ll 1$) holds in the
background region $\theta_{\rm inn}\le \theta \le \theta_{\rm
max}$, we have the following iterative equation for $\zeta_{\rm
c}(\theta)$: 
\begin{eqnarray}
\zeta_{\rm c}^{(k+1)}(\theta)
&\approx&
2\int_{\theta}^{\theta_{\rm inn}}\!d\ln\theta' 
\langle g_+(\theta)\rangle [1-\kappa^{(k)}(\theta)] \nonumber\\
&+& \frac{2}{1-(\theta_{\rm inn}/\theta_{\rm out})^2}\int_{\theta_{\rm
inn}}^{\theta_{\rm out}}\! d\ln\theta'  
\langle g_+(\theta')\rangle,
\end{eqnarray} 
where $\zeta_{\rm c}^{(k+1)}$ 
represents the 
aperture densitometry in the $(k+1)$th 
step of the iteration
$(k=0,1,2,...,N_{\rm iter})$; 
the $\kappa^{(k+1)}$ is
calculated from $\zeta_{\rm c}^{(k+1)}$ using equation (\ref{eq:zeta2kappa}).
This iteration is preformed by
starting with $\kappa^{(0)}=0$ for all radial bins, and repeated 
until convergence is reached at all radial bins. For a fractional
tolerance of $1\times 10^{-5}$, this iteration procedure converges
within $N_{\rm iter} \sim 10$ iterations.
We compute errors for $\zeta_{\rm c}$ and $\kappa$ 
with the linear approximation.

\begin{figure*}[!htb]
 \begin{center}
  \includegraphics[width=100mm, angle=270]{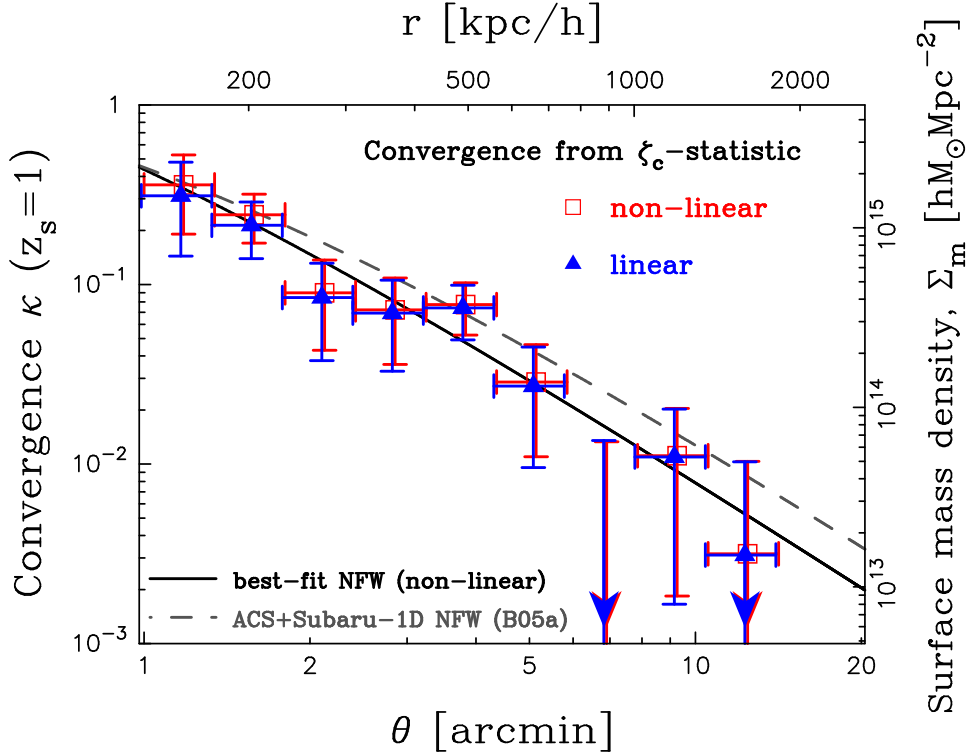}
 \end{center} 
\caption{
%Importance of non-linear corrections in 
%the 
Shear-based 1D mass reconstruction
utilizing the $\zeta_{\rm c}$-statistic.
As an outer boundary condition,
the mean background density $\bar\kappa_b$
in the range $16'\simlt \theta\simlt 19'$ is set to
$\bar\kappa_b=4\times 10^{-3}$ according to the ACS+Subaru-1D best-fit
 NFW model (B05a).
%%%
The square symbols represent the results
with the non-linear corrections. The triangle symbols 
show the reconstruction with linear approximation.
The mass profiles are scaled to a fiducial source redshift
of $z_s=1$.
Decorrelated error bars are shown.
Downwards-pointing arrows 
are used  where the lower error bar drops below zero.
%%%
Without the non-linear corrections, central bins are underestimated
by $\sim 15\%$ at maximum.
%%%%
The solid curve shows 
an NFW profile with a high concentration, $c_{\rm vir}=30$,
matching well the overall profile obtained with the non-linear
corrections.
%the best-fitting NFW profile 
%with high concentration $c_{\rm vir}=30$
%for the $\kappa$ profile obtained with the non-linear corrections.
%%%%
For comparison
an NFW model 
%from on the 1D analysis (B05a)
based on 
the combined 
ACS and Subaru distortion and depletion profiles (B05a)
is shown as a dashed curve.
}
\label{fig:zeta2kappa} 
\end{figure*}

Figure \ref{fig:zeta2kappa} shows the resulting model-independent mass
profile (squares) of A1689 with decorrelated error bars, along with the
results without the non-linear corrections (triangles).
Without the non-linear corrections, central bins at $\theta \simlt 3'$
are underestimated by $\sim 15\%$ at maximum.
%%% revised version
%The outer radial bins at $\theta \approx 7\arcmin$ and $12\arcmin$ are
%consistent with a null signal to within $1\sigma$.
%%% submitted version
%The outer profile at $\theta \simgt  6'$ is consistent with a null signal
%to within $1\sigma$. 
The reconstructed convergence in the first bin is 
%%% revised version :: kappa(1\farcm 2) = 0.322 \pm 0.15
$\kappa(1\farcm 2)= 0.32 \pm  0.15$ 
%%% submitted version :: wrong background aperture used
%$\kappa(1\farcm 2)= 0.316 \pm  0.122$ 
(decorrelated $1\sigma$ error),
which is consistent at the $\sim 1\sigma$ level
with the previous Subaru 1D results (B05a):
$\kappa(1\farcm 2)=0.44^{+0.11}_{-0.12}$ ($1\sigma$) for the red
background population with $\langle D_{ds}/D_s\rangle\approx 0.693$.
%based on the combined Subaru distortion and depletion profiles.
%%%%%
We note that B05a utilized an inner boundary condition on the mean
$\kappa$ interior to $\theta_{\rm min}=1'$ based on the ACS mass profile.

Force fitting an NFW profile to the derived $\kappa$ profile yields
%% @@submitted version
%% $M_{\rm vir}=(1.51\pm 0.26)\times 10^{15} M_{\odot}$ ($r_{\rm vir}=
%% 1.87\pm 0.10 {\rm Mpc}/h$) with the adopted prior $c_{\rm vir}\le
%% 30$, 
%% @@revised version (2008/04/08)
 $M_{\rm vir}=(1.48\pm 0.27)\times 10^{15} M_{\odot}$ and $c_{\rm
 vir}=27.3^{+2.7}_{-19.3}$ 
with the adopted prior $c_{\rm vir}\le 30$,
%($r_{\rm vir}=
% 1.87\pm 0.10 {\rm Mpc}/h$) with the adopted prior $c_{\rm vir}\le
where $\chi^2_{\rm min}/{\rm dof}=5.2/7 (7)$.
%%%
It is interesting to compare the reconstructed $\kappa$ profile with  
the tangential shear profile, $\langle g_{+}(\theta)\rangle$,
which is a direct observable with uncorrelated measurement errors
(see equation [\ref{eq:gt}]). In B05a we show the tangential distortion
profile for the same red background sample as used in the present study
(see Figure 1 of B05a, or Figure \ref{fig:gt} below).  
%%%
The best-fitting NFW model for the shear profile is
given by $M_{\rm vir}=(1.51\pm 0.26)\times 10^{15}M_{\odot}$ and
a high concentration,
$c_{\rm vir}=20^{+8.8}_{-5.3}$ ($\chi^2_{\rm min}/{\rm dof}=5.0/8$), 
which is in good agreement with the results from the 1D reconstruction
based on the aperture densitometry.  
%%%%
Such a detailed agreement ensures the validity of the boundary condition
for a shear-based mass reconstruction. 
%%%
We note that simply assuming $\bar\kappa_b=0$ yields similar results, 
 $M_{\rm vir}=(1.51\pm 0.3) \times 10^{15}M_{\odot}$
($c_{\rm vir}\le 30$) 
with $\chi^2_{\rm min}/{\rm dof}=5.5/7 (5)$,
being consistent within the $1\sigma$ uncertainty.

\subsection{Combining Strong and Weak Lensing Results}
\label{subsec:acs+subaru}

\subsubsection{ACS Constraints}
\label{subsubsec:acs}

As demonstrated in B05a and the previous sections,
it is crucial to have information on the 
central mass distribution in order to 
derive useful constraints on the concentration parameter,
and hence to distinguish the NFW profile from others.
To do this, we constrain the two NFW parameters
from $\chi^2$ fitting to the 
combined Subaru weak lensing and ACS
strong lensing observations following B05a and Oguri et al. (2005):
\begin{equation}
\label{eq:chi2_combined}
\chi^2=\chi^2_{\rm WL} + \chi^2_{\rm SL},
\end{equation}
where the $\chi^2_{\rm WL}$ for the Subaru observations is defined by
equation (\ref{eq:chi2_wl}).
For the ACS data we use an azimuthally averaged profile of the $\kappa$
map well constrained by the strong lensing observations of B05b; 
this profile is given in $N_{\rm SL}=12$ bins linearly spacing 
over the range $\theta = [0\farcm 077, 0\farcm 97]$
(see Figure 22 of B05b, or Figure 3 of B05a), and the amplitude
is scaled to $\langle D_{ds}/D_s\rangle=0.693$ of our red background
sample from $D_{ds}/D_s=0.881$ at $z_s=3$. 
It is important to note that the ACS strong lensing analysis of B05b
unveiled the secondary mass clump associated with a small clump of
galaxies (Teague et al. 1990; Czoske 2004). However, the mass 
contribution of this subclump is only a small fraction of 
the cluster mass component (see Figures 21 and 22 of B05b)
and it only slightly perturbs the tangential critical curve
(see B05b). The ACS mass profile is corrected for the subclump, where
the secondary clump region is locally masked out when taking an azimuthal
average (B05b). 
%%%
By combining the Subaru and ACS lensing analyses,
we can trace the cluster mass distribution over a 
large range in amplitude $\kappa\sim [10^{-3},1]$ and
in projected radius $r=[10^{-2},2] {\rm Mpc}/h$. 
%%%
The $\chi^2$ function for the ACS observations is expressed as
\begin{equation}
\chi^2_{\rm SL} = \sum_{i=1}^{N_{{\rm SL}}}\frac{
(\kappa_i-\hat\kappa_i)^2
}
{\sigma_i^2}
\end{equation} 
where $\hat{\kappa}(\btheta_i)$ is the model prediction of the 
NFW halo for the $i$th pixel ($i=1,2,...,N_{\rm pix}$), 
and $\sigma_i$ is the $1\sigma$ error for $\kappa_i$; 
the bin width of the
ACS-derived convergence profile is sufficiently broad to 
ensure that the errors between different bins are independent (B05b)

\subsubsection{Einstein-Radius Constraint}
\label{subsubsec:ein}

Alternatively, as a model independent constraint,\footnote{We
note that this is based on the assumption of the circular symmetry in
the projected lens mass distribution, and that a rather tight prior
(estimated from data) is applied to one of the ``parameters'' (i.e.,
$\theta_{\rm E}$). }  
we can utilize the observed location of tangential critical curves
traced by giant arcs and multiply imaged background galaxies, defining
an approximate Einstein radius, $\theta_{\rm E}$.
This radius is determined from strong lensing modeling of many multiple
images visible around these clusters in deep HST/ACS images (B05b),
and corresponds to the theoretically extreme value of the ellipticity,
$g_+(\theta_{\rm E})=1$. This Einstein-radius constraint
complements in a model independent manner the weak lensing shape
measurements.

We constrain the NFW model parameters ($c_{\rm vir}, M_{\rm vir}$) 
by combining weak lensing profiles and the Einstein-radius
constraint. We define the $\chi^2$ function for combined weak lensing
distortion and Einstein-radius constraints by 
\begin{equation}
\label{eq:chi2_g_rein}
\chi^2 = \sum_m
\frac{\left[ \langle g_{+}(\theta_m) \rangle - \hat{g}_{+}(\theta_m)
      \right]^2} 
  {\sigma_{+}^2(\theta_m)} + 
\frac{\left[1-\hat{g}_+(\theta_{\rm E},z_{\rm E})\right]^2}
 {\sigma_{+,{\rm E}}^2},
\end{equation}
where the first term is the $\chi^2$-function for the Subaru
tangential shear measurements (\S \ref{subsec:zeta}) and the second term
is the $\chi^2$ function for the Einstein radius constraint;
$\hat{g}_{+}(\theta_m)$ is the NFW model prediction
for the reduced tangential shear at $\theta=\theta_m$ calculated for the
red background sample (see \S  \ref{subsec:red}),
$\hat{g}_{+}(\theta_{\rm E}, z_{\rm E})$ is the  
model prediction for the reduced tangential shear at $\theta=\theta_{\rm
E}$, evaluated at the arc redshift, $z_s=z_{\rm E}$, and $\sigma_{+,{\rm
E}}$ is the rms error for $g_+(\theta_{\rm E}, z_{\rm E})$. 
Following B05b, we take $\theta_{\rm E}=45\arcsec$ at $z_E=1$, and
assume conservatively a 10\% error for the Einstein radius:
$\sigma_{\theta_{\rm E}}/\theta_{\rm E}=0.1$. We then
propagate this error to $g_+$ assuming a singular isothermal sphere (SIS)
model; At $r\simlt r_s$, the density slope of NFW is shallower than that
of SIS (see Figure 1 of Wright \& Brainerd 2000), so that this gives a
fairly conservative estimate of $\sigma_{+,{\rm E}}$. 
For the SIS model, $\partial\ln g_+/\partial\ln \theta_{\rm E}=2$
%$\Delta\ln g_+ = 2\Delta\ln \theta_{\rm E}$ 
at $\theta=\theta_{\rm E}$, so that $\sigma_{+,{\rm E}} =
2\sigma_{\theta_{\rm E}}/\theta_{\rm E}=0.2$.
%for our assumed $10\%$ uncertainty in $\theta_{\rm E}$.

Alternatively we can combine the $\kappa$ profile reconstructed from
weak lensing distortion measurements (\S \ref{subsec:zeta}) with the
Einstein-radius constraint. For an axially symmetric lens, the
Einstein-radius constraint is simply written as
$\bar{\kappa}(\theta_{\rm E},z_{\rm E})=1$. Thus, the $\chi^2$ function
is now expressed as
  \begin{equation}
\label{eq:chi2_k_rein}
\chi^2 = \sum_{m,n}
\left[\kappa(\theta_m) - \hat\kappa(\theta_m) \right]
 \left(C^{-1}\right)_{mn}
\left[\kappa(\theta_n) - \hat\kappa_n(\theta_n) \right]
+ 
\frac{
\left[ 1-\hat{\bar{\kappa}}(\theta_{\rm E},z_{\rm E}) 
\right]^2 
} 
{\sigma_{\bar\kappa,{\rm E}}^2},
\end{equation}
where $\hat{\bar{\kappa}}(\theta_{\rm E}, z_{\rm E})$ is the model
prediction for the average convergence interior to the Einstein radius 
$\theta_{\rm E}$ evaluated at the arc redshift of $z_s=z_{\rm E}$, and
$\sigma_{\bar\kappa, {\rm E}}$ is the rms error for
$\bar\kappa(\theta_{\rm E}, z_{\rm E})$,
which we take as
$\sigma_{\bar\kappa,{\rm E}} = \sigma_{\theta_{\rm E}}/\theta_{\rm
E}=0.1$.

%normalize but leave a degree of freedom corresponding to the
%uncertainty in the Einstein radius.

\subsubsection{Comparison} 
\label{subsubsec:comparison}

Table \ref{tab:nfwfit} summarizes for each mass reconstruction
the best-fitting NFW parameters
$(M_{\rm vir},c_{\rm vir})$, the minimized $\chi^2$ value with respect
to the degrees of freedom, and the predicted Einstein radius
$\theta_{\rm E}$ for a fiducial source at $z_s=1$.
Here for each of the Subaru reconstructions, we compare the
best-fit NFW parameters with and without the ACS inner profile combined,
as indicated in the third column of Table \ref{tab:nfwfit}.
When combined with the inner ACS profile,
all of the Subaru reconstructions yield a virial mass
in the range $M_{\rm vir}\sim (1.8-2.1)\times 10^{15}M_{\odot}$
and a high concentration in the range $c_{\rm vir}\sim 13-15$,
properly reproducing the observed Einstein radius of $\sim 45\arcsec$ 
at $z_s=1$ (B05b). In particular, fitting an NFW
profile to  the combined ACS and Subaru-2D data based on the MEM+ mass
reconstruction (see Table \ref{tab:memmethod}) yields $M_{\rm
vir}=(2.10\pm 0.17)\times 10^{15}M_{\odot}$ and  $c_{\rm
vir}=12.7^{+1.0}_{-0.9}$  ($\chi^2_{\rm min}/{\rm dof}=335/846$;
effective dof of $433$ without including upper limit bins with
$\kappa<0$), with the predicted Einstein radius of $\theta_{\rm
E}=45.3^{+5.9}_{-6.2}$ arcsec.  
%%%%

The 2D-based results here can be compared with the corresponding results
from the Subaru 1D analysis of B05a (see Table \ref{tab:nfwfit}) based
on the combined distortion and magnification measurements. The
combined ACS and Subaru-1D convergence profile is well fitted by an NFW
profile with $M_{\rm vir}=(1.9\pm 0.2)\times 10^{15}M_{\odot}$ and
$c_{\rm vir}=13.7^{+1.4}_{-1.1}$ (B05a), with the predicted Einstein radius of 
$\theta_{\rm E}=45.4^{+7.6}_{-6.9}$ arcsec, in good agreement with
the present full 2D analysis within the statistical uncertainties;
this agreement between the 1D and 2D analyses supports
the assumption of quasi-circular symmetry in the projected
mass distribution.
It is interesting to compare these results with different combinations
of lensing measurements having different systematics.
Another combination of the ACS and Subaru 1D convergence profile,
derived from the shear-based $\zeta$-statistic measurements (\S
\ref{subsubsec:ein}), yields fairly 
consistent results: $M_{\rm vir}=1.91^{+0.24}_{-0.20}\times
10^{15}M_{\odot}$ and $c_{\rm vir}=13.7^{+1.5}_{-1.3}$. This consistency
clearly demonstrates that our results here are insensitive to systematic
errors in the lensing measurements, such as the shear calibration
error. 
%%%%

In Figure \ref{fig:mass} we make a direct comparison 
between model-independent convergence profiles of A1689 from
different weak lensing techniques, along with the ACS-derived
inner profile obtained by B05b (filled triangles).
The filled circles with error bars show
the results based on the 2D $\kappa$ map reconstructed
from an entropy-regularized maximum-likelihood combination
of the lens magnification and distortion of red background galaxies
(MEM+, Table \ref{tab:memmethod}).
%the error bars are correlated,
%and are given by the square root of the diagonal part of the 
%bin-bin covariance matrix (equation [\ref{eq:bin2bincovar}]).
%%%%%%%%%%%%%%%%
The open triangles represent the $\kappa$ profile
from the non-linear $\zeta_{\rm c}$-statistic measurements.
%where decorrelated error bars are shown.
The filled circles show the results from the Subaru 1D analysis
by B05a based on the combined magnification and distortion
profiles of the same red background population as 
in the present study.
%measured from the same red background sample as
%in the prsent study.
%The 1D- and 2D-based NFW models from the respective combined
%ACS and Subaru data are also shown as solid and dashed curves,
%respectively.
It is clearly seen from Figure \ref{fig:mass} that
the Subaru-derived $\kappa$ profiles are all in good agreement
within the statistical uncertainties, over the full range of radii up to
$\sim 2{\rm Mpc}/h$. 
%%%%
The entire mass profile traced by the combined ACS and Subaru
information is well described by a single NFW profile (solid and dashed)
with a high concentration ($c_{\rm vir}\sim 13-14$).
For comparison an NFW profile with a low concentration $c_{\rm vir}=5$
normalized to the observed Einstein radius of $45\arcsec$ (at $z_s=1$)
is shown as a dotted curve.
Such a low concentration profile as favored by 
the standard  $\Lambda$CDM model
clearly overpredicts the outer
profile constrained by the Subaru weak lensing observations.

Figure \ref{fig:joint} shows the 
$68\%$, $95\%$, and $99.7\%$ confidence levels
($\Delta\chi^2=2.3, 6.17$, and $11.8$) in 
the $(c_{\rm vir},M_{\rm vir})$-plane for 
each of the three-different 2D MEM reconstructions 
(from top to bottom: MEM-S, MEM, and MEM+,
see Table \ref{tab:memmethod})
with (right) and without (left) the central ACS profile 
at $10{\rm kpc}/h \simlt r \simlt 180 {\rm kpc}/h$
combined.
As shown in the left panels, $M_{\rm vir}$ is well constrained
by the Subaru data alone, while the Subaru constraint on 
$c_{\rm vir}$ is rather weak, allowing a wide range of 
the concentration parameter, $c_{\rm vir}$. The complementary
ACS observations, when combined with the Subaru observations,
significantly narrow down the uncertainties on $c_{\rm vir}$
(right panels), placing stringent constraints on the inner mass
profile. 
For each sub-panel, the observed constraints on 
the Einstein radius,
$\theta_{\rm E}\simeq 45\arcsec$ at $z_s=1$,
are shown as a dashed curve.
This clearly demonstrates that the ACS-derived constraints
on the $\kappa$-field ensure the correct size of the observed
Einstein radius.

Table \ref{tab:nfwref} lists the best-fitting NFW parameters
obtained in different studies. Here we quote as the NFW parameters  
$(c_{200},M_{200})$ evaluated at a specific fractional overdensity of
$\Delta_c=200$  with respect to the critical density $\rho_{\rm
crit}(z_d)$ for closure of the universe at $z_d=0.183$  as well as those
in terms of the virial properties, $(c_{\rm vir},M_{\rm vir})$.
We convert a given set of  $(c_{200},M_{200})$ into $(c_{\rm vir},
M_{\rm vir})$ assuming a spherical  NFW density profile (see, e.g.,
Appendix A of Shimizu et al. 2003). 
%%%%%%%%%%%%%%%%%%%
We show in Figure \ref{fig:joint_gt} the same confidence levels as in
Figure \ref{fig:joint} but for the Subaru $g_+$ profile of B05a.
Also shown are best-fit sets of $(c_{\rm vir}, M_{\rm vir})$
taken from
%% the combined strong and weak lensing analysis of
%Halkola et al. (2006) and the CFHT weak lensing analysis of Limousin et 
%al. (2007) 
Halkola et al. (2006) and Limousin et al. (2007) as well as from our
combined ACS and MEM+ results. 
%%%%
%%%%
The best-fitting NFW parameters for the Subaru $g_+$ profile (cross)
are $M_{\rm vir}=1.51^{+0.27}_{-0.24}\times 10^{15}M_{\odot}$
and $c_{\rm vir}=20.0^{+8.8}_{-5.3}$ (Table \ref{tab:nfwfit}),
being marginally consistent with the very high concentration 
of $c_{\rm vir}=27.2^{+3.5}_{-5.7}$
derived  by Medezinski et al. (2007)
\footnote{
Unlike our analysis here, Medezinski et al. (2007) used
the observed constraints on the Einstein radius 
when fitting an NFW profile to 
the $g_{+}$ profile measured from 
the combined red and blue 
sample of the background.}.
The $\kappa$ field derived from the present full 2D analysis,
in conjunction with the central ACS profile, favors 
a slightly lower, but still high, concentration of $c_{\rm vir}\sim 13$
(triangle), reproducing the observed Einstein radius of
$\sim 45\arcsec$ at $z_s=1$.
%%%%
On the other hand,
Limousin et al. (2007) found from their CFHT weak lensing analysis
a concentration of $c_{\rm vir} =  9.6\pm 2.0$
%($c_{200}=7.6\pm 1.6$) 
and a virial mass of 
$M_{\rm vir} =  1.5^{+0.3}_{-0.2}\times 10^{15}M_{\odot}$
(filled circle),
corresponding to the Einstein radius of $24\pm 11$ arcsec (at $z_s=1$),
%%completely 
inconsistent with the observed Einstein radius.
%%%%
The results from Halkola et al. (2006) with  
$c_{\rm vir} =  9.6\pm 0.4$
and $M_{\rm vir} = 2.6^{+0.2}_{-0.1}\times 10^{15}M_{\odot}$
(square) 
come closer to the observed Einstein radius,
being marginally consistent within the $1\sigma$ uncertainty
($\theta_{\rm E}=39^{+6}_{-3}$ arcsec at $z_s=1$),
based on almost exactly the same shear data used in B05a but with a
different weighting which prefers their inner strong-lensing based
profile where the ACS data imply a shallower
slope, as discussed in Medezinski et al. (2007) and Limousin et
al. (2007).

It is useful to compare the results from different lensing studies
%%(see Table \ref{tab:nfwref})
%%model predictions 
in terms of the tangential distortion, $g_+$, that is directly
observable in weak lensing.
Figure \ref{fig:gt} shows the radial profiles of $g_+$ and $g_\times$
%the tangential
%distortion, $g_+$, and the $45^\circ$-rotated component, $g_\times$,
derived in B05a
from the same red background sample as used in this work, 
where we have added the observed Einstein-radius constraint of
$\theta_{\rm E}=45\pm 5$ arcsec ($z_s=1$), translated to the mean depth
of the red background sample (\S \ref{subsec:red}), marking the point of
maximum distortion, $g_+=1$ ($\theta_{\rm E}=39\pm 4$ arcsec). The
ACS+Subaru-2D NFW model (solid curve) 
fits reasonably well with the entire distortion profile, 
$r\sim [80,2000] {\rm kpc}/h$, although it slightly overpredicts the
outer profile at $\theta \simgt 9\arcmin$, meaning that the observed
$g_+$ profile is steeper at large radii, as pointed out in
Medezinski et al. (2007).
%%%%
In contrast the best-fit NFW model of Limousin et al. (2007),
shown with the dashed curve, is in agreement 
with the Subaru outer profile, particularly
at $10\arcmin \simlt \theta \simlt 18\arcmin$,
but underpredicts significantly the inner $g_+$ profile,
leading to a significant underprediction of the Einstein radius.
%%($\theta_{\rm E}=24\pm 11$  arcsec for $z_s=1$).
On the other hand,
the dotted curve in Figure \ref{fig:gt} shows the NFW profile of
Halkola et al. (2006),
%for a simultaneous fit to their ACS inner mass
%profile and the Subaru distortion profile of B05a. 
which is overall in good agreement with the observed $g_+$ profile,   
and with the NFW prediction of this work, 
but increasingly overestimates the distortion signal with radius, 
%being too shallow at large radii $\simgt 10\arcmin$,
and slightly underpredicts the Einstein radius. 
This is again consistent with that the discrepancy between the derived
NFW parameters of this work and Halkola et al. (2006) is due to
the relative weights in the fitting procedure
assigned differently to the strong and weak lensing
measurements; that is, a relatively higher
weight is given by Halkola et al. (2006) 
to the shallower inner profile at $r\simlt 40 {\rm kpc}/h$ constrained
by the radial arcs,\footnote{The location of the observed radial critical
curve is at $\theta\approx 17\arcsec$ as found in B05b.}  
and consequently, the location of the outer critical curve is slightly
underpredicted.  However, we argue this does not indicate a discrepancy
between our strong and weak lensing results, but is simply that the form
of the NFW profile is not entirely consistent with our data ranging from
$10$ to $2000 {\rm kpc}/h$; the best-fit NFW profile is either too
shallow at large radii or too steep at small radii, depending on the
radial limits being examined.

\subsection{Systematic Errors of the Concentration Parameter}
\label{subsec:sys}

In this subsection we address the issues of systematic errors on the
halo concentration $c_{\rm vir}$ inherent in our lensing measurements and
analysis methods. Here we discuss the following potential sources of
systematic uncertainty: (i) selection criteria for the red
background sample (\S \ref{subsec:red}), (ii) background redshift
distribution (\S \ref{subsec:red}), (iii)
magnification analysis (\S \ref{subsec:mag}, \S
\ref{subsec:declustering}), (iv) inner boundary condition (\S
\ref{subsec:acs}),
(v) strong lensing model (\S \ref{subsec:acs} and \S \ref{subsubsec:ein}),  
(vi) entropic prior (\S \ref{subsec:massrec}), and
(vii) shear calibration error.
In Table \ref{tab:sys} we summarize the systematic errors of 
the halo concentration, $c_{\rm vir}$,
given in fraction of the best-fit value of
$c_{\rm vir}=12.7$ obtained from the combined ACS and Subaru-2D (MEM+)
data. Adding all potential sources of error
in quadrature, the total uncertainty in our determination of
the halo concentration is $c_{\rm vir}=12.7 \pm 1 ({\rm statistical})
\pm 2.8 ({\rm systematic})$.  
%a joint fit to the ACS and Subaru-2D (MEM+) data. 
%e use the 2D MEM+ method for reconstructing the
%$\kappa$ map, and combine it with the ACS inner mass profile to deduce
%the best-fit NFW parameters, unless otherwise noted.

\subsubsection{Background Selection Criteria}
\label{subsubsec:sys_zs}

As clearly demonstrated by B05a and Medezinski et al. (2007), a secure
background selection is critical in the cluster weak lensing
analysis, in order to avoid dilution of the distortion signal by
contamination of unlensed cluster member galaxies. The degree of
dilution, which is proportional to the fraction of cluster membership,
is particularly prominent in the central region of rich clusters, 
as in the case of A1689 (see Medezinski et al. 2007). Practically, a  
reliable background sample can be defined by selecting galaxies with
colors redder than the cluster E/S0 sequence (see \S \ref{subsec:red}). 
Figure \ref{fig:dilution} shows that  the dilution effect becomes
significant when the lower (bluer) color limit of the entire red sample
is decreased below a color of $\Delta(V-i')\sim 0.1$, while no signature
of systematic variations is seen when the lower color limit is increased
above $\Delta(V-i')\sim 0.1$, ensuring that the dilution effect is
almost negligible there. Based on this, we defined our red background
sample by choosing a conservative color limit as $\Delta(V-i')>0.22$.

Here we vary the lower color limit of the red sample in the interval
of $[0.1,0.5]$ where the dilution effect is negligible. At the lower
color limit of $\sim 0.5$, however, the number of red galaxies is
decreased by about $40\%$ (see Figure \ref{fig:dilution}), when compared
with our fiducial sample of $\Delta(V-i')>0.22$, leading to noisier
results. 
We generate red samples at the lower color limit of $0.1$,
$0.15$, $0.2$, ..., $0.5$, and assess the scatter in the best-fit
concentration parameter, $c_{\rm vir}$.  
We find the error distribution is fairly random, with a small spread of
$\Delta c_{\rm vir}\sim \pm 0.5$, corresponding to a $\sim 4\%$
fractional systematic error. 

\subsubsection{Background Redshift Distribution}
\label{subsubsec:sys_pz}

% redshift distribution plays a crucial role in 

Redshift information of background galaxies plays a crucial role in the
determination of cluster mass profiles
In order to convert the observed lensing signal into physical mass
units, one needs to evaluate the mean distance ratio  
$\langle D_{ds}/D_s\rangle$ over the redshift distribution of background
galaxies (\S \ref{subsec:red}).  An overestimate of the source redshift
will systematically lead to an underestimate of the cluster mass, and
vice versa. 
In this way, the uncertainty in the
background redshift distribution will lead to systematic errors of the
cluster mass determination. The level of uncertainty depends on the
%%relative distance between the lens and the source, 
lensing geometry, and is less
significant for low-$z$ clusters (say, $z_d\simlt 0.2$). 
For purely weak lensing based data, this effect is less important for
the determination of the halo concentration, because the identification
of the inner characteristic radius ($r_s$) is basically independent of 
the background redshift. However, when the weak lensing measurements are
combined with inner strong lensing information,
this depth information becomes crucial
for the determination of halo concentration as well, because it
determines the relative normalization between the inner and outer
profiles. That is, an overestimate of the background redshift will cause
an underestimate of the surface mass density, which will increase the
difference between the inner profile derived from strong lensing and the
outer profile from weak lensing,
leading to 
an overprediction of the halo concentration. 
%This degeneracy of the NFW
%parameters ($c_{\rm vir}$, $M_{\rm vir}$) is clearly demonstrated in
%Figure \ref{fig:joint}. 

Here we turn to assess the level of systematic error arising from the
uncertainty in the background redshift distribution. Based on the
multicolor photometry of Capak et al. (2004) in the HDF-N, we obtain
a mean distance ratio of  $\langle
D_{ds}/D_s\rangle = 0.693 \pm \pm 0.02$, or a distance-equivalent
redshift of $z_{s,D}=0.68\pm 0.05$, 
for our color-magnitude selected red background sample
(\S \ref{subsec:red}; see also Medezinski et al. 2007).
A good agreement has been also found using the COSMOS deep multicolor
photometric catalog of Capak et al. (2007), yielding a similar depth of   
$\langle D_{ds}/D_s\rangle \approx 0.703$ (Medezinski et al. 2007 in
preparation), suggesting that the field-to-field variance (cosmic
variance) is not significant, and as small as the statistical uncertainty
obtained here.
We find this level of depth uncertainty will lead to
only a $\sim 1\%$ fractional systematic error of $c_{\rm vir}$.
%%%
Further, 
%for a more conservative error estimate of the concentration, 
we assign a conservative uncertainty in the 
% a generous range of systematic errors in the
(distance-equivalent) source redshift, $z_{s,D}=[0.7,1.0]$, and find a 
fractional systematic error of about $10\%$ in the concentration
parameter.  It is interesting to note that B05a obtained the best-fit
concentration parameter
of $c_{\rm vir}=13.7^{+1.4}_{-1.1}$ assuming a source redshift of
$z_s(=z_{s,D})=1$. This concentration is slightly higher ($\sim +8\%$) than
the best-fit value of $c_{\rm vir}=12.7\pm 1$ found in this work, but
this level of discrepancy can be easily reconciled by the systematic
error in the assumed depth of B05a:  Applying this bias correction to
the results of B05a yields $c_{\rm vir} \approx 12.3$, 
which agrees quite well with the present results of $c_{\rm vir}=12.7\pm
1$.

Lensing magnification influences the observed
surface density of background galaxies (Broadhurst et al. 1995), by
expanding the 
observed solid angle of the background (area distortion),
and decreasing the effective flux limit of the survey (flux
amplification). Thus the latter effect of magnification bias may change
the redshift distribution of background galaxies depending on the
distance from the cluster center,
which could be a potential source of the
systematic errors in the determination of the halo concentration.
Our red background sample, however, is highly depleted (see Figure
\ref{fig:magbias}), meaning that the area distortion effect is 
dominating over the flux amplification of fainter, distant background
galaxies. Indeed, the unlensed count slope measured at our magnitude
limit $i'=25.5$ is $s=d\log N_0(m)/dm\approx 0.22$, being fairly flat as
compared to the blue sample with $s\approx 0.4$, close to the lensing
invariant slope (see equation [\ref{eq:magbias}]).
Consequently, relatively few fainter objects are magnified into the
sample even in the central cluster region, so that 
the magnification effect on the source redshift distribution
is fairly negligible for the red background defined 
at fainter magnitude limits
For our sample,
magnification at $\theta\sim 2\arcmin$ is $\mu\simeq 1+2\kappa \approx
1.2$, or about $0.2 {\rm mag}$ of increased depth, corresponding to 
the fractional increase of only $\sim 5\%$ in the number of red
background galaxies.
Further, given 
the weak dependence of the redshift distribution of faint galaxies on
apparent magnitude (e.g., Medezinski et al. 2007), we do not need to
take this effect into serious consideration for our red background
sample. 

%increase of only $\sim 5\%$ 
%The magnification bias arises because gravitational lensing changes the
%apparent solid angle of the background but conserves the surface
%brightness, leading to an increase or decrease of the observed 
%%%Gravitational deflection of the light ray will 
%Magnification bias arises because gravitational lensing changes the
%solid angle of a source but conserves its surface brightness 
%The magnification effect will enlarge the sky solid angle, thus
%modifying the source density by a factor 1/mu, .  

Still, it is instructive to consider the potential systematic bias
caused by the magnification effect on the background redshift
distribution.  The net effect of depth correction is to reduce the
central surface mass density of the lens, since we attribute the high
lensing signal to the geometric information of the background.
For purely weak-lensing based data, this will lead to a lower
concentration. However, when the strong lensing information in the inner
region is taken into account as well, this depth correction will further
increase the difference between the inner and outer profiles derived
from strong and weak lensing, respectively, thereby enhancing the
concentration.  However, the amount of correction is negligible in
practice when the weak lensing analysis is based on the red background
galaxies as discussed above.

\subsubsection{Magnification Analysis}

Here we address the systematic uncertainties arising from a particular
treatment and various cuts in the weak lensing magnification analysis.  

\begin{itemize}
\item[(1)]
{Mask area correction} % 7\%
\end{itemize}

The masking effect due to cluster member galaxies and bright foreground
objects acts to reduce the
apparent number of background galaxies, and this reduction increases
towards the cluster center, leading to an overestimate of the central
depletion signal without the masking correction. 
%taken into account. 
%%%%%%
In the present study the mask area 
of bright objects
is evaluated as the area inside the ellipse of 
$\nu_{\rm mask}$-times the major ({\tt A\_IMAGE}) and minor ({\tt
B\_IMAGE}) axes 
computed from SExtractor (see Cobb et al. 2006 for a similar
discussion),  
where the multiplier is chosen as $\nu_{\rm mask}=3$ so that the
ellipse is visually 
consistent with the isophotal detection limit in our SExtractor
configuration (\S \ref{subsec:data}).
Here we adopt a conservative uncertainty of $\nu_{\rm mask}=3\pm 1$ on the
masking factor, and find the corresponding systematic uncertainty of
$\pm 7\%$ in 
the halo concentration. The lower limit on $c_{\rm vir}$ is obtained for
$\nu_{\rm mask}=2$, when the mask area correction is underestimated and
hence the depletion signal is overestimated.
%Note that the direction of the correction for $c_{\rm vir}$ is opposite
%when only weak lensing data are used (i.e., wihtout the inner strong
%lensing information; see the discussion in \S \ref{subsubsec:sys_pz}).
%%%%
%Here we vary the value of $\nu_{\rm mask}$ in the range of 2 to 4,
%and assess the corresponding uncertainties in $c_{\rm vir}$. 
%For $\nu_{\rm mask}=3\pm 1$, 
%As found in B05a, the mask area correction is small (less than several
%percent) and negligible at large radii, but becomes
%increasingly significant at $\theta\simlt 4'$.
%%When $\nu_{\rm mask}=2$, the mask area 
%Here we assess the uncertainties in $c_{\rm vir}$ due to a particular
%choice of the mask correction multiplier. 

\begin{itemize}
\item[(2)]
{Clustering rejection}
%%: 1.5\%
\end{itemize}

Similarly, we adopt a conservative uncertainty of $\nu_{\rm clust}=4\pm 1$
in the rejection
threshold $\nu_{\rm clust}$, which has been introduced to downweight
locally the
intrinsic clustering noise 
which otherwise perturbs the depletion signal (\S
\ref{subsec:declustering}).  
This yields a fractional uncertainty of only $\pm 1.5\%$ in $c_{\rm
vir}$. 
We note that the intrinsic clustering of background galaxies is a local
effect, and 
hence it does not affect significantly the radial profile fitting.

\begin{itemize}
\item[(3)]
{Unlensed count slope}
%%: 3.5\%
\end{itemize}

The conversion from the observed counts of the red background sample
into the magnification bias $\delta_\mu$ (equation [\ref{eq:magbias}])
depends on the slope parameter $s=d\log{N_0(m)}/dm$ of the unlensed
number counts (\S \ref{subsec:mag}), 
which was estimated as $s=0.22\pm 0.03$ from the outer region $\simgt
10\arcmin$ where the magnification effect is negligibly small
($\kappa, \delta_\mu \simlt 0.01$). We find that this level of
uncertainty in $s$ will cause an uncertainty in $c_{\rm vir}$ of about
$3.5\%$.

\subsubsection{Inner Boundary Condition}
%% 10\%

Any mass reconstruction technique based on the gravitational shear field
involves a non local process (see equation [\ref{eq:local}]), and hence
it is crucial to have a proper boundary condition for an accurate
determination of the cluster mass profile.  
In particular, our MEM+ method is based on the ACS strong lensing
information, which is incorporated as an inner boundary condition
on the central pixel, $\kappa_{\rm c}$ (\S \ref{subsec:acs}). 
%It is threfore important to assess the systematic errors in our mass
%reconstruction introduced by uncertainties in the adopted strong lensing
%mass model.  
B05a showed that the combined ACS and Subaru mass profile can be well
fitted by a high concentration NFW profile ($c_{\rm vir}\sim 14$) over 
the full range of ACS and Subaru data, $r=[10^{-2},2]$ Mpc$/h$. However,
it is also found in B05a that this high-concentration model somewhat
overestimates the inner slope at $r\simlt 40$ kpc$/h$ (see Figure
\ref{fig:mass}). 
This model yields a central surface mass density of 
$\kappa_{\rm c}(z_s=1)\approx 0.810$, which is
slightly higher than, but still consistent with, the prediction based on
B05b adopted in the present work, $\kappa_{\rm c}(z_s=1)=0.781\pm 0.1$,
within the $1\sigma$ statistical uncertainty.   
%%%%%
The mass profile of A1689 has also been examined by Limousin et
al. (2007) using independent weak lensing shape measurements from
CFHT12K data. 
%From this Limousin et al. (2007) obtained the best-fit
%parameters of $c_{\rm  vir}\approx 10$ and $M_{\rm vir}\approx 1.5\times
%10^{15}M_{\odot}$, 
Their best-fit NFW model, however, underpredicts the observed Einstein
radius (\S \ref{subsubsec:comparison}), and accordingly yields a much
lower central value of
$\kappa_{\rm c}(z_s=1)\approx 0.527$, which we take as the lower limit
on $\kappa_{\rm c}(z_s=1)$.
%%%
Allowing for a conservative uncertainty of
$\kappa_{\rm c}(z_s=1)=0.78^{+0.1}_{-0.25}$,
corresponding to $\kappa_{\rm c}(z_{s,D}=0.68) = 0.70^{+0.09}_{-0.22}$ 
for the effective depth of our red sample,
we find a $\pm 10\%$ fractional systematic uncertainty in the
halo concentration, $c_{\rm vir}$. 

\subsubsection{Inner Strong Lensing Information}
\label{subsubsec:inner}

%% 10%

The inner strong lensing information 
%plays a crucial role in the determination of the halo
%concentration parameter 
provides strong constraints on the halo concentration parameter
as demonstrated in Figure \ref{fig:joint}.
In the present work, the ACS-derived inner mass profile of B05b
is specifically used to determine the NFW halo parameters
of A1689 in the strongly lensed region, $r=[10,130]$ ${\rm kpc}/h$.
It is practically difficult to assess potential systematic errors 
introduced in strong lensing modeling because of the complex, non-linear
error propagation (B05b).
Instead, here we simply estimate the level of uncertainty
%%due to  the inner strong lensing model, 
by a comparison with
the result obtained with the model-independent constraint on the
Einstein radius $\theta_{\rm E}$ (see \S \ref{subsubsec:ein}) 
%$\theta_{\rm E}=45\arcsec$ for $z_s=1$,  
based on multiply lensed images identified
in the ACS observations of B05b.
Both the strong lensing analysis of B05b and Limousin et al. (2007)
yield a consistent value for the projected mass interior to
$45\arcsec$ of $M_{2{\rm D}}(45\arcsec)\approx 2\times 10^{14} M_\odot$,
or equivalently, $\bar\kappa(45\arcsec) = 1$ for a
source redshift of $z_s=1$: i.e., $\theta_{\rm E}=45\arcsec$ at
$z_s=1$. 
In contrast to the fit to the inner mass profile, this Einstein-radius
information provides an integrated constraint on the inner mass profile
interior to $45\arcsec$, or $r\approx 100{\rm kpc}/h$ in projected
radius. 
%%%
We find that a joint fit of the Subaru $\kappa$ map and the
Einstein-radius constraint yields a slightly higher concentration of
$c_{\rm vir}=14.0^{+2.5}_{-2.1}$  (see Table \ref{tab:nfwfit}),
corresponding to a fractional increase of about $10\%$. 
This tendency is also found for the
results with the shear-based 1D mass reconstruction from the 
$\zeta_{\rm c}$-statistic measurements (see Table \ref{tab:nfwfit}).
%in which case a fractional increase of $20\%$ is found. We thus adopt a
%conservative systematic error of $20\%$ in $c_{\rm vir}$ associated with
%the strong lensing model.
%%We note again this does not simply reflect the actual uncertainty 
%%in the strong lensing mass model, but is likely caused by 
%% error arising from the way in which the model is specified. 
%% error arising from the simplification

\subsubsection{Entropic Prior}

% 10\%

A particular choice of the regularization could be a
potential source of the systematic errors in the cluster mass
reconstruction.  In our mass reconstructions the model parameter
$m$ of the entropy prior is
fixed, but the Bayesian value of the
regularization parameter $\alpha$ that satisfies equation
(\ref{eq:alpha}) is obtained for a given value of $m$. 
When we vary the value of $m$ over a
relevant range of the cluster lensing signal, $m=[0.1,0.9]$ ($m=0.5\pm
0.4$),  we find the error distribution in the resulting value of $c_{\rm
vir}$ is almost random with a small spread of $\pm 10\%$.

Furthermore, a particular choice of the entropy function may lead to
some systematic bias in the determination of cluster mass profiles.
To check this possibility, 
%here we simply compare the present results
%obtained with the entropy regularization and those from other standard
%reconstructions  (\S \ref{sec:massprofile}).  
here we simply compare 
the present results from the MEM+ reconstruction 
with earlier 1D maximum-likelihood results of B05a,
both of which are based on the same distortion and magnification data,
and adopt the ACS-based inner boundary conditions.
After the correction for the systematic bias (\S
\ref{subsubsec:sys_zs}) the best-fit concentration of B05a is $c_{\rm
vir}\approx 12.3$,  which is in good agreement with $c_{\rm vir}=12.7\pm
1$ obtained with the entropy regularization. Thus, it is likely that 
the level of systematic uncertainty due to the particular choice of the
entropy regularization is negligibly small as compared to other sources
of the systematic errors. 

Our use of the entropy prior in the low S/N regime might potentially induce
some slight bias in the 
regularized maximum likelihood solution, and/or some slight
non-Gaussianity in the error distribution, underestimating the actual
error bars for the mass reconstruction and the NFW halo
parameters. However, our results show good consistency between the
entropy-regularized reconstructions and other standard reconstructions
within the statistical uncertainties. Thus, it seems this bias is not
significant for this work.

\subsubsection{Shear Calibration Error}
\label{subsubsec:calib}

A shear calibration error is one of the systematic errors that could
bias the weak lensing shape measurements (Haymans et al. 2006; Massey et
al. 2007) and
potentially have some influence on recovered mass profiles. To assess
this possibility we have measured the strength
of the weak lensing signal as a function of magnitude limit for our
background galaxy sample.
We found no particular tendency towards a
loss of the weak lensing signal with apparent magnitude
for red background galaxies, which span over a wide range of sizes.
This is comforting and
consistent with the expectations of the model-independent
KSB+ based technique for which the recovered signal should match reality
within the noise. 
We note that at the very weak lensing limit, our distortion
measurements are quite consistent with an independent estimate of the
weak lensing signal by Limousin et al. (2007).

However we have found that for blue background galaxies there is a
significant loss of the signal at faint magnitudes and this raises the
worrying question of unresolvable HII regions which we know do become
prevalent at faint blue magnitudes, acting effectively as point
sources and hence reducing the weak lensing signal. Such objects are
not included in our analysis, so as not to bias our lensing measurements.
The STEP project, aimed at assessing signal which may be lost in
ground based data, described in Heymans et al (2006) and Massey et
al. (2007), does not allow 
for unresolvable sources within galaxies, as it is
inherently assumed that galaxies are continuously resolvable, so that
stretched HST/ACS images of faint galaxies are taken to be perfectly
empirical representations of reality for the purpose of calibrating
galaxies dominated by bright HII regions.

Furthermore, we have found a good consistency between the purely
shear-based results (e.g., $\zeta$-statistic based 1D reconstruction)
and the results based on the combined distortion and magnification data
(e.g., MEM+ results, B05a results), implying that any shear calibration
error is not noticeable at the level of our analysis, otherwise we would
see inconsistency with our magnification analysis.

%%%%%%%%%%%%%%%%%%%%%%%%%%%%%%%%%%%%%%%%%%%%%%%%%%%%%%%%%%%%%%%%%%%%%%

\section{Summary and Discussion}\label{con}

In this paper, we have derived a projected 2D mass map of the well
studied lensing cluster A1689 ($z=0.183$) based on an
entropy-regularized maximum-likelihood combination of the lens
magnification and distortion of red background galaxies registered in
deep Subaru images.  The combination of distortion and
magnification data breaks the mass sheet degeneracy inherent in all
reconstruction methods based on distortion information alone.
%%%
The method is not restricted to the weak lensing regime but applies to
the whole area outside the tangential critical curve, where
non-linearity between the surface mass-density and the observables
extends to a radius of a few arcminutes.  
%%%
The strong lensing information from ACS observations was also readily
incorporated in this maximum likelihood approach (\S
\ref{subsec:acs}). 
We also utilized the distortion measurements to locally downweight 
the intrinsic clustering noise in the magnification measurements,
which otherwise perturbs the depletion signal (\S
\ref{subsec:declustering}).  
%%%%
The resulting 2D map showed that the projected surface density of A1689
is smoothly varying and symmetric, similar to the distribution of
cluster members. 
%%%%
The 2D mass map is well fitted by the Navarro-Frenk-White model, with a
continuously steepening profile, but the concentration parameter much
higher than expected for its virial mass ($\sim 2\times
10^{15}M_{\odot}$),  according to the clear predictions of standard
$\Lambda$CDM (Bullock et al. 2001; Neto et al. 2007).
%%%%
For consistency we have compared the best-fitting
NFW parameters obtained from different combinations of datasets,
boundary conditions, and weak lensing techniques (\S
\ref{subsubsec:comparison}).  
%%%
We find that all of the reconstructions tested here are consistent with
a virial mass in the range, $M_{\rm vir}=(1.5-2.1)\times
10^{15}M_{\odot}$, and the combined ACS and Subaru-2D mass
reconstruction yields a tight constraint on the concentration parameter,
$c_{\rm vir}=12.7^{+1.0}_{-0.9}$ ($c_{200}\sim 10$), improving upon the
statistical accuracy of our earlier 1D analysis (B05a).
%%%
Very good agreement is found between the present full 2D reconstruction
and the earlier 1D reconstruction (B05a), both of which are based on the
same distortion and magnification measurements, supporting the
assumption of quasi-circular symmetry in the projected mass distribution. 
We have also explored potential sources of systematic error on the
concentration parameter, such as the uncertainties in background
redshift distribution, selection criteria for the red background sample,
and strong lensing modeling. Taking into account all of the systematic
errors as well as the statistical uncertainty, our constraint on
the concentration is $c_{\rm vir}=12.7 \pm 1 ({\rm stat.}) \pm 2.8 ({\rm
systematic})$. 
%% where the second set of errors include the systematic uncertainties
%% Spergel et al. 2003
%% where the error includes random and systematic uncertainties
% we assess carefully various
%sources of potential systematic error in the halo concentration
%parameter derived from the lensing observations.
%We further explored an alternative combination of the ACS
%profile with the 1D convergence profile based solely on the distortion
%measurement, and obtained  

%%% Comparison with Limousin et al. (2007) etc.
For clusters well fitted by an NFW profile, the derived virial mass of
a cluster and the concentration parameter can be used to find the
Einstein radius $\theta_{\rm E}$, through the simple relationship 
$1=\bar\kappa_{\rm NFW}(\theta_{\rm E})$ (see Appendix \ref{app:nfw}).
%%%%
For A1689 with $M_{\rm vir}=(2.1\pm 0.2)\times 10^{15}_{\odot}$ and
$c_{\rm vir}=12.7^{+1.0}_{-0.9}$ (only statistical errors quoted),
this yields an Einstein radius of 
$45 \pm 6$ arcsec at $z_s=1$, or $53 \pm 7$ arcsec at $z_s=3$,
in very good agreement with the mean estimated radius of $\sim
50\arcsec$, based on the locations of the multiple images (Table 2 of
B05b).  
%%%%
The strong lensing mass model of Limousin et al. (2007), based on the
multiple images identified by B05b, properly reproduces the observed
Einstein radius of $45\arcsec$ at $z_s=1$,
%($\bar\kappa[45\arcsec]=1$ at
%$z_s=1$), or equivalently,  
%$M(45\arcsec)\approx 2\times 10^{14} M_\odot$ in terms of the 
%projected mass interior to $\theta_{\rm E}=45\arcsec$,
consistent with the strong lensing mass model of B05b (see \S
\ref{subsubsec:inner}).   
%%%%
In contrast, an Einstein radius of 
$24\pm 11$ arcsec at $z_s=1$ is implied by the NFW fit to this cluster
by Limousin et al. (2007), to independent weak lensing data from CFHT
($c_{\rm vir}\sim 9.6$, $M_{\rm vir}\sim 1.5\times 10^{15}M_\odot$; see
Table \ref{tab:nfwref}, Figures \ref{fig:joint_gt} and \ref{fig:gt}). 
This discrepancy may be attributed to a degree of contamination of the
sample of galaxies used to define the lensed background, which, 
as pointed out in B05a, can drag down the weak lensing signal 
if accidentally included in the background sample, and preferentially 
so at small radius (see Figure \ref{fig:gt}) where the
ratio of cluster members compared with background is much higher,
resulting in a shallow $g_+$ profile and hence a lower concentration
fit.

\begin{figure*}[!htb]
 \begin{center}
  \includegraphics[width=100mm, angle=270]{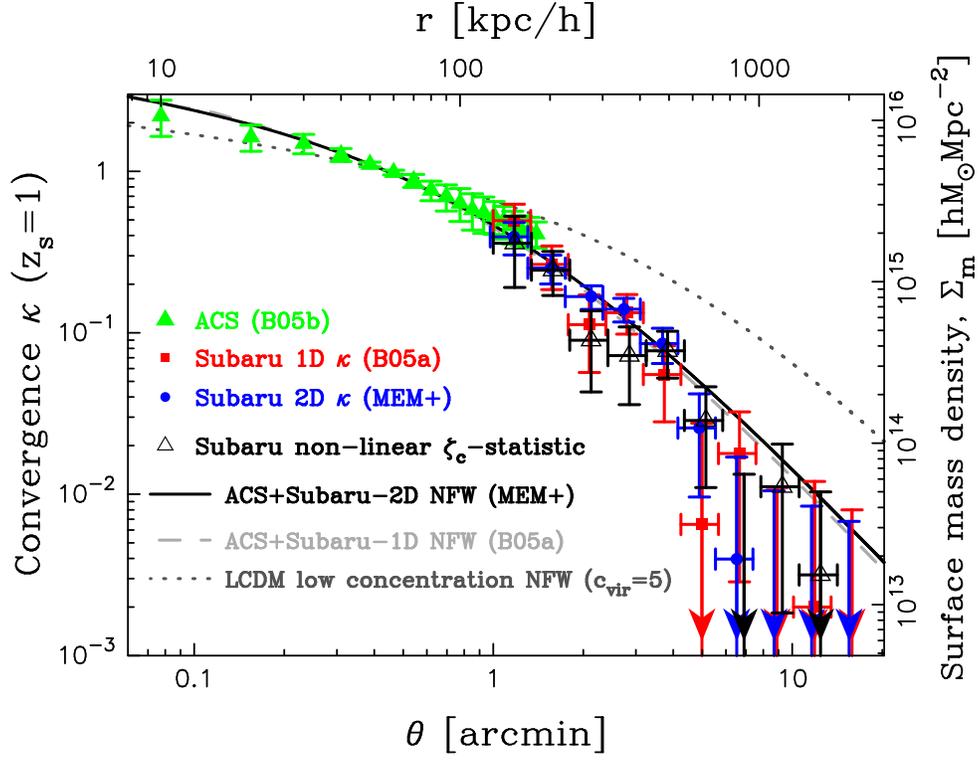}
 \end{center} 
\caption{
Comparison of 
model-independent
mass profiles of A1689.
All of the profiles are scaled to a fiducial source redshift of 
$z_s=1$.
%%%%
The filled circles represent the results
based on the 2D $\kappa$ map 
reconstructed from an entropy-regularized maximum-likelihood combination
of Subaru distortion and depletion data,
with the ACS constraint on the mean surface mass density
in the central pxiel (MEM+).
%using the ACS constraint on the central mass density.
%based on the MEM-reconstructed 2D $\kappa$ map using 
%the combined Subaru distortion and depletion data
%and the ACS constraints on the central mass density.
The error bars are correlated.
%(shear)
%(magnification bias) 
%measurements.
%%%%
The open triangles 
represent the mass profile from
the non-linear $\zeta_{\rm c}$-statistic measurements 
based on averaged tangential distortion data.
Decorrelated error bars are shown.
%%%
The filled triangles and circles show the results from the ACS strong
lensing analysis (B05b) and
%The filled circles with error bars 
%represent the mass profile from 
%the Subaru weak lensing analysis 
%with the 1D reconstruction method
from the Subaru 1D weak lensing analysis
based on the combined distortion and depletion profiles (B05a).
%tangential distortion and depletion measurements (B05a).
%Downwards-pointing arrows 
%are used  where the lower error bar drops below zero.
%%%% 
%The solid curve shows 
%the best-fitting NFW profile for the combined ACS+Subaru data
%by BTU05 ($M_{\rm vir}=1.93\times 10^{15} M_{\odot}, c_{\rm
% vir}=13.7$).
The solid curve shows 
the best-fitting NFW profile for the MEM-reconstructed
2D $\kappa$ map (MEM+).
The 1D- and 2D-based NFW models 
from the respective combined ACS+Subaru data
(B05a) are also shown as solid and dashed curves, respectively.
For comparison 
an NFW profile with a low concentration, $c_{\rm vir}=5$,
normalized to the observed Einstein radius ($\theta_{\rm E}=45''$), 
is shown as a dotted curve.
The low concentration model ({\it dotted}) predicted for $\Lambda$CDM
clearly overestimates the outer profile constrained by 
the Subaru weak lensing observations. 
%%%
The mass profiles are all 
in remarkable agreement 
over the full range of radii up to $\sim 2 {\rm Mpc}/h$.
}
\label{fig:mass}
\end{figure*}

\begin{figure*}[!htb]
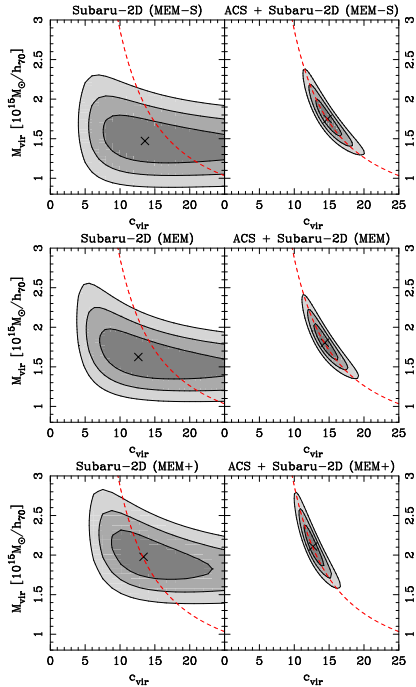

 \begin{center}
   \includegraphics[width=30mm, angle=270]{f14a.ps}\\
   \includegraphics[width=30mm, angle=270]{f14b.ps}\\
   \includegraphics[width=30mm, angle=270]{f14c.ps}
 \end{center} 
\caption{
Joint constraints on the NFW model parameters,
$(c_{\rm vir}, M_{\rm vir})$, derived from 
gravitational lensing observations of A1689.
%Subaru weak lensing
%observations of A1689. 
%%%
%The left panel shows 
Left panels show
the $68\%$, $95\%$,
and $99.7\%$ confidence levels 
($\Delta\chi^2=2.3$,  $6.17$, and $11.8$)
in the $(c_{\rm vir}, M_{\rm vir})$-plane 
for the 2D $\kappa$ map reconstructed from 
Subaru weak lensing observations.
%for the 2D $\kappa$ map reconstructed from an entropy-regularized
%maximum likelihood combination of Subaru distortion and depletion data
%(MEM+).
Right panels show the same confidence levels but for the joint
ACS+Subaru-2D NFW fitting, incorporating the inner mass profile
($10{\rm kpc}/h \simlt r \simlt 180 {\rm kpc}/h$)
constrained by ACS strong-lensing observations by B05b.
The cross in each panel shows 
the best-fit set of the NFW model parameters.
%the best-fitting model parameters.
The top, middle, and bottom panels 
correspond to the results 
 based on the 2D MEM+, MEM, and MEM-S
 reconstructions,
%from Subaru weak lensing observations, 
respectively (see Table \ref{tab:memmethod}).
%%%%
The virial mass $M_{\rm vir}$ is well constrained by the Subaru data 
alone, while the Subaru constraint on the concentration 
$c_{\rm vir}$ is rather weak.
The complementary ACS observations, when combined with the Subaru
observations, significantly narrow down the uncertainties on $c_{\rm
 vir}$, placing stringent constraints on the inner mass profile.
In each panel the observed constraints on the Einstein radius
($\theta_{\rm E}\simeq 45$\arcsec at $z_s= 1$)
are shown as a dashed curve.
} 
\label{fig:joint}
\end{figure*}

\begin{figure*}[!htb]
 \begin{center}
  \includegraphics[width=100mm, angle=270]{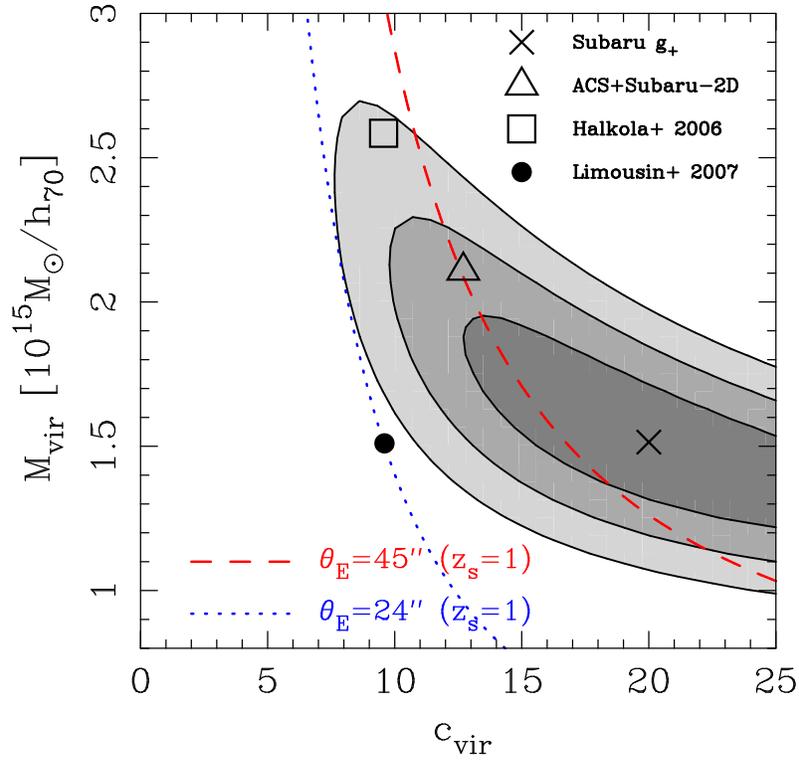}
 \end{center} 
\caption{
Joint constraints on the NFW model parameters,
$(c_{\rm vir}, M_{\rm vir})$
obtained from the Subaru tangential shear ($g_+$)
profile of A1689 (see Figure 1 of B05a).
%%%
The cross shows the best-fitting set of the NFW parameters, and 
the contours show the $68\%$, $95\%$, and $99.7\%$ confidence levels  
($\Delta\chi^2=2.3$,  $6.17$, and $11.8$) in the $(c_{\rm vir}, M_{\rm
 vir})$-plane. 
The observed constraints on the Einstein radius,
$\theta_{\rm E}\simeq 45$\arcsec at $z_s = 1$, are shown as a dashed
 curve. The dotted curve shows the $c_{\rm vir}-M_{\rm vir}$
relation for $\theta_{\rm E}=24\arcsec$ at $z_s=1$.
%%%
The triangle symbol shows the best-fit set of
$(c_{\rm vir}, M_{\rm vir})$ for the combined ACS and Subaru-2D (MEM+)
 results. The square and circle show the best-fit sets of $(c_{\rm vir},
 M_{\rm vir})$  
from the combined strong and weak lensing analysis of  Halkola et
 al. (2006) and the weak lensing analysis of Limousin et al. (2007),
 respectively. 
}  
\label{fig:joint_gt} 
\end{figure*}

\begin{figure*}[!htb]
 \begin{center}
  \includegraphics[width=.7\textwidth, angle=270]{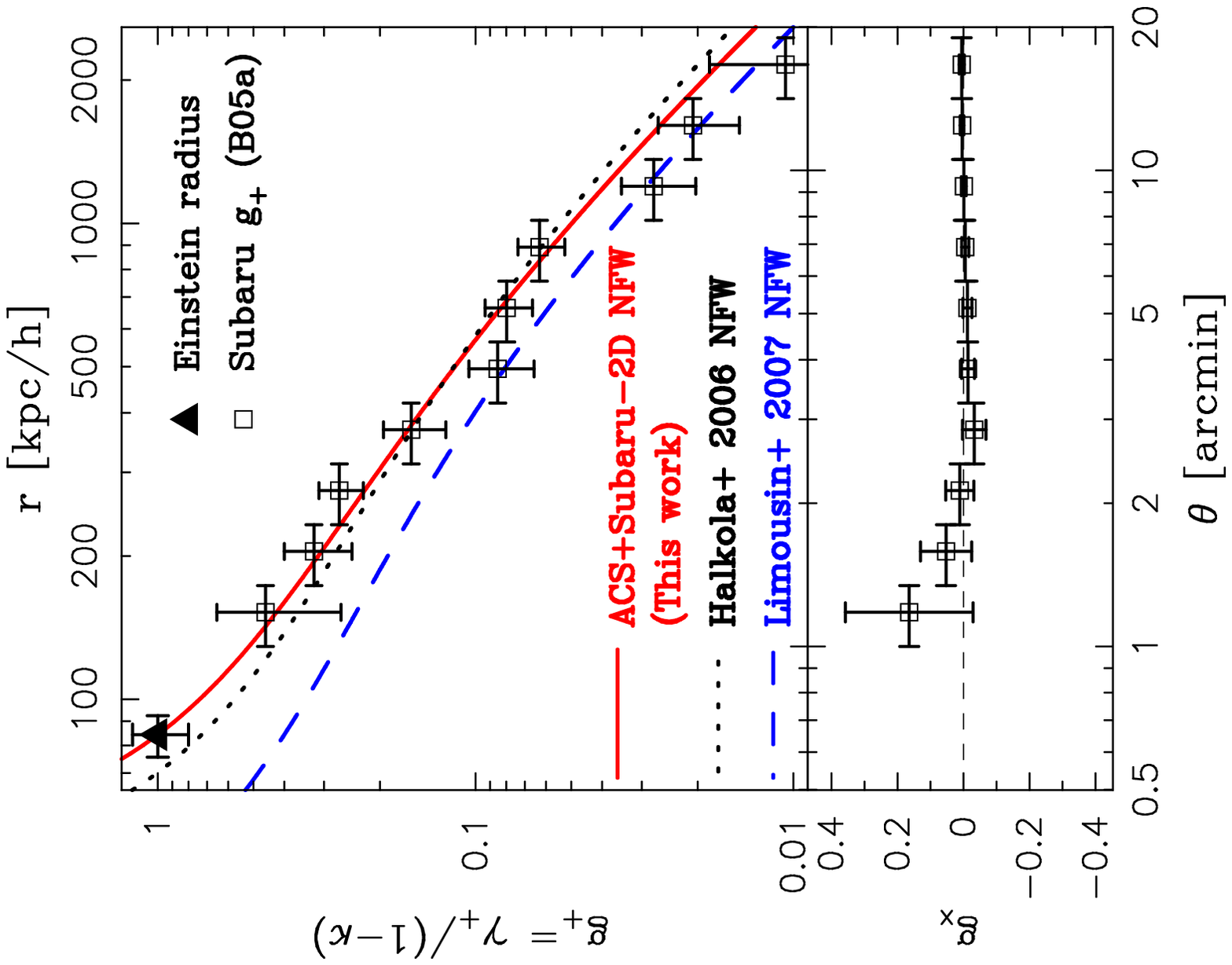}
 \end{center} 
\caption{
\label{fig:gt}
Tangential distortion profile $g_+(\theta)$ ({\it square}, upper 
 panel) from the Subaru weak lensing analysis of the red background sample
 (B05a).  The solid curve shows
 the best-fit NFW profile derived from the joint strong and weak lensing
 analysis of ACS and Subaru observations (this work), incorporating full
 distortion  and magnification information.
The Einstein radius constraint ({\it triangle}) of
 $\theta_{\rm  E}=45\arcsec$ ($z_s=1$), determined from multiply lensed
 images in ACS  observations (B05b), is translated to the corresponding
 depth 
%%$\langle  D_{ds}/D_s\rangle\approx 0.693$
of the Subaru red background sample using the ACS+Subaru-2D NFW 
 model ({\it solid}), and added to the distortion profile ($g_+=1$), 
 marking the point of maximum distortion. 
The ACS+Subaru-2D NFW model ({\it solid}) fits well with the combined
 ACS  and Subaru distortion information over the full range of data,
 $r=[80,2000] {\rm  kpc}/h$, but somewhat overpredicts the outer
 distortion profile at $\theta \simgt 9\arcmin$ ($r\simgt 1.2{\rm
 Mpc}/h$).
%%%% 
Also shown with the dashed curve is the best-fit NFW profile from the
 CFHT weak lensing analysis of Limousin et al. (2007), 
which, in contrast, is in good agreement with the Subaru outer profile
 at $\theta\simgt 4\arcmin$, 
%%($r\simgt 500 {\rm kpc}/h$), 
but underpredicts
significantly the inner distortion profile and hence the Einstein radius.
The dotted curve shows an NFW profile of Halkola et al. (2006)
 for a simultaneous fit to their ACS inner mass profile and the Subaru
 distortion profile of B05a shown here, but with a different weighting
 that prefers the inner strong-lensing based profile where the data
 imply a shallower slope (see Figure \ref{fig:mass}).
%%%
 The lower panel shows
  the radial profile of the $45^\circ$ rotated component
 $g_\times(\theta)$ for the same Subaru red background sample
 (B05a). The $\times$-component of the red galaxy sample is consistent
 with a null signal at all radii, indicating the reliability the Subaru
 distortion  analysis. 
} 
\end{figure*}

%%% Comparison with Limousin and Halkola results
Taking into account the systematic errors (\S \ref{subsec:sys}),
combined ACS and Subaru constraints allow a shallower (but still steeper
than theoretically expected) mass profile
with $c_{\rm vir}= 9-10$, similar to the values found in Halkola
et al. (2006) and Limousin et al. (2007). 
This, however, does not simply mean that the discrepancy 
between different lensing studies
%%% in the derived NFW halo parameters
%%% between the models 
is fully solved: When the NFW model is normalized to reproduce the
observed Einstein radius ($45\arcsec$ at $z_s=1$), then this
concentration would indicate a large virial mass of 
$M_{\rm vir}=(3-3.3)\times 10^{15}M_\odot$ (see Figures \ref{fig:joint}
and \ref{fig:joint_gt}), which however is considerably higher than the
mass estimates derived from the X-ray observations ($M_{\rm vir}\approx
10^{15}M_\odot$ in Andersson \& Madejski 2004, $M_{\rm vir}\approx
1.4\times 10^{15}M_\odot$ in Lemze et al. 2008).  
The discrepancy between the present results and the results by Halkola
et al. (2006) can be explained by the relative weights in the least
$\chi^2$ fitting, 
assigned differently to the ACS- and Subaru-based measurements,
indicating slight deviation of the observed profile from the NFW
predictions (see discussion in \S \ref{subsubsec:comparison}; also see
discussion in Medezinski et al. 2007). 
This is seen at
the innermost radii $r\simlt 40{\rm kpc}/h$ (Figure
\ref{fig:mass}),  where the ACS data indicate a shallower profile. 
Consequently, when one prefers the innermost region to fit the data,
this could lead to a lower concentration ($c_{\rm vir}=9-10$) as favored
by the shallower slope in the innermost region, and to a lower value for
the Einstein radius when the data at outer critical radius are less
weighted.  Nonetheless, our best-fitting NFW model provides a good
approximation to our data over the radii we have considered, $r=[10^{-2},
2]$ ${\rm Mpc}/h$.

%%% Importance of dilution 
B05a demonstrate that dilution of the lensing signal is certainly the
cause of the very low concentration fit ($c_{\rm vir}\sim 4.5$, Table
\ref{tab:nfwref}) obtained by Bardeau et al. (2005), due to the
inclusion of relatively blue cluster members in the definition of the
background sample of the same 
CFHT weak lensing data as Limousin et al. (2007), and for which the
equivalent Einstein radius is only $\sim 4\arcsec$
(at $z_s=1$, see Table \ref{tab:nfwref}).
%The study by Halkola et al. (2006) comes closer to the observed radius 
%with a value of $39^{-3}_{+6}$ arcsec (at $z_s=1$, see Table
%\ref{tab:nfwref}),  as shown in 
%Figure \ref{fig:joint_gt}, implied by their published best-fit NFW
%parameters for the joint combination of the weak lensing results of B05a
%and a parametric model of the strong lensing region based on the
%multiple images identified in B05b. The difference here seems to be due
%to the relatively higher weighting given to the strong lensing region by
%Halkola et al. (2006) where the slope of the projected mass profile is
%shallower than at large radius (see also Figure \ref{fig:gt}).
%%%% Application of dilution 
On the other hand,
the dilution of the lensing signal caused by cluster members can be 
used to derive the proportion of galaxies statistically belonging
to the cluster by comparing the undiluted red background distortion
signal with the radial distortion profile of color-magnitude space
occupied by the cluster members, but including inevitable background
galaxies falling in the same space (Medezinski et al. 2007). 
This technique allows the light profile of the cluster to be determined in
a way which is independent of the number density fluctuations
in the background population, which otherwise limit the
calculation of the cluster light profiles and luminosity functions
from counts of member galaxies.
The resulting light profile can be compared with the mass profile 
to examine the radial behavior of $M/L$ (Medezinski et al. 2007).

%%% Lemze et al.'s X-ray analysis
A recent joint X-ray and lensing analysis of A1689 by Lemze et al.
(2008) also produces very similar concentration, 
$c_{\rm vir}=12.2^{+0.9}_{-1}$, and virial mass, 
$M_{\rm vir}\sim (1.4\pm 0.4)\times 10^{15}M_{\odot}$, 
to that derived here in our analysis (Table \ref{tab:nfwref}). 
This is derived from a model-independent approach to the X-ray emission
profile and the projected lensing mass profile, assuming hydrostatic
equilibrium, utilizing the mass profile derived in B05a.
Interestingly, the observed temperature profile falls
short of the predicted temperature profile derived from the joint fit
and this may imply that the gas distribution is clumpy on small scales, 
in the form of a higher density cold gas phase (Lemze et al. 2008).
%%%%%%
Moreover,
this anomaly seems to be consistent with conclusions regarding 
gas substructure in the recent detailed hydrodynamical simulations 
of cluster gas by Kawahara et al. (2007), 
implying that other similar detailed lensing-X-ray 
studies should also show a similar temperature discrepancy.

Great progress continues to be made in the detailed predictions of 
$\Lambda$CDM, particularly on cluster scales where gas cooling is not a
worry.  Recently the whole Millennium survey (Springel et al. 2005)  
has been converted into the
observer's frame following the full geodesics through the volume to
simulate the effect of structure on the light received by an observer
(Hilbert et al. 2007). This work has shown that although in general
clusters form in overdense regions, the material associated with a
given cluster in the form of extended groups and filaments outside the
virial radius of the cluster is of relatively low mass contrast
compared to the projected mass due to the cluster itself, and
therefore lensing based projected masses of clusters are not
overestimated by more that a few percent (Hilbert et al. 2007).

This simulation has also been used to better define the relationship
between the concentration parameter and the virial mass of halos, 
over the full range of mass from galaxies up to the most massive
cluster-sized halos, in the context of standard $\Lambda$CDM (Neto et al
2007). A clear prediction has emerged that most massive halos generated
in these simulations have the lower concentration ($c_{\rm vir}\sim 5$),
and  the cause of this is in part attributed to the generally later
collapse of the more massive halos reflecting the lower mean density of 
the universe. For example, at $z_{\rm vir}=0.183$, for the standard
choice of cosmological parameters the criterion for virialization is
$\bar\rho(<r_{\rm vir}) \sim 115 \rho_{\rm crit}(z_{\rm vir})
\sim 277 \bar\rho(z_{\rm vir})$. 
One possibility to achieve earlier formation of massive clusters is
to allow deviation from Gaussianity of the primordial density
fluctuations in the early universe (e.g., Grossi et al. 2006; Sadeh,
Rephaeli, \& Silk 2007).

%%%%%%%%%%%%%%%%%%%%%%%%%%%%%%%%%%%%%%%%%%%%%%%%%%%%%%%%%%
%%% Halo triaxiality in general
%%%%%%%%%%%%%%%%%%%%%%%%%%%%%%%%%%%%%%%%%%%%%%%%%%%%%%%%%
A degree of triaxiality is inevitable for collisionless
gravitationally collapsed structures. Discussion of the likely effect
of triaxiality on the measurements of lensing properties has been
examined analytically (Oguri et al. 2005; Sereno 2007;
Corless \& King 2007) and in numerical investigations (Jing \& Suto
2002; Hennawi et al. 2007).  
A bias in favor of prolate structure pointed to the observer is
unavoidable at some level, as this orientation boosts the projected
surface mass density and hence the lensing signal. This effect 
has been evaluated in the context of the CDM model and serves as a
guide to the likely degree of bias which may affect lensing work.
%%%%%%%%%%%%%%%%%%%%%%%%%%%%%%%%%%%%%%%%%%%%%%%%%%%%%%%%%%%
%%% Theoretical efforts to examine the triaxiality effect
%%%%%%%%%%%%%%%%%%%%%%%%%%%%%%%%%%%%%%%%%%%%%%%%%%%%%%%%%%%
Hennawi et al. (2007) conducted a detailed study of the properties of 
lensing cluster population identified in $\Lambda$CDM cosmological
$N$-body simulations. 
The level of bias in terms of the concentration
%The effect of this bias in terms of the concentration
parameter derived from 2D lensing measurements
was explicitly estimated and found to amount to an $\sim 34\%$
increase, which results from a combination of two effects, namely
the orientation bias ($\sim 19\%$) due to halo triaxiality
and 
the selection effect towards higher 3D concentrations ($\sim 18\%$).
%that favors projections along the major axis of triaxial CDM halos.
The level of correction for the orientation bias is also derived from
semi-analytical 
representation of simulated CDM triaxial halos by Oguri et al. (2005).
The anomalously high concentrations of $c_{\rm vir}\simgt 13$, such as
found for A1689, CL0024+1654 (Kneib et al. 2003), and MS2137-23 (Gavazzi
et al. 2003), appear inconsistent with the distribution of
concentrations found in detailed simulations of Hennawi et al. (2007),
which predict that only $<2\%$ of lensing clusters should have such high
concentrations. 
It is also unlikely that the baryonic component in these massive
clusters increases the concentrations over the $\Lambda$CDM prediction
for dark matter halos (see discussion in Hennawi et al. 2007, B05b, and
Broadhurst \& Barkanna 2008).

A chance projection of foreground/background structure along the
line-of-sight can potentially influences projected lensing observations,
boosting the surface mass density locally and hence the
concentration. The ACS strong lensing analysis of B05b revealed the
secondary mass clump in the central region of A1689 
associated with a small clump of galaxies. The existence of this
subclump has been suggested by earlier observations, such as 
the spectroscopic study of Teague et al. (1990) and Czoske (2004),
and the X-ray study of Andersson \& Madejski (2004). 
No obvious substructure is visible in a large spectroscopic sample of
525 cluster members identified in Czoske (2004), in contrast to 
earlier work of Teague et al. (1990) based on a smaller sample of 176
identified cluster members. 
%who suggested a complex dynamical
%state of A1689 based on 176 identified cluster members. 
%%%
%%% no obvious substructure visible in
%%% large sample of czoske etal 2004, something like that.
Recently, this
secondary mass clump has also been directly detected by the weak lensing
flexion analyses by Leonard et al. (2007) and Okura et al.
(2008) based on the ACS and Subaru data, respectively. However, the
detailed ACS strong lensing analyses based on multiply imaged background
galaxies showed that the mass contribution of the secondary mass clump
is only a small fraction of the main cluster component (Figures 21 and
22 of B05b),
implying a lower $M/L$ for the subclump than for the main cluster,
which is tightly constrained by the geometrical positions of sets of
multiply lensed images, or the location of critical curves.

%%%%%%%%% Practical method for examining the triaxiality
In the near future, 
the question of the effect of triaxiality on lensing based cluster
mass profiles may be examined empirically. For example, the total
X-ray luminosity or the total lensing based mass of a cluster should
not depend on the orientation with respect to the line of sight,
whereas the concentration parameter and the Einstein radius will be
affected and hence expanded lensing studies could in principle reveal
whether relaxed clusters of fixed mass or fixed X-ray luminosity tend
to have consistent lensing based concentrations, or instead a broader
distribution may be uncovered, with a mean concentration smaller than
derived for A1689, indicating triaxiality produces a significant
bias. Current indications based on several massive clusters favor the
NFW profile, but with consistent concentrations
(Medezinski et al. 2007, in preparation), 
similar in value to A1689, underscoring the tension between 
detailed lensing based mass profiles and the predictions of standard 
$\Lambda$CDM.

%%%%%%%%%%%%%%%%%%%%%%%%%%%%%%%%%%%%%%%%%%%%%%%%%%%%%%%%%%%%%%%%%%%%%%%%%
%%%
%%%          Acknowledgments
%%% 
%%%
%%%%%%%%%%%%%%%%%%%%%%%%%%%%%%%%%%%%%%%%%%%%%%%%%%%%%%%%%%%%%%%%%%%%%%%%%

\acknowledgments

We thank the anonymous referee for a careful reading of the manuscript
and and for providing invaluable comments.
We are grateful to Masahiro Takada for valuable discussions and
comments. 
%We thank Masahiro Takada for valuable discussions and suggestions.
We thank 
J.-H. Proty Wu,
Elinor Medezinski,
Guo-Chin Liu, 
Ue-Li Pen, 
and Sandor Molnar for fruitful discussions.
We thank Nick Kaiser for making the IMCAT package publicly available. 
KU thanks Ludovic Van Waerbeke for kindly providing
his code for a maximum likelihood mass reconstruction.
%We thank the Suprime-Cam team for their support during
%the observation. 
%KU acknowledges fruitful discussions with xxx.
%KU thanks Patrick Koch for providing helpful comments.
%valuable discussions.
% We thank the annonymous referee for invaluable comments and suggestions
% which improved the paper significantly,
Part of this work is based on  
data collected at the Subaru Telescope,
which is operated by the National Astronomical Society of Japan.
%%%
%The work is partially supported by the COE program at Tohoku University.
This work in part supported by 
the National Science Council of Taiwan
under the grant NSC95-2112-M-001-074-MY2.

%%%%%%%%%%%%%%%%%%%%%%%%%%%%%%%%%%%%%%%%%%%%%%%%%%%%%%%%%%%%%%%%%%%
%%%%%%%%%%%%%%%%%%%%%%%%%%%%%%%%%%%%%%%%%%%%%%%%%%%%%%%%%%%%%%%%%%%
%%%
%%% Tables
%%%
%%%%%%%%%%%%%%%%%%%%%%%%%%%%%%%%%%%%%%%%%%%%%%%%%%%%%%%%%%%%%%%%%%%
%%%%%%%%%%%%%%%%%%%%%%%%%%%%%%%%%%%%%%%%%%%%%%%%%%%%%%%%%%%%%%%%%%% 

%%\input{tab.tex}

%%%%%%%%%%%%%%%%%%%%%%%%%%%%%%%%%%%%%%%%%%%%%%%%%%%%%%%%%%%%%%%%%%%
%%%%%%%%%%%%%%%%%%%%%%%%%%%%%%%%%%%%%%%%%%%%%%%%%%%%%%%%%%%%%%%%%%%
%%%
%%% Tables
%%%
%%%%%%%%%%%%%%%%%%%%%%%%%%%%%%%%%%%%%%%%%%%%%%%%%%%%%%%%%%%%%%%%%%%
%%%%%%%%%%%%%%%%%%%%%%%%%%%%%%%%%%%%%%%%%%%%%%%%%%%%%%%%%%%%%%%%%%%

%%% Table 1

\begin{deluxetable*}{lllll}
\tabletypesize{\scriptsize}
\tablecaption{
\label{tab:memparam}
Parameters used in the 2D MEM reconstruction}
\tablewidth{0pt}
\tablehead{
\colhead{$m$ \tablenotemark{a}} &
\colhead{$\alpha$ \tablenotemark{b}}  &
\colhead{$N_{\rm pix}$ \tablenotemark{c}}&
\colhead{$N_{\rm data}$ \tablenotemark{d}} &
\colhead{$\kappa_{\rm c}$ \tablenotemark{e}} 
} 
\startdata
%% ** Submitted Version **
%% $0.5$ & $96.6$ &  $31\times 27$ & 1012 & 0.700 \\
 $0.5$ & $96.2$ &  $31\times 27$ & 1012 & 0.700 \\
\enddata
\tablenotetext{a}{MEM model parameter.}
\tablenotetext{b}{Bayesian value of $\alpha$.}
\tablenotetext{c}{This includes the ACS-constrained central pixel.}
\tablenotetext{d}{Number of usable measurements.}
\tablenotetext{e}{ACS constraint on the central $\kappa$ pixel assuming
 $\langle D_{ds}/D_s\rangle=0.693$.}
\end{deluxetable*}

%%% Table 2

\begin{deluxetable*}{lllrrrrrr}
\tabletypesize{\scriptsize}
\tablecaption{
\label{tab:memmethod} 
2D MEM reconstructions 
with different datasets and boundary conditions}
\tablewidth{0pt}
\tablehead{
\colhead{Method  \tablenotemark} &
\colhead{Dataset \tablenotemark{a}} &
\colhead{Boundary Conditions \tablenotemark{b}}  &
\colhead{$N_{\rm data}${c}} &
\colhead{$\alpha$\tablenotemark{d}} &
%% ** Submitted Version **
%%\colhead{$F_{\rm min}$ \tablenotemark{e}} &
%% ** Version 2 (revised) **
\colhead{NDF \tablenotemark{e}} &
\colhead{$\chi^2_{\rm min}$ \tablenotemark{f}} & 
\colhead{RMS \tablenotemark{g}} &
\colhead{S/N \tablenotemark{h}} 
} 
\startdata 
%% ** Submitted **
%2D MEM-S & shear           & MEM + ACS &  710 &  88.7  & 322  & 448  &
%0.082 & 14.7\\ 
%2D MEM & shear + magbias  &  MEM       & 1012 & 100.1  & 625  & 1039 &
% 0.075 & 18.3 \\ 
%2D MEM+ & shear + magbias  & MEM + ACS & 1012 &  96.6  & 651  & 1089 &
% 0.076 & 19.4\\ 
2D MEM-S & shear           & MEM + ACS &  710 &  91.2  & 515 & 448  & 0.081   
 & 14.4\\ 
2D MEM & shear + magbias  &  MEM       & 1012 &  97.3  & 785 & 1014 & 0.076
 & 18.2 \\ 
2D MEM+ & shear + magbias  & MEM + ACS & 1012 &  96.2  & 784 & 1065 & 0.077   
 & 19.4\\ 
\enddata
\tablecomments{
The following three sets of combinations of datasets and
boundary conditions are considered: 
(i) 2D MEM+ method using shear and magnification
 data with the ACS constraint on the central pixel, 
(ii) 2D MEM method using shear
 and magnification data without the central ACS constraint, 
and (iii) 2D MEM-S method using shear data with the central ACS constraint. 
}
\tablenotetext{a}{Dataset used for weak lensing analysis.}
\tablenotetext{b}{With or without the ACS constraint on 
the central pixel in the strong lensing region.}
\tablenotetext{c}{Number of usable measurements.}
\tablenotetext{d}{Bayesian value of $\alpha$ ($m=0.5$).}
%\tablenotetext{e}{Minimum functional value of $F$.}
\tablenotetext{e}{Classical number of degrees of freedom, ${\rm NDF} \equiv N_{\rm data}-N_{\rm good}$ (see equation [\ref{eq:alpha}]).}
\tablenotetext{f}{Minimum functional value of $\chi^2$.}
\tablenotetext{g}{Average rms error of $\kappa$ in the observed region
 $(30'\times 24')$.}
\tablenotetext{h}{Detection significance of the convergence signal
 defined by equation (\ref{eq:sn}).}
\end{deluxetable*}

%%% Table 3
 
\begin{deluxetable*}{lllrr}
\tabletypesize{\scriptsize}
\tablecaption{
\label{tab:mrec}
Summary of the Methods for Mass Profile Reconstructions
}
\tablewidth{0pt}
\tablehead{
\colhead{Method  \tablenotemark{a}} &
\colhead{Dataset \tablenotemark{b}} & 
\colhead{Boundary Conditions \tablenotemark{c}}  &
\colhead{$(\theta_{\rm min},\theta_{\rm max})$ \tablenotemark{d}} & 
\colhead{$N_{\rm bin}$ \tablenotemark{e}}
} 
\startdata 
$\zeta_{\rm c}$-statistic  & tangential shear &  $\bar{\kappa}_b=4\times
 10^{-3}$\tablenotemark{\dag} & ($1',16'$) & 9 \\%& 6.6\\
2D MEM-S  & shear          & MEM + ACS & ($1',18'$) & 10 \\
2D MEM  & shear + magbias  & MEM       & ($1',18'$) & 10 \\
2D MEM+  & shear + magbias & MEM + ACS & ($1',18'$) & 10 \\
Subaru 1D\tablenotemark{*} & tangential shear + magbias & ACS  &  ($1',18'$) & 10 \\% & 6.9\\
\enddata
\tablenotetext{a}{Weak lensing mass reconstruction method. All methods
 apply to the non-linear but subcritical regime.}
\tablenotetext{b}{Dataset used for weak lensing analysis.}
\tablenotetext{c}{With or without the ACS constraint on the central
 pixel in the strong lensing region.}
\tablenotetext{d}{Lower and upper radial limits.}
\tablenotetext{e}{Number of radial bins in the range of $(\theta_{\rm
 min},\theta_{\rm max})$.}
\tablenotetext{\dag}{This employs an outer boundary condition 
%that the
on the 
mean convergence $\bar\kappa_b$ within 
$16'<\theta<19'$. The mean background level $\bar\kappa_b$ is calculated
 to be $\bar\kappa_b=4\times 10^{-3}$ using the ACS+Subaru-1D
 best-fit NFW model by B05a.
 }
\tablenotetext{*}{1D maximum likelihood analysis by B05a 
based on the joint measurements of weak lensing distortion and depletion profiles.}
\end{deluxetable*}

%%%%%%%%%%%%%%%%%%%%%%%%%%%%%%%%%%%%%%%%%%%
%% Table 4
%%%%%%%%%%%%%%%%%%%%%%%%%%%%%%%%%%%%%%%%%%%

\begin{deluxetable*}{llllllll}
\tabletypesize{\scriptsize}
\tablecaption{
\label{tab:nfwfit}
Summary of the best-fitting NFW parameters
for Subaru weak lensing observations
} 
\tablewidth{0pt} 
\tablehead{ 
\colhead{Method  \tablenotemark{a}} &
\colhead{Data} &
\colhead{ACS \tablenotemark{b}} &
\colhead{ER \tablenotemark{c}} &
\colhead{$M_{\rm vir}$ \tablenotemark{d}} & 
\colhead{$c_{\rm vir}$\tablenotemark{e}}  &
\colhead{ $\chi^2_{\rm min}/{\rm dof}$\tablenotemark{f}} & 
\colhead{ $\theta_{\rm E}$\tablenotemark{g}} 
} 
\startdata 
%%%
tangential shear & 1D $g_+$ &--- &--- & $1.51^{+0.27}_{-0.24}$ &
 $20.0^{+8.8}_{-5.3}$ &  5.0/8 (8) & 50.9 \\
tangential shear & 1D $g_+$ &--- & yes & $1.59^{+0.24}_{-0.22}$ &
 $15.7^{+3.4}_{-2.5}$ &  9.1/9 (9) & 44.5 \\
$\zeta_{\rm c}$-statistic & 1D $\kappa$  & --- & --- & $1.48 \pm 0.27 $ & $27.3^{+2.7}_{-19.3}$  &  5.2/7
 (7) & 59.1 \\
$\zeta_{\rm c}$-statistic & 1D $\kappa$ &--- & yes & $1.51^{+0.25}_{-0.22}$ & $16.5^{+4.0}_{-3.1}$
 &  11.4/8 (8) & 44.8  \\ 
$\zeta_{\rm c}$-statistic & 1D $\kappa$ &yes & --- & $1.91^{+0.24}_{-0.20}$ & $13.7^{+1.5}_{-1.3}$
 &  12.3/19 (19) & 45.3  \\ 
2D MEM-S & 2D $\kappa$ & --- & --- & $1.48^{+0.20}_{-0.18} $ & $14.1^{+10.3}_{-4.8}$ & 361/834 (406)
 & 38.2 \\
2D MEM-S  & 2D $\kappa$ & yes & --- &$1.75^{+0.17}_{-0.16} $ & $14.6^{+1.3}_{-1.1}$ & 369/846 (418)
 & 45.1 \\
2D MEM   & 2D $\kappa$ & --- &--- & $1.60^{+0.21}_{-0.17} $ &
 $14.9^{+11.9}_{-5.2}$ & 287/834 (478) & 42.7 \\  
 2D MEM   & 2D $\kappa$ & yes & --- & $1.81^{+0.21}_{-0.14} $ &
 $14.3^{+1.2}_{-1.1}$  & 294/846 (490) & 45.1 \\ 
2D MEM+  & 2D $\kappa$ & ---  & --- & $1.76^{+0.20}_{-0.17} $  & $15.5^{+7.4}_{-4.2}$
 &  323/836 (423) & 51.0 \\ 
%% new:: aho-kprof.txt, aho-kprof.cov
2D MEM+  & 2D $\kappa$ & --- & yes & $1.93^{+0.22}_{-0.19} $  &  $14.0^{+2.5}_{-2.1}$
 &  387/837 (424) & 46.4 \\
2D MEM+  & 2D $\kappa$ & yes & --- & $2.10 \pm 0.17 $         &  $12.7^{+1.0}_{-0.9}$
 &  327/848 (435) & 45.3 \\
Subaru 1D\tablenotemark{*} & 1D $\kappa$ & --- & --- &
 $1.69^{+0.30}_{-0.28}$ & $\le 30$ & $5.36/8$ (6) & 66.9  \\ 
Subaru 1D\tablenotemark{*} & 1D $\kappa$ &yes & --- & $1.93\pm 0.20$  &
 $13.7^{+1.4}_{-1.1}$ & $13.3/20$ (18) & 45.4  \\ 
\enddata
%%%%
\tablecomments{
A flat prior of $c_{\rm vir}\le 30$ is adopted in the model fitting.
}
\tablenotetext{a}{Mass reconstruction method (see Table
 \ref{tab:mrec}).}
\tablenotetext{b}{Fitting with or without the ACS-derived inner mass
 profile (\S \ref{subsubsec:acs}).}
\tablenotetext{c}{Fitting with or without the strong lensing
 Einstein-radius (ER) constraint, $\theta_{\rm E}=45\arcsec$ at $z_s=1$
 (\S \ref{subsubsec:ein}).} 
\tablenotetext{d}{Virial mass and $1\sigma$ error in units of $10^{15}M_{\odot}$.}
\tablenotetext{e}{Concentration parameter, $c_{\rm vir}=r_{\rm 
 vir}/r_s$, and $1\sigma$ error.}
\tablenotetext{f}{Values in parentheses refer to effective degrees of
 freedom excluding upper limit bins with $\kappa<0$.}
\tablenotetext{g}{Einstein radius in units of arcsec for a fiducial
 source at $z_s=1$, defined as $1=\bar\kappa(\theta_{\rm E})$.}
\tablenotetext{*}{Taken from B05a. A fiducial source redshift of $z_s=1$
 is assumed in B05a.}
\end{deluxetable*}

%% Table 5 :: Mvir-Cvir constraints from the literature

\begin{deluxetable*}{llllllll}
\tabletypesize{\scriptsize}
\tablecaption{
\label{tab:nfwref}
Comparison between best-fitting NFW parameters 
for A1689 from different observations and methods
%by different authors
%taken from the literature
} 
\tablewidth{0pt} 
\tablehead{ 
\colhead{Reference} &
\colhead{Method \tablenotemark{a}} &
\colhead{$M_{\rm vir}$ \tablenotemark{b}} &
\colhead{$c_{\rm vir}$ \tablenotemark{c}} &
\colhead{$M_{200}$ \tablenotemark{d}} & 
\colhead{$c_{200}$\tablenotemark{e}}  &
\colhead{ $\theta_{\rm E}$\tablenotemark{f}} & 
\colhead{ Remarks}
} 
\startdata 
King, Clowe, \& Schneider 2002 &  WL   & $1.0$ & $6.1$ & $0.84$ & $4.8$  &
 11  & ESO/MPG\\%% (fit to the shear pattern)\\
%%% Clowe 2003 
Clowe 2003      &  WL   & $1.3$ & $9.9$ & $1.1 $ & $ 7.9$  &
 22  & ESO/MPG\\%% (shear profile)\\
%%% Bardeau 2005
Bardeau et al. 2005 &  WL & $ 1.72^{+0.8}_{-0.6} $  & $ 4.5^{+0.6}_{-0.4} $ 
                          & $ 1.41^{+0.63}_{-0.47}$ & $3.5^{+0.5}_{-0.3}$ 
 & 3.5  & CFH12K\\
%%% ACS strong lensing
%%% Mvir=2.62e15Msun/h
B05b      & SL   & $3.7$ & $8.2^{+2.1}_{-1.8}$ & $3.2$ & $6.5^{+1.9}_{-1.6}$  &
 44  & ACS\\
%%% BTU05
B05a      & SL+WL   & $1.93\pm 0.20$ & $13.7^{+1.4}_{-1.1}$ & $1.72\pm
 0.19 $ & $ 10.9^{+1.1}_{-0.9}$  &
 45  & ACS + Subaru ($\kappa$ profile)\\
%%% Halkola+ 2006 (SL)::      
%%% r200 = 2.82Mpc/h70 +/- 0.11 
%%% C200 = 6.0 \pm 0.5
%%% M200 = 2.13954868143346 [1.89881528994971 -2.39981469631041]1e15Msun/h
%%% M200 = 3.05e15Msun/h70
%%% Mvir = 2.4902 [2.1973--2.8116] = 3.55e15Msun/h70
%%% Cvir = 7.592  [6.9746 -- 8.2089]
Halkola et al. 2006  & SL 
 & $3.55\pm 0.4$  & $7.6 \pm 0.6$ 
 & $3.05\pm 0.3$  & $6.0 \pm 0.5$  & 37 & ACS  \\
%% Halkola+ 2006 (SL + WL):: 
%% r200=2.55Mpc/h70 \pm 0.07 
%% C200=7.6 \pm 0.5
%% M200(1e15Msun/h)   = 1.5799 [1.4552 -- 1.71584] 
%% Mvir(1e15Msun/h)   = 1.8093 [1.6599 -- 1.9736]
%% Cvir               = 9.5647 [8.9486 -- 10.181]
%% rE: 38.889220 [35.474812 -- 44.9133]-3.414408 +6.024080
Halkola et al. 2006  & SL+WL    
  &  $2.58 \pm 0.2$  & $9.6 \pm 0.6 $  
  &  $2.25 \pm 0.2$  & $7.6 \pm 0.5 $   & 39  
 & ACS + Subaru ($g_+$ profile) \\
%% Bardeau+ 2007
%% M200 = 1.379e15Msun/h \pm 0.238  [1.141--1.617]
%% C200 = 4.28 \pm 0.82
%% Mvir = 1.65e15 Msun/h [1.3449 -- 1.9760]
%% Cvir = 5.466 [4.449-6.480]
Bardeau et al. 2007  & WL 
 & $2.35\pm 0.4$  & $5.5  \pm 1.0$ 
 & $1.97\pm 0.3$  & $4.28 \pm 0.8$  &
 12 & CFH12K \\%%(fit to the shear pattern) \\
%%% Limousin et al. 2007
%%% M200 = 0.924 e15Msun/h [0.784--1.064]
%%% C200 = 7.6\pm 1.6      [6.0 -- 9.2]
%%% Mvir = 1.51^{+0.25}_{-0.24}1e15Msun/h70
%%% Mvir=1.057e15Msun/h [0.88768--1.2384E]
%%% Cvir=9.6\pm 2.0     [7.6--11.6]]
%%% rE  = 24.087869     [12.741739 -- 35.662879]-11.34+11.574931
Limousin et al. 2007  & WL 
 & $1.51^{+0.3}_{-0.2}$  & $9.6 \pm 2.0$ 
 & $1.32\pm 0.2$         & $7.6 \pm 1.6$  
 & 24 & CFH12K \\%%(shear profile) \\
%%% Lemze et al. 2007 (X-ray)
%%% Mvir = 1.5661e15Msun/h
%%% Cvir = 12.2   [11.2--13.1]
%%% M200 = 1.3875e15Msun/h
%%% C200 = 9.741  [8.9282--10.554]
Lemze et al. 2008\tablenotemark{*} & SL+WL+X
 &  $2.23\pm 0.6^\dag$ &  $12.2^{+0.9}_{-1}$ 
 &  $1.98$ &  $9.7 \pm 0.8$      
 & 45 & ACS + Subaru + Chandra\\  
%%% This work
%%% M200=1.086e15Msun/h [1.086-1.384]
%%% C200=10.709 [7.96-15.17]]
This work  & WL & $  1.97 \pm 0.20 $  & $13.4^{+5.4}_{-3.3}$ &
 $1.76 \pm 0.20$  & $10.7^{+4.5}_{-2.7}$       &
 45 & Subaru ($\kappa$ map) \\
%%% 
%%%
This work  & SL+WL & $2.10 \pm 0.17$  & $12.7^{+1.0}_{-0.9}$ &
 $ 1.86 \pm 0.16$ & $ 10.1^{+0.8}_{-0.7} $  &
 45 & ACS + Subaru ($\kappa$ map) \\
\enddata
\tablecomments{
A similar table of best-fitting NFW parameters 
is found in Comerford \& Natarajan 2007 (Table 1) 
which also include the results for other clusters as well as A1689.
}
\tablenotetext{a}{Analysis method.}
\tablenotetext{b}{Virial mass $M_{\rm vir}$ and $1\sigma$ error in units
 of $10^{15}M_{\odot}$.} 
\tablenotetext{c}{Virial concentration $c_{\rm vir}=r_{\rm vir}/r_s$ and
 $1\sigma$ error.} 
\tablenotetext{d}{$M_{200}$ and $1\sigma$ error in units of
 $10^{15}M_{\odot}$.} 
\tablenotetext{e}{Specific concentration $c_{200}=r_{200}/r_s$ and 
 $1\sigma$ error.}
\tablenotetext{f}{Einstein radius in units of arcsec for a fiducial
 source at $z_s=1$, defined as $1=\bar\kappa(\theta_{\rm E})$.} 
\tablenotetext{*}{Based on the Chandra X-ray data and the 
projected mass profile from the joint ACS and Subaru-1D analysis by
 B05a.  Hydrostatic equilibrium assumed.}
\tablenotetext{\dag}{Note that the outermost radius point of Lemze et
 al. (2008)  is at $1.5 {\rm Mpc}/h$, which is smaller than the virial
 radius of A1689, $r_{\rm vir}\approx 2 {\rm Mpc}/h$.
As compared to the NFW-based prediction in the table,
their model-independent
reconstruction of the total mass density $\rho(r)$ 
%%with an extrapolation index of $-3$ 
yields a virial mass of 
$M_{\rm  vir} = (1.4\pm 0.4)\times 10^{15} M_\odot$,
assuming an extrapolation index of -3.}
\end{deluxetable*}

%% Table 6 :: systematic uncertainties

\begin{deluxetable*}{llr}
\tabletypesize{\scriptsize}
\tablecaption{
\label{tab:sys}
 Sources of systematic error and their effects on the determination of
 the halo concentration parameter $c_{\rm vir}$.
} 
\tablewidth{0pt} 
\tablehead{ 
\colhead{Source of error} &
\colhead{Uncertainty range} &
\colhead{Fractional error in $c_{\rm vir}$} 
} 
\startdata 
 Lower color limit for the red sample & $[0.1,0.5]$ & 4\% \\
 Source redshift\tablenotemark{*}   & $z_{s,D}=[0.7,1]$ & 10\% \\
 Mask area correction  & $\nu_{\rm mask}=3\pm 1$ & 7\% \\
 Clustering noise rejection & $\nu_{\rm clust}=4\pm 1$ & 1.5\% \\
 Slope of unlensed number counts\tablenotemark{\dag} & $s=0.22\pm 0.03$ & 3.5\%  \\
 Inner boundary condition & $\kappa_{\rm c}(z_s=1)=0.78^{+0.1}_{-0.25}$ &
 10\% \\ 
 Strong lensing modeling\tablenotemark{\dag\dag}  & B05b, ER & 10\% \\
 Entropy prior & $m=0.5\pm 0.4$ & 10\% \\ 
\enddata
\tablecomments{
 The systematic errors are presented in fraction of $c_{\rm vir}=12.7$
 derived from a joint fit to the ACS-based inner $\kappa$ profile of
 B05b and the Subaru-based $\kappa$ map reconstructed with the MEM+ method.
}
\tablenotetext{*}{Effective source redshift $z_{s,D}$ equivalent to the
 mean distance ratio $\langle D_{ds}/D_s\rangle$ defined by equation
 (\ref{eq:zD}).} 
\tablenotetext{\dag}{$s=d\log N(m)/dm$.} 
\tablenotetext{\dag\dag}{Einstein-radius constraint ($\theta_{\rm
 E}=45\arcsec$ for $z_s=1$) used instead of the ACS inner mass profile
 of B05b.} 
\end{deluxetable*}

\clearpage

%%%%%%%%%%%%%%%%%%%%%%%%%%%%%%%%%%%%%%%%%%%%%%%%%%%%%%%%%%%%%%%%%%%
%%%%%%%%%%%%%%%%%%%%%%%%%%%%%%%%%%%%%%%%%%%%%%%%%%%%%%%%%%%%%%%%%%%
%%%
%%% Figures
%%%
%%%%%%%%%%%%%%%%%%%%%%%%%%%%%%%%%%%%%%%%%%%%%%%%%%%%%%%%%%%%%%%%%%%
%%%%%%%%%%%%%%%%%%%%%%%%%%%%%%%%%%%%%%%%%%%%%%%%%%%%%%%%%%%%%%%%%%%

%% the file, uncomment the \plotone, \plottwo, and
%% If you need a layout that cannot be achieved with \plotone or
%% \plottwo, you can invoke the graphicx package directly with the
%% \includegraphics command or use \plotfiddle. For more information,
%% please see the tutorial on "Using Electronic Art with AASTeX" in the
%% documentation section at the AASTeX Web site,
%% http://www.journals.uchicago.edu/AAS/AASTeX.
%%
%% The examples below also include sample markup for submission of
%% supplemental electronic materials. As always, be sure to check
%% the instructions to authors for the journal you are submitting to
%% for specific submissions guidelines as they vary from

%% This example uses \plotone to include an EPS file scaled to
%% 80% of its natural size with \epsscale. Its caption
%% has been written to indicate that additional figure parts will be
%% available in the electronic journal.

%%%%%%%%%%%%% Figure 1

%%%%%%%%%%%%% Figure 2

%%%%%%%%%%%%% Figure 3

%%%%%%%%%%%%% Figure 4

%%%%%%%%%%%%% Figure 5

%%%%%%%%%%%%% Figure 6

%%%%%%%%%%%%% Figure 7

%%%%%%%%%%%%% Figure 8

%%%%%%%%%%%%% Figure 9

%%%%%%%%%%%%% Figure 10

%%%%%%%%%%%%% Figure 11

%%%%%%%%%%%%% Figure 12

%%%%%%%%%%%%% Figure 13

%%%%%%%%%%%%% Figure 14

%%%%%%%%%%%%% Figure 15

%%%%%%%%%%%%% Figure 16

%%%%%%%%%%%%%%%%%%%%%%%%%%%%%%%%%%%%%%%%%%%%%%%%%%%%%%%%%%%%%%%%%%%
%%%%%%%%%%%%%%%%%%%%%%%%%%%%%%%%%%%%%%%%%%%%%%%%%%%%%%%%%%%%%%%%%%%
%%%
%%% Appendix
%%%
%%%%%%%%%%%%%%%%%%%%%%%%%%%%%%%%%%%%%%%%%%%%%%%%%%%%%%%%%%%%%%%%%%%
%%%%%%%%%%%%%%%%%%%%%%%%%%%%%%%%%%%%%%%%%%%%%%%%%%%%%%%%%%%%%%%%%%%  

\appendix

\section{Discretized Estimator for the Lensing Convergence}
\label{app:kappad}        
  
In this Appendix, we aim to derive an expression 
for the discrete convergence profile  
%in terms of the   
using the  
weak lensing 
aperture densitometry 
$\zeta_{\rm c}(\theta)$
given by 
equation (\ref{eq:zeta}).   
%We then derive the azimuthally averaged convergence, 
%$\kappa(\theta)$,  
%from the cumulative mass estimator,
%$\zeta_{\rm c}(\theta)$.
%%%%
In the continuous limit,
the averaged convergence $\bar\kappa(\theta)$
and the convergence $\kappa(\theta)$ 
are related by
\begin{eqnarray}
\bar{\kappa}(\theta)&=&\frac{2}{\theta^2}\int_0^{\theta}
\!d\ln\theta'\theta'^2\kappa(\theta'),\\
\kappa(\theta) &=&
 \frac{1}{2\theta^2}\frac{d(\theta^2\bar{\kappa})}{d\ln\theta}.
\end{eqnarray}
For a given set of annular radii $\theta_m$
$(m=1,2,...,N)$,
discretized estimators can be written in the following way:
\begin{eqnarray}
\label{eq:avkappa_d}
\bar\kappa_m &\equiv&
\bar{\kappa}(\theta_m)=
\frac{2}{\theta_m^2}\sum_{l=1}^{m-1}
\Delta\ln\theta_l
%\frac{\Delta\theta_m}{\left<\theta_m\right>}
%%% 
\bar\theta_l^2
\kappa(\bar\theta_l),\\
\label{eq:kappa_d}
\kappa_l&\equiv&
\kappa(\bar\theta_l) =
% \frac{ 
%\theta^2_{l+1}\bar\kappa_{l+1}-
%\theta^2_{l}  \bar\kappa_l
%}
%{2\bar\theta_l^2 \Delta\ln\theta_l}
%\equiv \alpha^{l}_2\bar\kappa_{l+1}
%-\alpha^l_1\bar\kappa_l
\alpha^l_2\bar\kappa_{l+1}-\alpha^l_1\bar\kappa_l
\ \ \ \ \ (l=1,2,...,N-1),
\end{eqnarray}
where 
\begin{equation} 
\alpha_1^l = \frac{1}{2\Delta\ln\theta_l} 
\left( 
  \frac{\theta_{l}}{ \overline{\theta}_l }
\right)^2, \, \,  
\alpha_2^l = \frac{1}{2\Delta\ln\theta_l} 
\left(\frac{\theta_{l+1}}{ \overline{\theta}_l }\right)^2,
\end{equation}
with
$\Delta\ln\theta_l \equiv (\theta_{l+1}-\theta_l)/\bar\theta_l$
and $\bar\theta_l$
being the area-weighted center of the $l$th
annulus defined by $\theta_l$ and $\theta_{l+1}$;
in the continuous limit, we have
\begin{eqnarray}
\bar\theta_l
&\equiv& 
2\int_{\theta_l}^{\theta_{l+1}}\!d\theta'\theta'^2/
(\theta_{l+1}^2-\theta_{l}^2)\nonumber\\ 
&=&
\frac{2}{3}
\frac{\theta_{l}^2+\theta_{l+1}^2+\theta_{l}\theta_{l+1}}
{ \theta_{l}+\theta_{l+1} }. 
\end{eqnarray} 
 
The technique of the aperture densitometry
(Fahlman et al. 1994; Clowe et al. 2000) allows us
to measure the azimuthally averaged convergence $\bar\kappa(\theta)$
up to an additive constant $\bar{\kappa}_b$, corresponding to 
the mean convergence in the outer background annulus
with inner and outer radii of $\theta_{\rm inn}$ and $\theta_{\rm out}$,
respectively (\S \ref{subsec:zeta}):
\begin{equation}
\label{eq:z2k}
\bar\kappa(\theta)=
\zeta_{\rm c}(\theta)
+\bar\kappa_b.
\end{equation}
Substituting equation (\ref{eq:z2k}) into equation (\ref{eq:kappa_d})
yields the desired expression 
as
\begin{equation}
\kappa(\theta_l)=\alpha^{l}_2\zeta_{\rm c}(\theta_{l+1})
-\alpha^l_1\zeta_{\rm c}(\theta_l)
+ (\alpha_2^l-\alpha_1^l)\bar\kappa_b.
\end{equation}

\section{The NFW Lens Model}
\label{app:nfw}

The NFW universal density profile 
has a two-parameter functional form as
\begin{eqnarray}
 \rho_{\rm NFW}(r)= \frac{\rho_s}{(r/r_s)(1+r/r_s)^2} \label{eq:nfw}
\end{eqnarray}
where $\rho_s$ is a characteristic inner density, and $r_s$ is a
characteristic inner radius.
The virial properties are related 
%%%
In stead of using $\rho_s$ and $r_s$, 
we introduce for an NFW halo
the virial mass $M_{\rm vir}$
and the concentration parameter,
$c_{\rm vir}\equiv r_{\rm vir}/r_s$, defined as the ratio of the
virial radius $r_{\rm vir}$ to the scale radius.
The virial mass and virial radius are related through the
following equation:
\begin{eqnarray}
\label{eq:mvirrvir}
M_{\rm vir} = \frac{4\pi}{3}\bar{\rho}(z_{\rm vir})\Delta_{\rm
vir}r_{\rm vir}^3, 
\end{eqnarray}
where
$\Delta_{\rm vir}$
is the mean overdensity with respect to the mean cosmic density
$\bar\rho(z_{\rm vir})$
%$\bar{\rho}(z_{\rm vir}) = \Omega_m(z_{\rm vir})\rho_{\rm crit}(z_{\rm vir})$ 
at the virialization epoch $z_{\rm vir}$,
predicted by the dissipationless spherical
tophat collapse model (Peeebles 1980; Eke, Cole, \& Frenk 1996). 
We assume the cluster redshift $z_d$ is
 equal to the cluster virial redshift $z_{\rm vir}$.
We use the following fitting formula in a flat 3-space with cosmological
constant (see Oguri, Taruya, \& Suto 2001):
\begin{eqnarray}
\Delta_{\rm vir} &=& 18 \pi^2 (1+0.4093 \omega_{\rm vir}^{0.9052}), 
\end{eqnarray}
where $\omega_{\rm vir}\equiv 1/\Omega_m(z_{\rm vir})-1$.

%%%
The inner density $\rho_s$ can be then
expressed in terms of other virial properties of the NFW halo:
\begin{equation}
\rho_s = \bar{\rho}(z_{\rm vir}) 
\frac{\Delta_{\rm vir}}{3}\frac{c_{\rm vir}^3}{\ln(1+c_{\rm vir})-
c_{\rm vir}/(1+c_{\rm vir})}.
\end{equation}
%where $\rho_{\rm cr}$ is the
%critical density of the universe at the cluster redshift $z_d$,
%and $r_s$ is the NFW scale radius.
%%%
Hence, for a given cosmological model and a halo virial redshift,
we can specify the NFW model with the halo virial mass
$M_{\rm vir}$ and the halo concentration parameter $c_{\rm vir}$.

For an NFW profile, it is useful to decompose the convergence 
$\kappa(\theta)$ and the
averaged convergence $\bar{\kappa}(\theta)$ as
\begin{eqnarray}
\kappa_{\rm NFW}(x) &=& \frac{b}{2} f(x),\\
\bar{\kappa}_{\rm NFW}(x) &=& \frac{b}{x^2}g(x),
\end{eqnarray}
where $b=4\rho_s r_s/\Sigma_{\rm crit}(z_d,z_s)$ is the dimensionless
 scaling convergence, 
 $x=\theta/(r_s/D_d)$ is the dimensionless angular radius,
and $f(x)$ and $g(x)$ are dimensionless functions. 
We have analytic expressions for $f(x)$ and $g(x)$
as (Bartelmann 1996):
\begin{eqnarray}
\label{eq:nfw_f}
f(x) &=&
\left\{
  \begin{array}{ll}
  \frac{1}{1-x^2}
   \left(
    -1+ \frac{2}{\sqrt{1-x^2}} {\rm arctanh}\sqrt{\frac{1-x}{1+x}}
   \right)  & \ \ \ (x<1), \\
\frac{1}{3} & \ \ \ (x=1), \\
  \frac{1}{x^2-1}
  \left(
   +1- \frac{2}{\sqrt{x^2-1}} \arctan\sqrt{\frac{x-1}{x+1}}  
  \right)  & \ \ \ (x>1),
  \end{array}
\right. \\
%%%
\label{eq:nfw_g}
g(x) &=& \ln\left(\frac{x}{2}\right)
+
\left\{
  \begin{array}{ll}
  \frac{2}{\sqrt{1-x^2}} {\rm arctanh}\sqrt{\frac{1-x}{1+x}}
   & \ \ \ (x<1), \\
 1 & \ \ \ (x=1),\\
  \frac{2}{\sqrt{x^2-1}} \arctan \sqrt{\frac{x-1}{x+1}} 
   & \ \ \ (x>1).
  \end{array}
\right. 
\end{eqnarray}
The tangential shear $\gamma_{+, {\rm NFW}}(\theta)$ is then 
evaluated by
\begin{equation}
\gamma_{+,{\rm NFW}}(\theta) = \bar\kappa_{\rm NFW}(\theta)-\kappa_{\rm
 NFW}(\theta). 
\end{equation}

For a given source redshift $z_s$, the Einstein radius
is then readily calculated by $1=\kappa_{\rm NFW}(\theta)$;
or more explicitly, 
using equation (\ref{eq:nfw_g}) we have
\begin{equation}
\theta_{\rm E}^2 =  b \theta_s^2
g(\theta_{\rm E}/\theta_s),
%\theta_{\rm E}^2 =  b\left(\frac{\theta_{\rm vir}}{c_{\rm vir}}\right)^2
%g(\frac{c_{\rm vir}\theta_{\rm E}}{\theta_{\rm vir}})
\end{equation}
where $\theta_s\equiv r_s/D_d=r_{\rm vir}/(c_{\rm vir}D_d)$ 
is the angular size of the 
NFW scale radius.
This equation for $\theta_{\rm E}$ can be solved
numerically, for example, by the Newton-Raphson method.

%%%%%%%%%%%%%%%%%%%%%%%%%%%%%%%%%%%%%%%%%%%%%%%%%%%%%%%%%%%%%%%%%%%
%%%%%%%%%%%%%%%%%%%%%%%%%%%%%%%%%%%%%%%%%%%%%%%%%%%%%%%%%%%%%%%%%%%
%%%
%%% References 
%%%
%%%%%%%%%%%%%%%%%%%%%%%%%%%%%%%%%%%%%%%%%%%%%%%%%%%%%%%%%%%%%%%%%%%
%%%%%%%%%%%%%%%%%%%%%%%%%%%%%%%%%%%%%%%%%%%%%%%%%%%%%%%%%%%%%%%%%%%

\end{document}